\documentclass[useAMS,usenatbib]{mnras}
\usepackage{epsfig}
\usepackage{amssymb}
\usepackage{upgreek}
\usepackage{subfig}
\title[Anomalous H$\beta$ responses] {Anomalous Broad-Line Region Responses to Continuum Variability in Active Galactic Nuclei. I. H$\upbeta$ Variability}
\author[C. M. Gaskell et al.]{C. Martin Gaskell$^{1}$\thanks{E-mail:
mgaskell@ucsc.edu}, Kayla Bartel$^{1}$, Julia N.~Deffner$^{1,2}$, Iris Xia$^{1,3}$ \\
\\$^1$Department of Astronomy and Astrophysics, University of California, Santa Cruz, CA 95064
\\$^2$Menlo School, 50 Valparaiso Ave, Atherton, CA 94027
\\$^3$Earth System Science Department, Stanford University, Stanford, CA 94305-4216}

\begin{document}

\date{Revised 2021 August 23}

%\pagerange{\pageref{firstpage}--\pageref{lastpage}} \pubyear{2017}

\maketitle

\label{firstpage}

\begin{abstract}

In the standard AGN reverberation-mapping model, variations in broad-line region (BLR) fluxes are predicted from optical continuum variability (taken as a proxy for the ionizing continuum) convolved with a response function that depends on the geometry.  However, it has long been known that BLR variability can deviate from these predictions.  We analyze both extensive long-term H$\upbeta$ and continuum monitoring of NGC~5548 and a large sample of high-quality H$\upbeta$ light curves of other AGNs to investigate the frequency and characteristics of anomalous responses of the BLR.  We find that anomalies are very common and probably occur in every object.  Onsets can be on a timescale only slightly longer than the light-crossing time and durations are of the order of the characteristic timescale of variability of the optical continuum to several times longer. Anomalies are larger when NGC~5548 is in a low state, but otherwise there is no correlation with continuum variability.  There is abundant evidence for the optical continuum of AGNs varying independently of the higher-energy continua and this is sufficient to explain the anomalous responses of the total BLR flux.  There are good reasons for believing that the frequent lack of correlation between spectral regions is due to anisotropic and non-axisymmetric emission.  Rapid changes in line profiles and velocity-dependent lags are consistent with this.  Motion of compact absorbing clouds across the line of sight is another possible cause of anomalies. The prevalence of anomalies should be considered when planning reverberation-mapping campaigns.
\end{abstract} % 246 words


\section{Introduction}

\citet{Lyutyi+Cherepashchuk72} and \citet{Cherepashchuk+Lyutyi73} demonstrated that the lag in the response of the broad-line region (BLR) of active galactic nuclei (AGNs) to changes in the continuum could be used to infer the size of the BLR.  This technique is now called ``reverberation mapping'' \citep{Blandford+McKee82}.  Starting with \citet{Antonucci+Cohen83} and \citet{Ulrich+84} there have now been a large number of monitoring campaigns using modern detectors to study BLR variability.  \citet{Gaskell+Sparke86} introduced the interpolated cross-correlation function method for determining lags from irregularly-sampled data and showed that BLR sizes were much smaller than thought at the time.  The cross-correlation method has now been applied to many hundreds of time series to determine sizes of regions reprocessing radiation in AGNs.  \citet{Gaskell88} and \citet{Koratkar+Gaskell89} used the method to perform velocity-resolved reverberation mapping to show that the BLR is predominantly virialized and hence can be used to estimate black hole masses. The cross-correlation method has been used to determine lags for hot dust IR emitting regions \citep{Clavel+89}, {lags across the optical and near UV continuum \citep{Wanders+97,Collier+98},} lags for extreme ultraviolet (EUV) and X-ray emitting regions \citep{Chiang+00}, and lags for scattering regions producing polarized light \citep{Gaskell+12}.

Fundamental assumptions of reverberation mapping recognized from the outset are that

(i) variations of the observed intensity of reprocessed radiation {at a given location} only depend on the {instantaneous} intensity of the radiation being reprocessed,

(ii) changes in the observed continuum (in the optical, say) are a good proxy for the radiation being reprocessed (ionizing radiation in the UV for, example) and 

(iii) the structure of the emitting and reprocessing regions does not change on the timescale of the observing campaign. 

It has long been recognized that the response of the BLR to continuum changes is not necessarily going to be {linear}.  \citet{Pronik+Chuvaev72} found that the response of the broad H$\upbeta$ line saturates {at high flux levels}, something expected when there are matter-bounded clouds. { \citet{Korista+Goad04} showed that the Pronik-Chuvaev effect could also be reproduced from a distribution of ionization-bounded clouds.  \citet{Wamsteker+Colina86} found a Pronik-Chuvaev effect for C\,IV.} \citet{Sparke93} pointed out that the response {of a line to continuum increases} can be negative in some circumstances.  

As for the second assumption, using the optical {continuum} flux as a proxy for the ionizing flux has also always been recognized as a major limitation.  However, as regards the third assumption, the structure of the emitting and reprocessing regions has generally been thought not to change on the timescale of observing campaigns, with the exception of the BLR getting {slightly} smaller, {as expected}, when an AGN is in a low state {\citep{Wanders95}.  \citet{Cackett+Horne06} refer to this as ``breathing".}

There are some additional assumptions that are not usually explicitly stated:

(iv) the driving radiation is only emitted from what is effectively a point source at the centre of the AGN

(v) the emission of the driving radiation is azimuthally symmetric about the axis of symmetry and

(vi) the reprocessing regions are also azimuthally symmetric (e.g., they have a spherical or disc-like distribution or are in a bi-polar wind).

If all of the above assumptions are correct, the light curve of the reprocessed radiation is the optical light curve convolved with a response function\footnote{As \citet{Uttley+14} point out (see their footnote 1), the signal processing term for the response of a system to a $\updelta$-function is ``impulse function".  {In acoustic and audio applications this is called the ``impulse response" or ``impulse response function" (IRF).} Following \citet{Blandford+McKee82} the {impulse} response function has been commonly called the ``transfer function" in reverberation mapping {literature}, but, as Uttley et al.\@ point out, in signal processing the latter term is used for the Fourier transform of the {IRF.  For clarity we will refer to the IRF as the ``response function."}} which only depends on the geometry. In this paper we show that there are {\em many} anomalous cases where observed line light curves {\em cannot} be explained by convolving the optical light curve with a response function and therefore that {\em one or more of the standard assumptions must break down}.  We first review previous evidence for anomalous BLR responses. We then examine well-observed H$\upbeta$ and continuum light curves of NGC~5548 and a large number of other AGNs to investigate the frequency and duration of anomalous responses.  We argue that anomalies are much more common than has hitherto been recognized.  We discuss the timescales of anomalies and their relationship to continuum variability.  Finally, we consider possible causes of the anomalies in the light of our results. {In Paper 2 \citep{Xia+Gaskell21} we examine the anomalous behaviour of higher-ionization lines.}

\section{Early observations and interpretations}

\subsection{Variability of the total line flux}

{A reason why} \citet{Gaskell+Sparke86} introduced the cross-correlation method to determine lags {was} {\em because observations then available were of relatively poor quality and irregularly sampled.} The goal was simply to determine the lag -- the first moment of the response function -- which gives a responsivity-weighted size of the reprocessing region, or, in the case of velocity-resolved reverberation mapping, gives kinematic information. A lot of early discussion (e.g.,  \citealt{Gaskell+Peterson87}) was about the reliability of lags determined by the cross-correlation method, especially when the time series were irregularly sampled.  Simply determining lags (especially for getting black hole masses) was, and often still is, a main focus of many reverberation-mapping campaigns.  Papers only rarely compared observed line fluxes with what was predicted from continua convolved with response functions.  Instead, a typical reverberation mapping paper showed raw light curves and for analysis showed just the cross-correlation functions as an indication of the reliability of the lag determinations.  

The first comparisons of line fluxes with predictions from continua convolved with a response function were shown by \citet{Krolik+91} (see their Figs.~10 -- 12) for the UV observations of \citet{Clavel+91} of NGC~5548, and by \citet{Horne+91} for the H$\upbeta$ observations of \citet{Peterson+91} of the same AGN.  Although there are some small systematic deviations, the agreement generally seems good.  Discrepancies were tacitly written off as observational errors.

There is an important problem that could not have been appreciated at the time that was giving a false sense of agreement: {\em the response functions used extend too far back in time.}  We now know that the IR emission from the inner wall of hot dust in NGC~5548 has a lag in the range of 40 -- 70 days depending on the level of activity \citep{Koshida+14} so the strong contributions to the response function at times of 150 -- 200 days needed to match the line light curves cannot be real because they would put the BLR beyond the wall of dust.  The times of the spurious peaks in the response functions correspond to the typical times between continuum events.  As \citet{Maoz94} pointed out, there is a sort of aliasing.  The inversion program used to estimate the response function (e.g., by \citealt{Krolik+91} and \citealt{Horne+91}) adds in a response to much earlier events in an effort to improve an imperfect fit.  

Something that makes spurious responses from too early times seem real is that the widely-used maximum entropy inversion method is constrained to produce a positive response function, a problem pointed out by \citet{Krolik+Done95}.  The Subtractive Optimally Localized Averages (SOLA) modification of the Backus Gilbert method \citep{Pijpers+Wanders94} and the regularized linear inversion method of \citet{Krolik+Done95} both allow the response function to go negative.  This gives further freedom in fitting the convolved continuum to line flux observations, but, because it gives a response function that goes negative it is more obvious that features in the response function at large times are not real. 

As spectroscopic monitoring of AGNs improved in quality and quantity, anomalies became more apparent. \citet{Maoz94} gave a valuable critical appraisal of the best reverberation mapping up to that time and pointed out that ``details of emission-line light curves cannot be accurately reproduced with only the simplest assumptions." In particular he showed that, for NGC~5548, the response differed from one continuum event to another.  The explanation \citet{Maoz94} favoured for the anomalous behaviour was variations in longer-wavelength continuum not reflecting variations in the ionizing continuum.

Given these changes from event to event, it is not surprising that the response functions of \citet{Pijpers+Wanders94} for NGC~5548 show significant variations from year to year, something that should not happen if the continuum emission and BLR geometry stay the same.

\subsection{Line profile variability}

\citet{Gaskell88} showed that the red and blue wings of C\,IV {in NGC~4151} varied together, thus excluding the BLR motions being dominated by outflow.  This result has been widely confirmed {for many objects} (see references in \citealt{Gaskell+Goosmann13}), but there have been exceptions.  Notably, from 2007 observations of NGC~3227, \citet{Denney+09b} found the {\em blue} wing of H$\upbeta$ appearing to lead the red wing, thus implying outflow of the H$\upbeta$-emitting BLR.  \citet{Denney+09b} suggested that different AGNs have different BLR kinematics.  However, \citet{Kollatschny+Dietrich96} had earlier shown that during intensive monitoring of NGC~5548 in 1989 (the same period considered by \citealt{Maoz94}) the C\,IV line had shown {\em a change from the blue wing leading to the red wing leading in only 100 days} (see their Fig.~7).  Since it is impossible for the whole BLR to change direction in 100 days, {\em the change in which line wing was leading the other cannot reflect a kinematic change}. This suggests that the apparent outflow \citet{Denney+09b} reported for NGC~3227 is not real and is due instead to a breakdown of reverberation mapping assumptions.  {\citet{Gaskell10} showed how a period of anisotropic continuum emission (i.e., a  breakdown of the assumption of azimuthally-symmetric continuum emission) could readily explain the apparent outflow \citet{Denney+09b} reported.  The non-axisymmetric continuum model predicts that the outflow signature found by \citet{Denney+09b} was temporary. Re-observation of NGC~3227 in 2012 and 2014 \citep{DeRosa+18} showed that the outflow signature of 2007 had indeed disappeared.}

A related result of \citet{Kollatschny+Dietrich96} was that only {\em some} parts of line profiles varied during continuum outbursts and that which part this was changed between outbursts.  A study of the H$\upbeta$ line profile variability of NGC~5548 over 5 years \citep{Wanders+Peterson96} similarly showed that only part of the line profile varies strongly (see their Fig.~5) and which part this is varies from year to year.  In the high-temporal-resolution monitoring of NGC~5548 by \citet{Denney+09b} the velocity range over which H$\upbeta$ varies the most is remarkably narrow -- notice the large sharp spike in the blue wing of the RMS variability spectrum in their Fig.~1c. Even more interestingly, \citet{Sergeev+01} had shown that for another well-monitored AGN, NGC~4151, during each observing season {\em there were one or more very narrow velocity ranges of the line profile that did not vary with the continuum} (see their Fig.~9).  Furthermore, {\em the velocity of these uncorrelated regions varies from year to year.}

\section{The statistics of anomalous behaviours}

To investigate how common anomalous responses of total line fluxes are, we compared the observed H$\upbeta$ fluxes with predictions from the optical continuum light curves for the best-studied AGN, NGC~5548, and a large number of other AGNs.  We chose H$\upbeta$ because it is by far the most widely observed broad line in reverberation-mapping campaigns.

\subsection{Sample}

The best reverberation-mapping results are summarized in the on-line data base described by \citet{Bentz+Katz15}.  We omit objects where the data did not permit a black hole mass estimation and restrict ourselves to cases where the observed H$\upbeta$ and continuum fluxes are readily available in the literature or on-line.  The objects we studied are given in alphabetical order in Table 1 with the reference(s) to the data and the lags derived by the authors.  When authors give multiple estimates of the lag, we used the centroid lags given by the standard interpolated cross-correlation function method of \cite{Gaskell+Sparke86}.  Details of this are given in \cite{Gaskell+Peterson87}.  As noted in Table 1, we have sometimes averaged different lag estimates.

\subsection{Analyses}

We linearly interpolated the optical continuum points and convolved the resulting light curve with a response function.  For the response function we used a simple boxcar centered on the lag derived by the cross-correlation method by the observers, and we adopted a half-width of half the lag.  As noted by \citet{Maoz94}, {and confirmed by our own experiments,} the fits are insensitive to the assumed shape of the response function, and the anomalous responses we find (see below) are far larger than any differences that can be explained by the shape of the response function.  Note that for objects with multiple years of observation, especially NGC~5548, {\em we have taken an average lag} even though different lags can be found for different years and the lag varies, as expected, with the activity level (so-called ``breathing" -- see \citealt{Cackett+Horne06}).  {We have done this for consistency so that different seasons (different panels in the figures) can readily be compared.} The fits on {the shortest timescales (a few days)} can, of course, be improved if the lag is a free parameter, but {\em deviations from what is predicted with a given lag are one of the things we are looking for.} { As will be seen below, such deviations occur in a manner that is not consistent with just the size of the BLR changing with luminosity. For the particularly well-studied case of NGC~5548, we also considered deviations from a luminosity-dependent lag (see Section 4 below).} 

In the absence of earlier observations, the curves showing the interpolated continuum convolved with the response function necessarily begin 1.5 times the lag (i.e., the width of our response function) after the first observed continuum point because we often have no knowledge of what the continuum was doing before the first observation. 

The next step in our analysis was to scale the convolved continuum to try to fit the observed H$\upbeta$ fluxes. For NGC~5548, where the observations cover a wide range of continuum levels and the curvature of the relationship is clear (see Fig.~2 of \citealt{Gilbert+Peterson03}, \citealt{Goad+04} and Fig.~2a of \citealt{Cackett+Horne06}), we used a {fourth-order} polynomial fit {to predict the H$\upbeta$ flux from the continuum flux}.  For other objects, {all of which have much smaller data sets,} the curvature {in the relationship between the H$\upbeta$ flux and the continuum flux } was not obvious, so we simply used a linear relationship.  For these objects our first attempt at scaling was to make the means and variances of the observed line fluxes and the convolved continuum fluxes for the same dates be the same.  The plots were then examined by eye and the slope of the linear scaling was adjusted to try to match the observed line fluxes better {so that there were not systematic differences in high- and/or low-flux states.  Note that while the scaling is arbitrary, it does not affect anomalies {\it at the same flux level} at different times.} We finally calculated the ratios of the observed H$\upbeta$ fluxes to the prediction for the same date.  

{Since the convolved, interpolated continuum for any date is almost always derived from multiple continuum flux measurements, the errors in the ratios are dominated by the errors in the line-flux measurements. The latter are usually given by the observers and are independent from night to night.  Although the errors in the convolved interpolated continua are expected to be small, they can be larger for some time periods when there is sparse sampling or poor data. To estimate the errors in our convolved interpolated continua arising from measurement errors in the continuum fluxes and from our interpolation we adopted the widely-used FR/RSS method (see \citealt{Peterson+98a} for a detailed description.)}

\section{Results}

\subsection{NGC 5548}

The 13-year monitoring of NGC~5548 by the {\it International AGN watch} makes it by far the best-monitored AGN.  For consistency and comparison with other objects, we analyzed NGC~5548 in {a similar manner to how} we analyzed other objects.\footnote{{As noted above, the NGC~5548 data are plentiful enough and cover a wide enough range in luminosity that we could fit the curvature in the relationship between H$\upbeta$ flux and continuum level.  We could not do this for other objects.}} We show the results in Figs.~1 and 2. Figure 1 shows the comparison between observed H$\upbeta$ fluxes and the predictions from the continuum for the whole 13-year period 1989-2001.  Details for each individual observing season can be seen in the pairs of panels in Fig.~2.  The upper panel of each pair shows the observed H$\upbeta$ fluxes (the points) and the predictions from the continuum (the curves {with the errors indicated}).  The black dots at the bottom show the observed continuum for reference.  Note, however, that the observed continuum points have been arbitrary shifted and scaled for display purposes.  The lower panel of each pair shows the ratios of observed H$\upbeta$ line fluxes to the predicted fluxes.  The upper left pair of panels show all 13 years of data; the other pairs show individual years. Note that for the panel labelled ``1989" the observations actually began in December 1988 and similarly for the other panels.  Note also that the ranges of the vertical scales are not always the same.  

It can be seen from Fig.~1 and the upper left panel of Fig.~2 that for the 13-year data set as a whole, H$\upbeta$ generally follows the optical continuum. However, it is clear from Fig.~1 and Fig.~2 that there are deviations on timescales of weeks to months or longer.  The panels for individual years show details of these deviations more clearly.  Only for one season (1995) is there agreement with the prediction to within $\pm 10$\% throughout the season.   Inspection of Figs.~1 and 2 shows many cases where the {\em same} predictions based on the continuum have different observed line fluxes (e.g., as in 1997) or where the observed line fluxes are the same, but the predictions differ (compare, for example, the beginning and end of the 1993 season).

{As noted, we use a fixed lag in Figs.~1 and 2. Allowance for the systematic variation of the lag with luminosity has little effect because the correlation of $\tau$ with luminosity is weak (See Fig.~6 of \citealt{Cackett+Horne06}). The standard deviations from a fixed $\tau$ of 17 days and from the predictions of a luminosity dependent $\tau$ are $\pm 5.6$ and $\pm 4.1$ days respectively. The median absolute deviation of the lags for each year from the fixed mean is $\pm 2.5$ days while from the luminosity-dependent prediction it is $\pm 4.2$ days (i.e., greater).  Year-to-year variations in $\tau$ are thus comparable to the effect of ``breathing".} 

\subsection{Timescale of anomalies in NGC 5548}

{The timescale of anomalies gives important clues to their cause.  We can identify two types of anomalies in the response of the total line flux to continuum changes.  The first is a change in the {\em lag} other than the change predicted by a change in the mean flux (i.e., not due to ``breathing".)  The second is the {\em flux} of a line being higher or lower than predicted from the continuum level.  These are probably related but the two types produce different signatures in our plots of the residuals.}

\subsubsection{Timescale of changes in the lag}

{As discussed (see section 4.1), the actual lag determined for an observing season can differ from the predicted lag by several days. Furthermore, inspection of Fig.~2 shows that the lag can change {\em within} an observing season. For example, in 1995 and 1996 the times of peaks and troughs in the H$\upbeta$ light curve match up well with the predictions.  However, in 1992, while the first trough (around MJD 48740) is on time, the following peak (around MJD 48790) is about 10 days too early as also is the peak after it around MJD 48820.  The lag has thus changed in less than $\sim 50$ days. The following year (1993) the trough at MJD 49100 is late by a week, while the following trough (around MJD 49160) is early by a week. The peak in between these two troughs is on time.  The timing of these 1993 events shows that the lag has again changed within $\sim 50$ days.  It is not possible to attribute these changes to luminosity-dependent changes in the BLR size (``breathing'').  For example, in 1993 there is no reason why the minimum in H$\upbeta$ around MJD 49170 should be a week early compared to the minimum around MJD 49100 when the optical continuum behaviour before each is similar.}

{The determination of a lag depends on variability of the continuum because one needs variation of the continuum to see variation of the line flux.  It is thus not possible to put a lower limit on how rapidly the lag changes because we do not have strong continuum changes on very short timescales.  We can say though, from the timing of peaks and troughs in 1992 and 1993 just discussed, that {\em the lag changes on a timescale of less than $\sim 50$} days.}

\subsubsection{Timescale of changes in the amplitude of the response of H$\upbeta$}

{Residuals for NGC~5548 appear to show short-term anomalies on a timescale of a couple of weeks to a month or two and gradual, longer-term changes on the timescale of a year.  The variability of AGNs has long been approximated as a damped random walk \citep{Gaskell+Peterson87} and this could well be the case for the anomalies as well. A damped random walk is characterized by a damping timescale, $\tau_{damp}$. This timescale can be estimated from $W_a$, the half width at half maximum of the auto-correlation function (see section III of \citealt{Gaskell+Peterson87}.) In Fig.~3 we show the auto-correlation function (ACF) of the residuals of the observed H$\upbeta$ fluxes from the predictions of a breathing model (model B2 of \citealt{Cackett+Horne06}). This ACF is calculated using the interpolated correlation function of \citet{Gaskell+Sparke86}.  From this we get $W_a \sim 46$ days.}

{An alternative way to characterize the variability timescale is to look at the lengths of well-observed individual anomalies.  This is more subjective because of limitations of sampling (gaps in the time series) and observational errors.  It is also highly likely that some or many of the longest duration anomalies are actually back-to-back or overlapping shorter duration anomalies.  Inspection of the essentially perfect sampling during the 2014 {\it AGNSTORM} campaign supports this. For example, looking at Fig.~9 of \citet{Pei+17} shows that, starting around MJD 6738, there is a rapid 5\% H$\upbeta$ anomaly lasting 18 days if one uses their adopted lag.  A series of short-timescale anomalies can be seen to follow during the {\it AGNSTORM} monitoring.  The structure of these would have largely been missed with the poorer sampling in the data we consider in our Figs.~ 1 and 2.}  

{In Fig.~4 we show a distribution of estimated durations of anomalies.  These were determined as follows.  The residuals of the H$\beta$ fluxes from the predictions of breathing model B2 of \citet{Cackett+Horne06} were smoothed with a boxcar function of 10-days width to minimize the effects of observational errors. To get estimates of the durations of events we determined the time intervals from each minimum in the residuals to the following maximum and from each maximum to the following minimum.  There are important caveats in interpreting the distribution in Fig.~4.  The first is that because of the necessity of smoothing the data to get round the problem of observational noise, we will be missing genuine events of duration less than about two weeks.  A second problem is the unavoidable subjectivity in deciding whether a long event is indeed a long event, or multiple short events.  Despite these clear limitations, the characteristic timescale of events indicated by inspection of Fig.~4 is consistent with the timescale inferred from the ACF.  This timescale (about 20-70 days) is similar to the timescale of the 2014 so-called``holiday" of NGC~5548 \citep{Goad+16,Pei+17}.}

{ Because we have a homogenized data set covering many observing seasons, NGC~5548 is the best case for looking for long-term anomalies.  The residuals for all 13 years (see the upper left panels at the start of Fig.~2) suggest that there is long-term variation in the H$\upbeta$ anomalies (i.e., the departures from the predictions are not white noise).   There is evidence for slow variations on the timescale of an observing season or so with some years showing a gradual linear trend over a year. For example, there are clear downward trends in the ratio of observed-to-predicted H$\upbeta$ fluxes in 1990 and 1991, and there are strong upward trends in 1993 and 1994.  At the start of the 1989 observing season the H$\upbeta$ flux was high relative to our predictions and it stayed high into the 1990 season.  It was thus high for somewhat more than a year.  The ACF in Fig.~3 shows, however, that there is no significant correlation for the whole data set on a timescale of longer than $\pm$ a year.}

{To summarize,it is quite likely for NGC~5548 that anomalies occur on all timescales from a couple of weeks to at least a year with a prefered timescale in the range 20-70 days, but there is no clear evidence yet for anomalies lasting more than a few years.}

\subsection{Frequency of occurrence of anomalies.}

{Inspection of Fig.~2 shows that we are seeing rapid anomalous behaviour (minima or maxima in the residuals) at least four or five times per observing season.  A couple of important caveats are needed here.  Firstly, it is clear that {\em a good fraction of rapid anomalous events must be being missed because of the limited sampling}  and, secondly, it is likely that some of the longest-duration anomalies are actually back-to-back or overlapping shorter duration anomalies.}

{There are 62 anomalies in Fig.~4 with an average duration of 34 days.  This adds up to 5.8 years during the 13 observing seasons.  Since the effective duration of good coverage during observing seasons is around half a year, we get that important result that {\em ``anomalies" are occurring all the time.}}

\subsection{The amplitude of anomalies in NGC~5548 and the relationship to continuum variability.}

{Fig.~5 shows the distribution of the absolute values of the amplitudes of the anomalies shown in Fig.~4 (data treatment as described in 4.2.2) .  It can be seen that the amplitudes of the anomalies are generally less than 20\%.  There are, for example, no large, factor-of-two anomalies.  The amplitude of the 2014 anomaly (\citealt{Goad+16,Pei+17} about 6\% -- see Figure 12c of \citealt{Pei+17}) is quite typical of the earlier anomalies that can be seen in Fig.~2.  Indeed, many of the anomalies in the 13 years of monitoring considered here are of {\em larger} amplitude than the 2014 one. Thus the latter} was neither unique nor unprecedented for NGC~5548. {It was merely particularly obvious because of the high quality of the {\it AGNSTORM} data.}

Interestingly, {for NGC~5548, the H$\upbeta$} anomalies do not seem to be correlated with events in the continuum in any obvious way.  {We find no correlations whatsoever between the durations of short-term anomalies shown in Fig.~3 with the flux level of the AGN at the time of the anomaly or with the H$\upbeta$ lag for the year in which the anomaly occurs.  The durations of the anomalies  are similar from year to year.} The only correlation we found is between the size of the anomaly and the level of activity.  {We show this in Fig.~6}  We show the absolute value of the logarithm of the residual because the logarithms of the residuals are symmetric about zero by construction (i.e., because of the calibration between mean line intensity and mean continuum flux).  Although anomalies are present at all activity levels, they are clearly larger when NGC~5548 is in a low state.  We checked whether this could be an artifact of the relative errors in the line fluxes being larger when NGC~5548 was fainter and did not find a significant effect.

\subsection{Other AGNs}

We show our results for other AGNs in Fig.~7.  The panels are organized as in Fig.~2, but note that the scales on the time axes differ significantly from object to object. For some {objects} the plots cover a number of years while for others it might be only a month or so.  For each pair of panels we have given the lag we adopted.  Whether this is an average of multiple studies is stated in Table 1.  The scaling between the convolved continuum and the H$\upbeta$ fluxes is not as well established for these other AGNs as for NGC~5548.  To interpret each panel of ratios it is necessary to look at the fit to the light curve directly above it to see if changes in the ratio are a consequence of the choice of scaling of the predicted flux from the continuum observations.  Scaling issues can be ruled out as the cause of anomalies when the residuals have opposite signs at similar line or continuum flux levels.

It can be seen from Fig.~7 that for some objects, such as WAS 61 and for the first period of monitoring of 3C~273 (= PG~1226+023), the agreement between the observed line fluxes with the predictions of simple theory is good and there are no systematic deviations over the period of monitoring. However, most other cases show systematic deviations (and 3C~273 shows them for the second monitoring period).  The most common changes in the ratio of observed-to-predicted fluxes are slow gradual changes during the course of the an observing season.  {The monitoring period is, unfortunately, often shorter that the timescale of these changes.}  As for NGC~5548 though, there are many cases where the ratio changes relatively abruptly or rapidly.  Again as for NGC~5548, the amplitude of the anomalies in the other AGNs is almost always less than 20\%.

\section{Possible Causes of Anomalies}

{We have shown for NGC~5548, and probably for other AGNs too, that what we have been calling ``anomalies" {\em are actually happening all the time}.\footnote{{Because of this ``anomaly" is probably not the best word to describe what is going on.}}  This is an important clue to their origin.  Because they are going on all the time, most of the anomalies are not caused by unlikely, one-of-a kind events.  We consider here various possible causes.}

\subsection{Instrumental effects?}

It is possible that abrupt changes are due to changes in instrumental setup. Most of the studies produced very homogeneous data sets (same telescope, same spectrograph settings) and in multi-observatory campaigns inter-telescope differences have been carefully calibrated out by the original authors.  Although some calibration issues might remain, we do not think they are the cause of the anomalies.

\subsection{Analysis artifacts?}

The precise shape of the changes in the ratio does obviously depend on our choice of scaling of the convolved continuum to the observed line fluxes.  For NGC~5548 there is a very extensive data set that permits a good determination of the average relationship between the H$\upbeta$ flux and the continuum flux, but for many other objects the relationship is uncertain and we have limited temporal coverage.  Overestimating or underestimating the amplitude of variations will produce changes in the ratio that are correlated with the light curves of AGN.  Nevertheless, it can be seen by inspecting pairs of panels in Figs.~2 and 4 that for most AGNs with non-constant ratios of residuals in the lower panels, {\em changing the scaling of the predictions will not make the discrepancies go away}.  This is obviously the case when there are different residuals at the same flux level.

\subsection{Emission from jets?}

We now turn to possible causes due to the AGNs themselves.  The observed continuum of a blazar is dominated by relativistically-beamed emission from a jet aimed close to our line of sight.  Therefore, {in this case,} much of the variable radiation we see would not impinge on the BLR.  3C 273 shows some blazar-like characteristics (see, for example \citealt{Ghisellini+10}).  We could therefore expect there to be times when we are seeing the variability of the jet and not of the accretion disc and corona.  However, we find that anomalous BLR responses are so common that blazar-like activity cannot be the explanation in general since most AGNs are not blazars.

\subsection{Independent variability of optical and high-energy continua}

It has always been recognized that using the optical flux as a proxy for the ionizing flux is a weak point in reverberation mapping.  \citet{Koratkar+Gaskell89} found an anomalous response of C\,IV compared with the $\lambda$1346 continuum and found that the C\,IV line variability could be explained by including the X-ray variability which did not track the UV variability.  The X-ray flux of AGNs is frequently unrelated on short timescales to the UV and optical flux.  For example, multi-wavelength monitoring of NGC~4151 \citep{Edelson+96} shows a powerful flare in the 1 -- 2 keV X-rays that is also seen at $\lambda$1370.  It has no effect in the $V$-band, however.  For NGC~3516 the 2--10 keV X-rays are impressively uncorrelated with the optical \citep{Maoz+02}.  For 3C\,390.3 \citet{Gaskell06} shows simultaneous events around JD 2449800 in the 0.1 -- 2 keV soft X-rays and in the UV at $\lambda$1370 but no major flare in the optical.  At JD 2449975 there is a soft X-ray flare with no counterpart in the UV or optical.  Then at JD 2449950 there is a strong flare in the UV that is not seen in the optical (but which might be followed by an X-ray flare 5 days later.)  

These multi-wavelength monitoring campaigns and others confirm that optical variability is a poor proxy for variability of the ionizing continuum {(see, for example, Figure 3 of \citealt{Gaskell06})}.  We therefore believe that {\em the often poor correlation between optical continuum variability and ionizing-continuum variability is the most likely cause of anomalous BLR responses.}  {Possible causes of the independent behaviour of X-ray and UV/optical continua have been considered as explanations of the 2014 anomaly during the AGNSTORM campaign. For example, \citet{Mathur+17} and \citet{Sun+18} proposed that the 2014 anomaly was due to a change in Comptonization.}  

{As noted, an important consequence of our finding that anomalies are very common  is that they are in general {\em not} caused by rare, special events. A more common explanation is needed.  This explanation must not only explain deviations in the total line fluxes of broad lines (our main focus here) but, as discussed in our introduction, changes in line profiles and velocity-dependent lags on similar timescales have also be explained.  We consider three possibilities below.}

\subsection{Anisotropic continuum emission}

\citet{Gaskell06} suggest that because continuum variability is so rapid, bulk relativistic or near-relativistic motions must be involved even in non-blazars and the associated emission will naturally be anisotropic.  This will lead to components of emission varying independently because different parts of reprocessing regions will be excited by different events (see Fig.~5 of \citealt{Gaskell06}).  Since the low-ionization BLR responsible for H$\upbeta$ emission is a flattened disc (see \citealt{Gaskell09}), different parts of the line profile come from different parts of this disc.  Anisotropic continuum emission means that the BLR disc is not illuminated uniformly and hence different parts of the line profile will show different correlations with the optical continuum as discussed above.  As we note in Section 2.2 above, this is commonly observed to be the case.

\subsection{Off-axis continuum emission}

Although, on average, energy generation from an accretion disc is strongest at small radii \citep{Lynden-Bell69}, it cannot peak exactly in the centre because this is where the black hole is.  Variability has to take place at least a few Schwarzschild radii away.  \citet{Gaskell08} argued that strong UV variability without soft X-ray or optical variability (for example, what happened around JD 2449950 in 3C\,390.3) requires that the source of the variable emission be off-axis (see cartoon in Fig.~5b of \citealt{Gaskell11}).  He pointed out that this would cause anomalous BLR responses. \citet{Gaskell10} and \citet{Gaskell11} show how off-axis variability naturally explains changes in BLR line profiles, their correlation or lack of correlation with continuum variability, and changes {over time} in kinematic signatures in velocity-resolved reverberation mapping. These changes are generally to be expected to be on the timescale of continuum variability (i.e., the time a region remains active). {This is because if continuum variability is due to active regions turning off and on, the timescale of continuum variability, gives the timescale of regions turning off and on.  If the various anomalies are also due to this turning off and on, then {\em we predict that the timescale of the anomalies will be similar to the timescale of variability.} The characteristic timescale of variability can be found from UV and optical light curves.  For NGC~5548 \citet{Collier+Peterson01} get this timescale to be $40^{+18}_{-12}$ days. This is in good agreement with the characteristic timescale of H$\upbeta$ anomalies in NGC~5548 we find here (see Figs.~3 \& 4).}  \citet{Gaskell10} and \citet{Goosmann+14} discuss how off-axis variability additionally explains the velocity and time dependence of polarization of the BLR.

\subsection{Absorbing clouds}

In general, anomalous BLR responses require departures from axial symmetry.  In addition to anisotropic and non-axisymmetric emission just discussed, another possible explanation of BLR anomalies is patchy obscuration \citep{Gaskell+Harrington18}.  This is most likely to happen in the AGNS most inclined our line of sight.  {When the obscuration is not affecting both the continuum and the BLR (see Fig.~4 of \citealt{Gaskell+Harrington18}), an anomalous response will be observed.  If it is the observer's line-of-sight to the continuum that is obscured, most the H$\upbeta$ emission will appear anomalously strong; if the BLR is obscured most, the H$\upbeta$ will appear anomalously weak.} Because the obscuration needs to cross our line of sight, the patches will only cause anomalies on timescales of months to years. \citet{Gaskell+Harrington18} show how such  patchy obscuration can readily explain the changes \citet{Pei+17} found  in the velocity-dependent lags of H$\upbeta$ in NGC~5548 over only a few months.  In Fig.~5 of \citet{Gaskell+Harrington18} it can be seen the the passage of patches of obscuration across the BLR change the average lag of H$\upbeta$.

\section{Limitations of reverberation mapping}

Anomalous BLR responses, are a major source of error in estimating BLR sizes. {A short-period anomaly before or after a dip or peak in a light curve shifts the estimated lag.  A large longer-timescale can give a spurious lag, especially when the light curve is undersampled. An anomalously strong H$\upbeta$ can be misinterpreted as a delayed line response to continuum variability.}  PG 2130 (see Fig.~7) provides a good illustration of this.  The continuum is in a high state around JD 2449700.  H$\upbeta$ is in a high state around JD 2449900, about 200 days later.  However, with the hindsight of knowing from \citet{Grier+08,Grier+12} that the true lag is probably $\thickapprox 15$ days, one can notice in Fig.~7 how the many rapid changes in the H$\upbeta$ flux of PG 2130+099 match similar changes in the continuum with the shorter lag.

Because of anomalous BLR responses, {\em getting better BLR {effective} radii is not a matter of getting better sampling.}  A short intensive campaign might give an apparently accurate lag for H$\upbeta$, but another campaign at a later time could give a different lag.  The abnormally short NGC~5548 H$\upbeta$ lag found in the 2014 \citep{Pei+17} is a good illustration.  If the aim of a monitoring program is to get reliable lag for a large emitting region (such as that producing H$\upbeta$), as opposed to determining smaller lags such as continuum lags or the lag of He\,II, the campaign needs to be longer than the typical duration of anomalies.  For a typical AGN with a lag of a week to a month (such as the majority of AGNs in Table 1) one needs observations covering a couple of years.

The cross-correlation method should not be used blindly for determining lags and their associated errors in lags. It is important to plot observed line variability with predicted variability, as we have done here, to spot anomalous behavior.

\section{Conclusions}

We find that anomalous BLR responses are common events found in the majority of reverberation-mapped AGNs.  ``Anomalies" are the rule rather than the exception.  This shows that the standard assumption of the optical continuum being a good proxy for a driving, central ionizing continuum is not a good assumption.

Mechanisms to explain anomalous responses of BLRs need to explain how common anomalies are.  All deviations of the total line flux, such as those shown here, can be explained by the ionizing radiation varying relatively independently {of the optical continuum}, especially on the timescale of typical optical variability, but variable obscuration is another {possible factor}. 

The evidence from changes in response functions, line profile variability, and velocity-resolved reverberation mapping points to anisotropic and/or off-axis continuum variability as the cause of the most rapid anomalous behaviour.  Compact absorbing clouds crossing our line of sight can also be a cause of changes.
 
The ubiquity of anomalous BLR responses is a major limitation to the reliability of reverberation-mapping campaigns studying total lines fluxes or parts of line profiles (velocity-resolved reverberation mapping).  For the H$\upbeta$ line, denser sampling will not {necessarily} lead to better results.  Instead, for obtaining the most accurate {effective sizes of regions} it is important that monitoring campaigns be longer that the timescale of typical anomalies.  For typical bright Seyferts this means more than one year. The cross-correlation method of determining lags should not be used blindly. The observed line fluxes should always be compared with the continuum light curve convolved with a response function and anomalous responses noted.

\section*{Acknowledgments}

KB, JND and IX carried out their work under the auspices of the Science Internship Program (SIP) of the University of California at Santa Cruz.  We wish to express our appreciation to Raja GuhaThakurta for his excellent leadership of the SIP program.  We are grateful to the anonymous referee for thoughtful and helpful comments that have improved the clarity of the paper and to Ski Antonucci and Jerry Kriss for useful discussions.

\section*{Data availability}

References to the papers containing the data used in this paper are given in Table 1.

%%%%%%%%%%%%%%%%%%%%%%%%%%%%%%%%%%%%%%%%%%%%%%%%%%%%%%%%%%%%%%%%%%%%%%%%%%%
%%%%%%%%%%%%%%%%%%%%%%%%%%%%%%%%%%%%%%%%%%%%%%%%%%%%%%%%%%%%%%%%%%%%%%%%%%%

\newpage
\clearpage

\begin{table*}%[p]
\centering
\caption{Light Curves Analyzed}
 %\begin{supertabular}{l l c c c l} 
\begin{tabular}{l l c l l l}											
\hline											
Object	&	References	&	$\tau_{cen}$	&		Notes	\\%&	Who?	\\
\hline	
1RXS J1858+4850	&	\citet{Pei+14}	&	13.53	&			\\%&	Iris	\\
3C 120 (1)	&	\citet{Peterson+98b}	&	30.63	&	Lag averaged	\\%&	Kayla	\\
3C 120 (2)	&	\citet{Kollatschny+14}	&	30.63	&			\\%&	Kayla	\\
%3C 120 (3)	&	\citet{Grier+12}	&	~	&		&		&	Kayla	\\
3C 390.3	&	\citet{Dietrich+98}	&	20	&			\\%&	Kayla	\\
Akn 120	&	\citet{Peterson+98b}	&	51.4	&		Lag averaged	\\%&	Kayla	\\
Akn 564	&	\citet{Shemmer+01}	&	30	&			\\%&	Kayla	\\
Fairall 9	&	\citet{Santos-Lleo+97}	&	23	&		\\%&	Kayla	\\
Mrk 6	&	\citet{Doroshenko+12}	&	15.8	&		Lag averaged	\\%&	Kayla	\\
Mrk 79	&	\citet{Peterson+98b}	&	13.7	&		Lag averaged	\\%&	Kayla	\\
Mrk 110	&	\citet{Peterson+98b}	&	25.33	&		Lag averaged	\\%&	Kayla	\\
Mrk 142	&	\citet{Du+14}	&	2.9	&		Lag averaged	\\%&	Julia	\\
%Mrk 279	&	\citet{Santos-Lleó+01}	&	16.7	&		\\%&	Julia	\\
Mrk 279 &   \citet{Santos-Lleo+01}  & 16.7 & \\%& Julia \\
Mrk 290	&	\citet{Denney+10}	&	8.72	&		\\%&	Iris	\\
Mrk 335	&	\citet{Peterson+98b}	&	14.6	&		Lag averaged	\\%&	Kayla	\\
Mrk 486	&	\citet{Wang+14}	&	20	&			\\%&	Iris	\\
Mrk 493	&	\citet{Wang+14}	&	12.2	&		\\%&	Iris	\\
Mrk 509	&	\citet{Carone+96}	&	80	&		\\%&	Julia	\\
Mrk 590	&	\citet{Peterson+98b}	&	23.18	&		Lag averaged	\\%&	Kayla	\\
Mrk 817	&	\citet{Denney+10}	&	20.5	&		Lag averaged	\\%&	Iris	\\
Mrk 1044	&	\citet{Wang+14}	&	4.8	&		Lag averaged	\\%&	Kayla	\\
NGC 3516	&	\citet{Denney+10}	&	11.7	&			\\%&	Julia	\\
NGC 3783	&	\citet{Stirpe+94}	&	8	&			\\%&	Julia	\\
NGC 4051 (1)	&	\citet{Peterson+00}	&	2	&			\\%&	Iris	\\
NGC 4051 (2) & \citet{Denney+09a} & 2 & \\
NGC 4151	&	\citet{Kaspi+96}	&	2	&			\\%&	Iris	\\
NGC 4593	&	\citet{Denney+06}	&	4.0	&		Lag averaged	\\%&	Julia	\\
NGC 5273	&	\citet{Bentz+14}	&	2.21	&		\\%&	Iris	\\
NGC 5548	&	\citet{Peterson+91}	&	17.83	&	Lag averaged	\\%&	Iris	\\
 & \citet{Peterson+92} \\
 & \citet{Peterson+94} \\
  & \citet{Korista+95} \\
  &\cite{Peterson+99} \\
  & \citet{Peterson+02} \\
NGC 7469 (1)	&	\citet{Collier+98}	&	8.1	&		\\%&	Iris	\\
NGC 7469 (2)    &   \citet{Peterson+14} & 8.1 &  \\
PG 0026+129	&	\citet{Kaspi+00}	&	111	&			\\%&	Kayla	\\
PG 0804+761	&	\citet{Kaspi+00}	&	146.9	&	\\%&		&	Kayla	\\
PG 0953+414	&	\citet{Kaspi+00}	&	150.1	&		\\%&	Kayla	\\
PG 1211+143	&	\citet{Kaspi+00}	&	103	&		\\%&	Julia	\\
PG 1226+023 (3C 273) (1)	&	\citet{Kaspi+00}	&	306.8	&		\\%&	Julia	\\
PG 1226+023 (3C 273) (2)	&  \citet{Xiong+17}  &	146.8	&			\\%&	Julia	\\
& \citet{Zhang+19}  \\	
PG 1229+204	&	\citet{Kaspi+00}	&	37.8	&			\\%&	Julia	\\
PG 1307+085	&	\citet{Kaspi+00}	&	105.6	&			\\%&	Iris	\\
PG 1411+442	&	\citet{Kaspi+00}	&	124.3	&		\\%&	Iris	\\
PG 1426+015	&	\citet{Kaspi+00}	&	95	&			\\%&	Iris	\\
PG 1613+658	&	\citet{Kaspi+00}	&	40.1	&		\\%&	Iris	\\
PG 1617+175	&	\citet{Kaspi+00}	&	71.5	&			\\%&	Iris	\\
PG 1700+518	&	\citet{Kaspi+00}	&	251.8	&		\\%&	Iris	\\
PG 2130+099 (1)	&	\citet{Kaspi+00}	&	16.25	&			\\%&	Iris	\\
PG 2130+099 (2)	&	\citet{Grier+08}	&	22.9	&		\\%&	Iris	\\
PG 2130+099 (3)	&	\citet{Grier+12}	&	9.6	&			\\%&	Iris	\\
PG0052+251	&	\citet{Kaspi+00}	&	89.9	&			\\%&	Kayla	\\
SDSS J1139+3355	&	\citet{Rafter+13}	&	5.4	&		\\%&	Julia	\\
WAS 61	&	\citet{Du+14}	&	11.4	&		\\%&	Julia	\\
Zw 229-015	&	\citet{Barth+11}	&	3.86	&		\\%&	Iris	\\
 \hline
 \end{tabular}
\end{table*}

\newpage
\clearpage

\begin{figure*} % FIgure 1
\centering
\subfloat{\includegraphics[width=.40\textwidth]{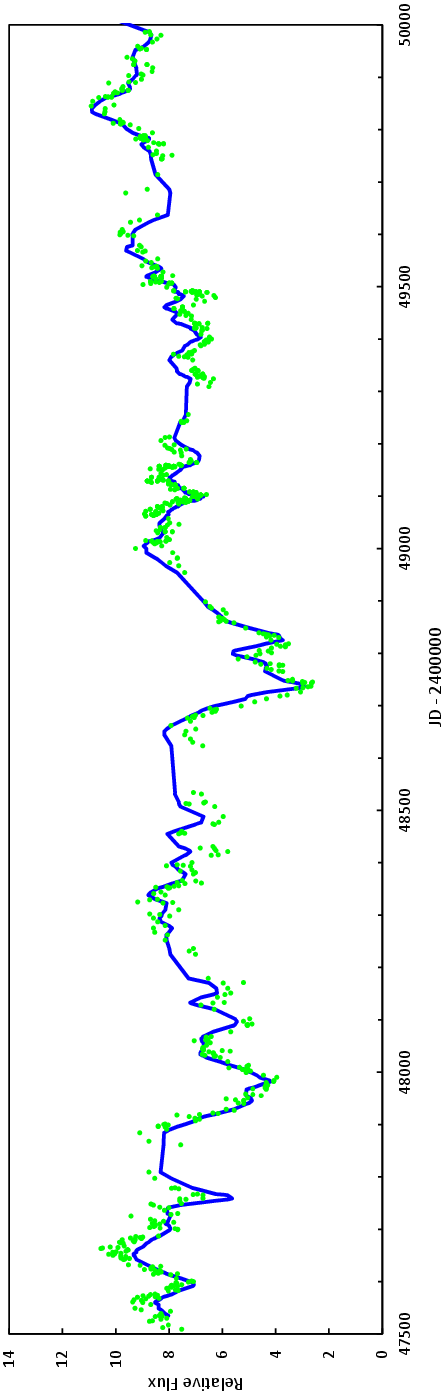}}\quad
\subfloat{\includegraphics[width=.40\textwidth]{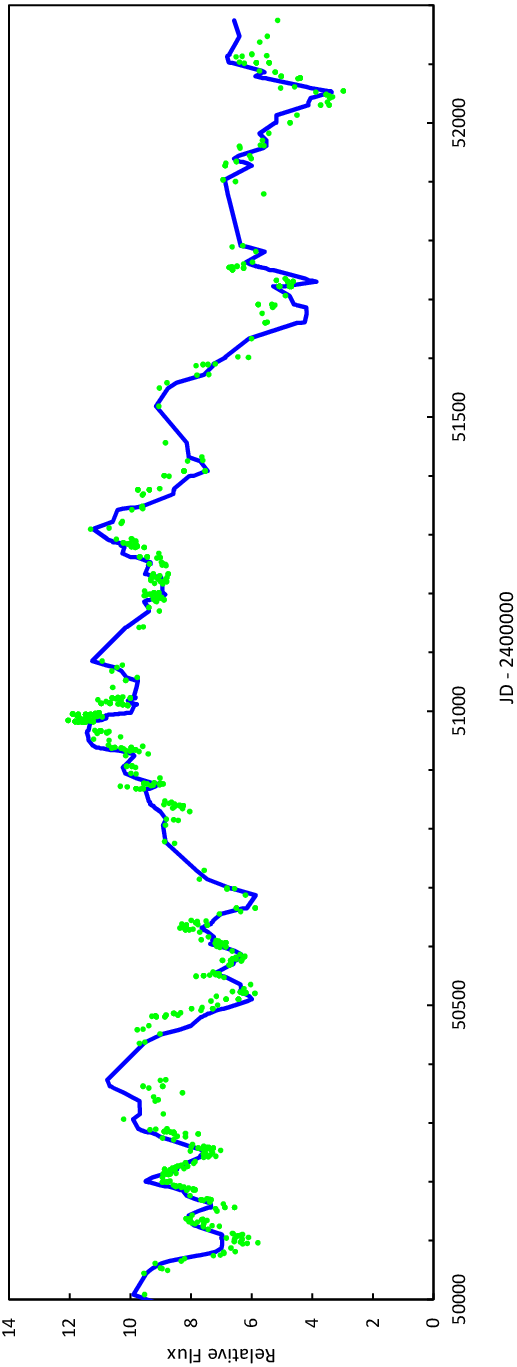}}\quad
\caption{Observed H$\upbeta$ fluxes  (points) for NGC 5548 from 1988 to 2001 versus the prediction from the optical continuum (curve).  See text for details.}
\end{figure*}
% This figures is from

\newpage
\clearpage
\begin{figure*} % Figure 2 - Part 1
   \centering
   \subfloat{\includegraphics[width=.45\textwidth]{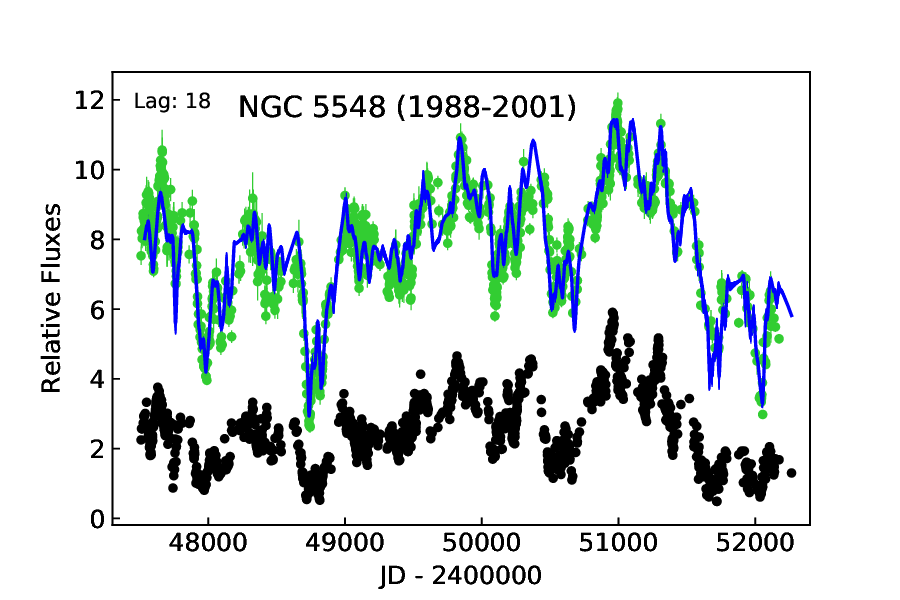}}\quad
   %data started on December 1988 so first year labeled 1989
   \subfloat{\includegraphics[width=.45\textwidth]{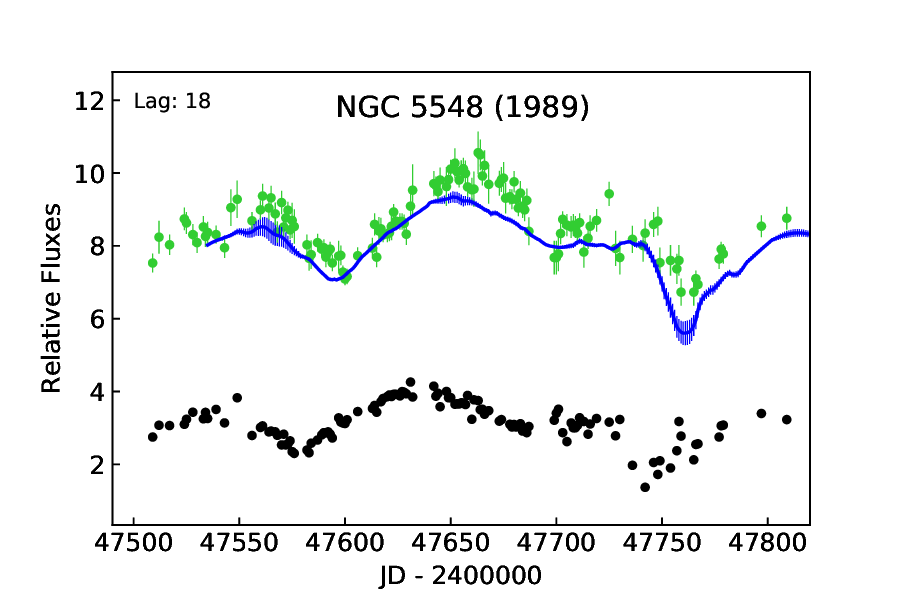}}\quad
   \subfloat{\includegraphics[width=.45\textwidth]{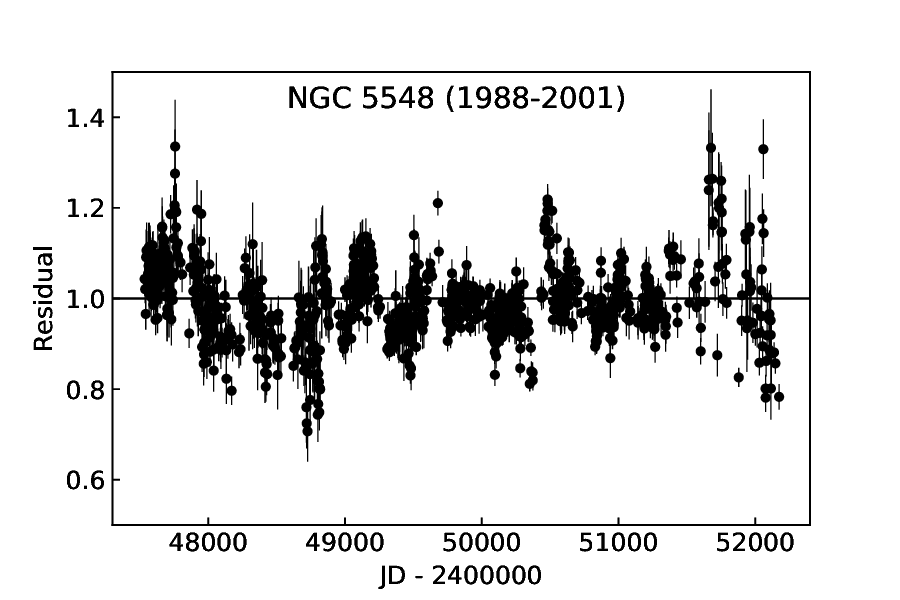}}\quad
   \subfloat{\includegraphics[width=.45\textwidth]{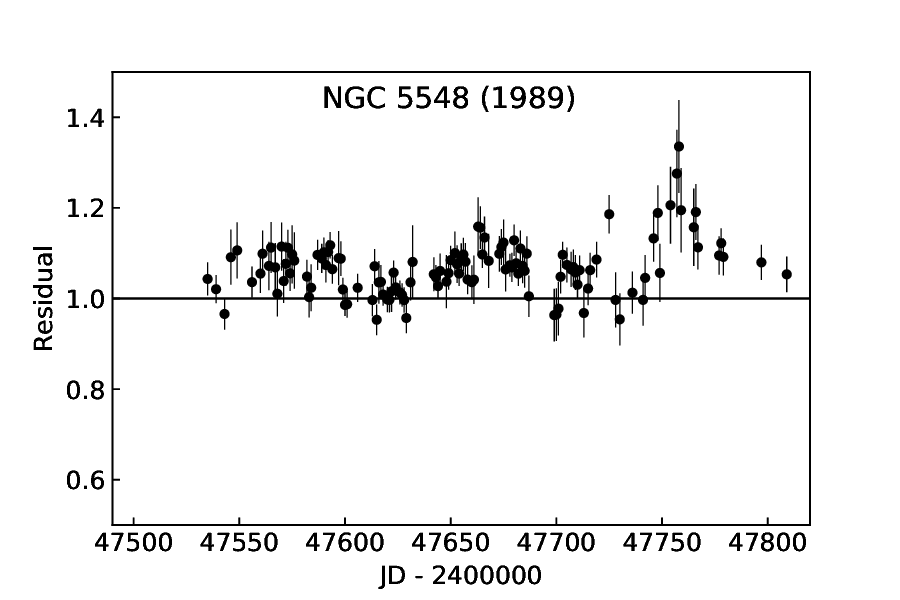}}\quad
   \subfloat{\includegraphics[width=.45\textwidth]{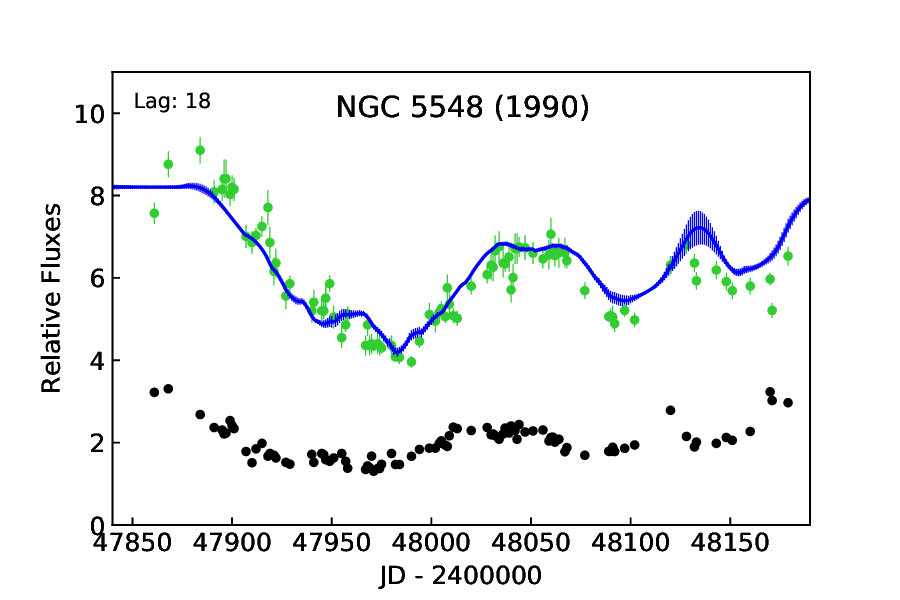}}\quad
   \subfloat{\includegraphics[width=.45\textwidth]{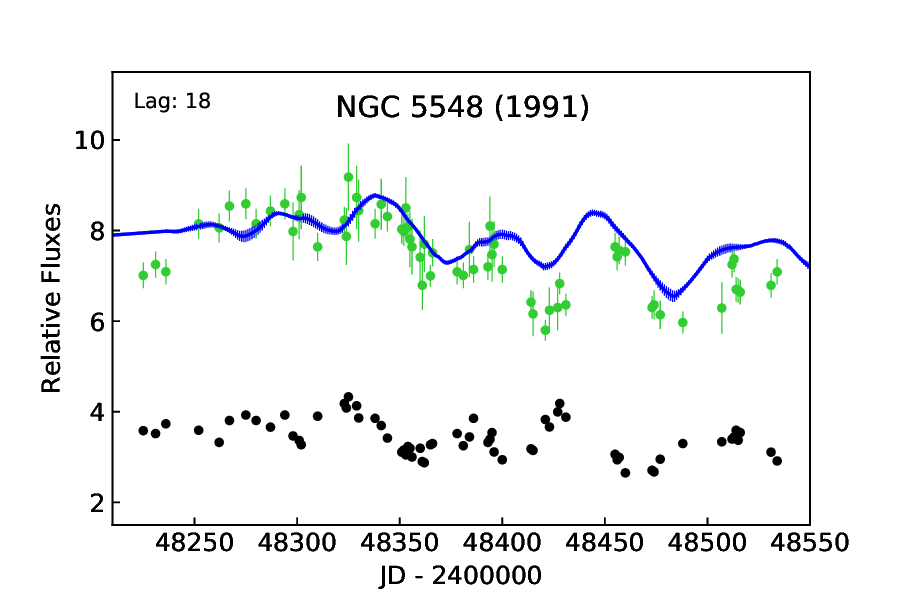}}\quad
   \subfloat{\includegraphics[width=.45\textwidth]{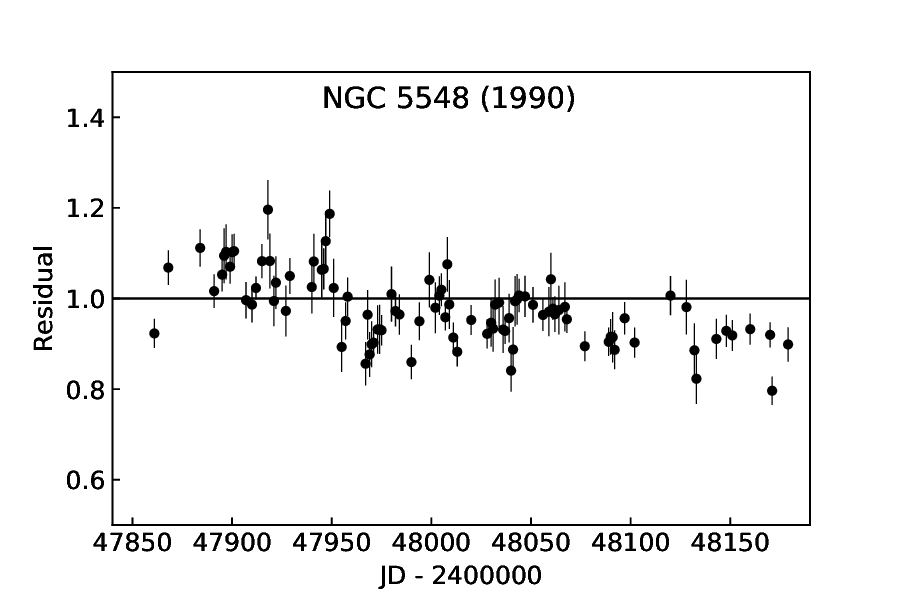}}\quad
   \subfloat{\includegraphics[width=.45\textwidth]{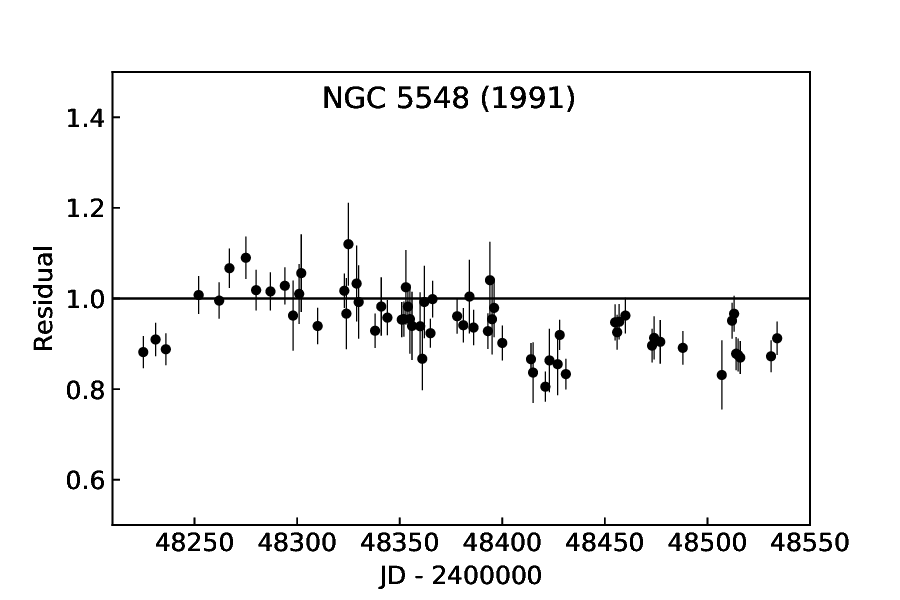}}\quad
   \caption{Optical continuum ($\uplambda$5100) and H$\upbeta$ flux variability for NGC~5548 for the years indicated (upper panels).  Observed H$\upbeta$ fluxes are shown in green; predictions from continuum light curves are shown as blue curves. Observed continuum fluxes with an arbitrary scaling and offset are shown as black dots at the bottom of each upper panel. Residuals (as a ratio) are shown in the lower panels.  See text for details.}
   \end{figure*}

\newpage
\clearpage
\begin{figure*}  % Figure 2 - Part 2
   \setcounter{figure}{1}
   \centering
   \subfloat{\includegraphics[width=.45\textwidth]{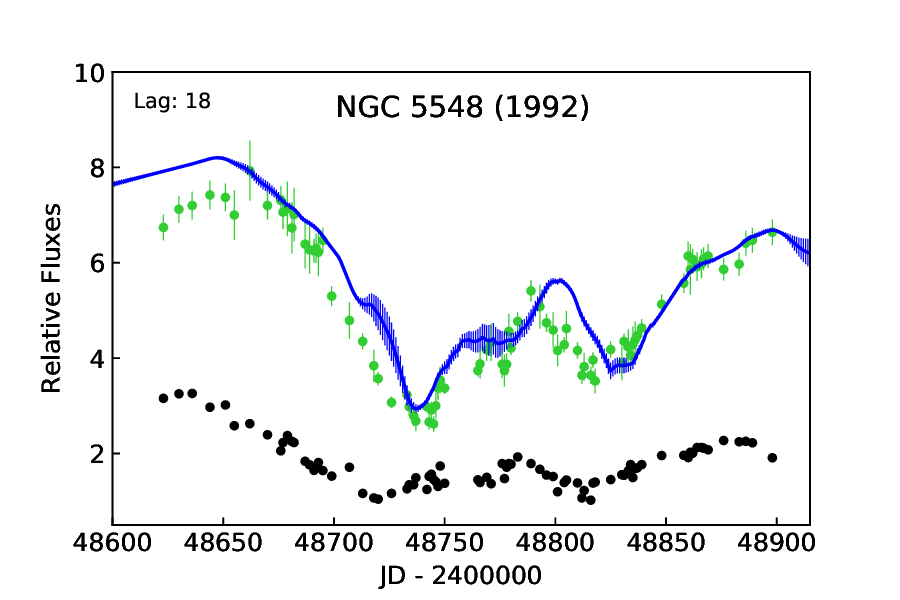}}\quad
   \subfloat{\includegraphics[width=.45\textwidth]{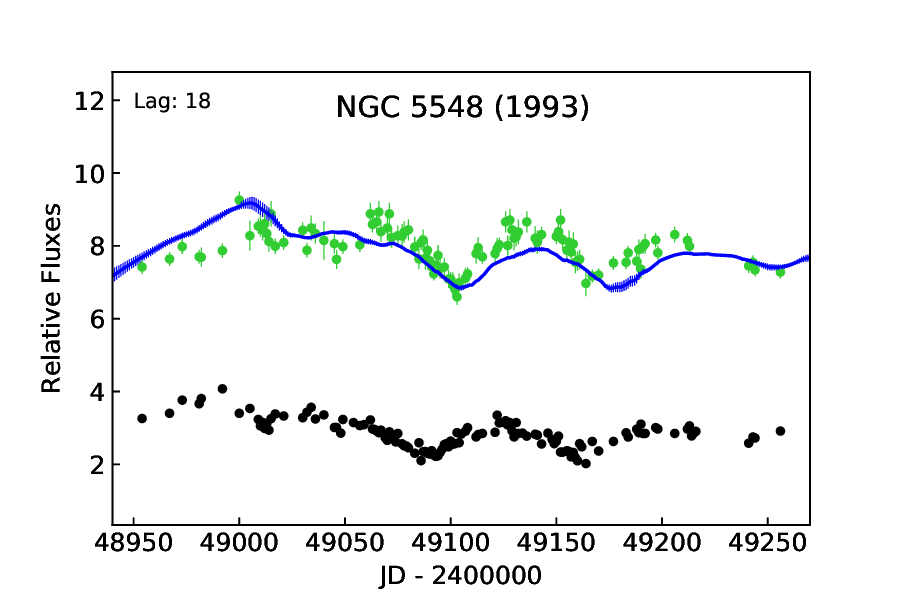}}\quad
   \subfloat{\includegraphics[width=.45\textwidth]{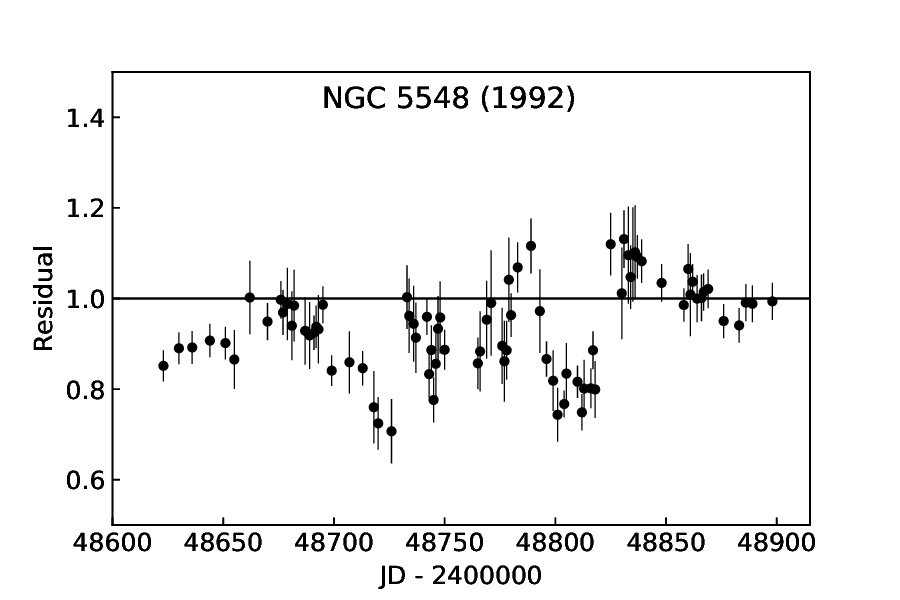}}\quad  
   \subfloat{\includegraphics[width=.45\textwidth]{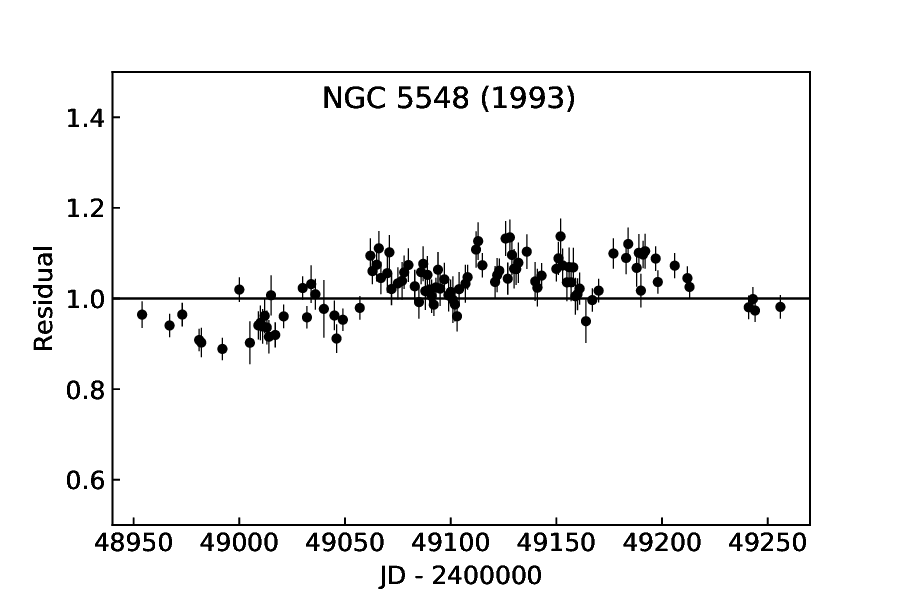}}\quad
   \subfloat{\includegraphics[width=.45\textwidth]{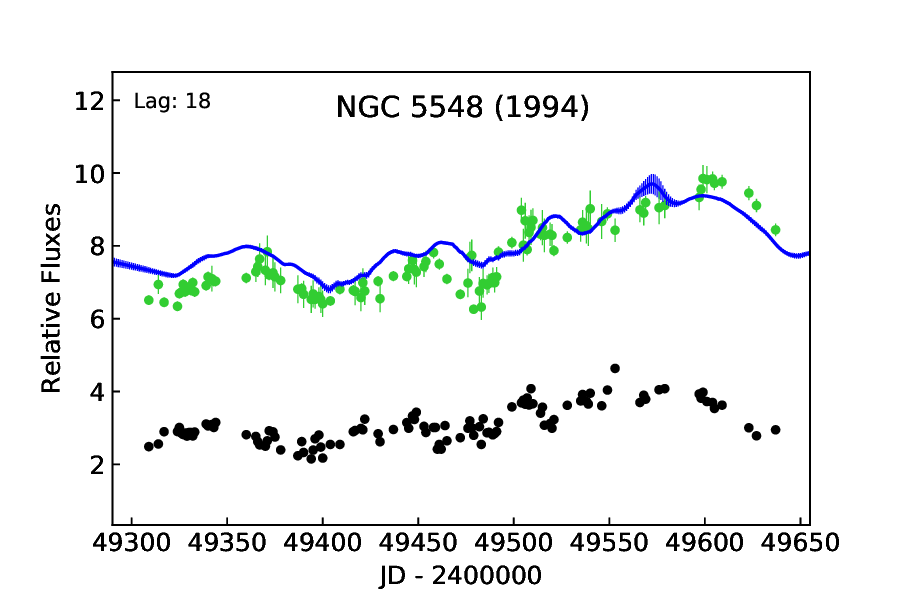}}\quad
   \subfloat{\includegraphics[width=.45\textwidth]{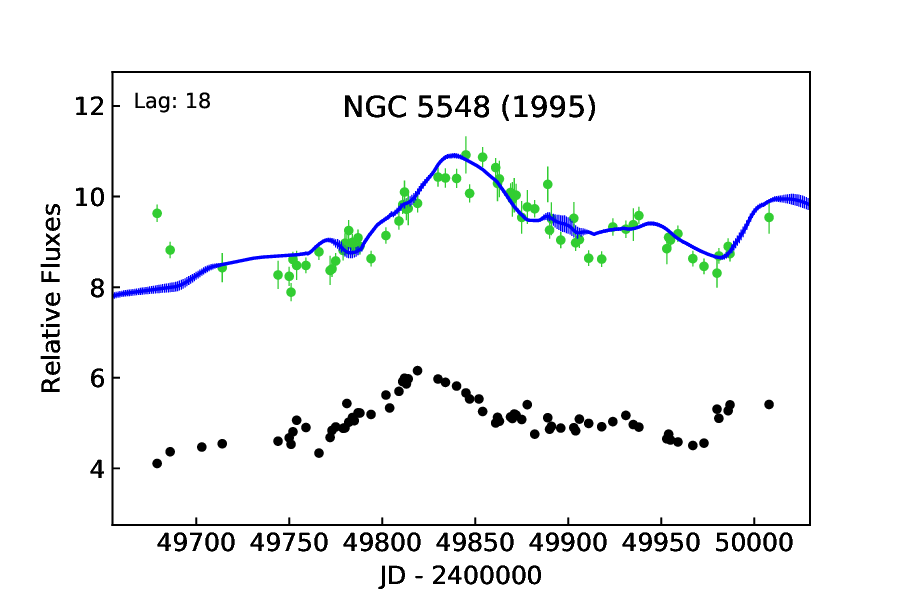}}\quad
   \subfloat{\includegraphics[width=.45\textwidth]{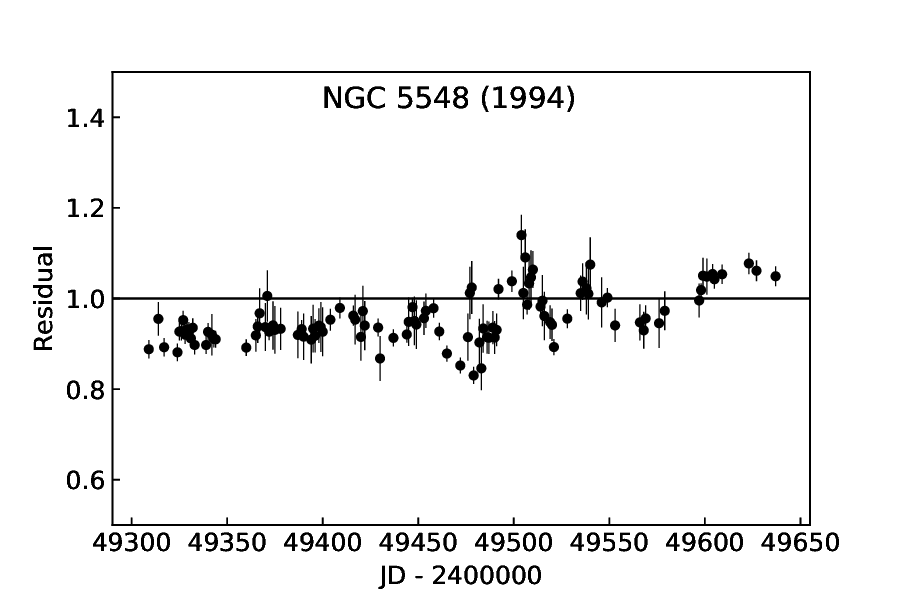}}\quad
   \subfloat{\includegraphics[width=.45\textwidth]{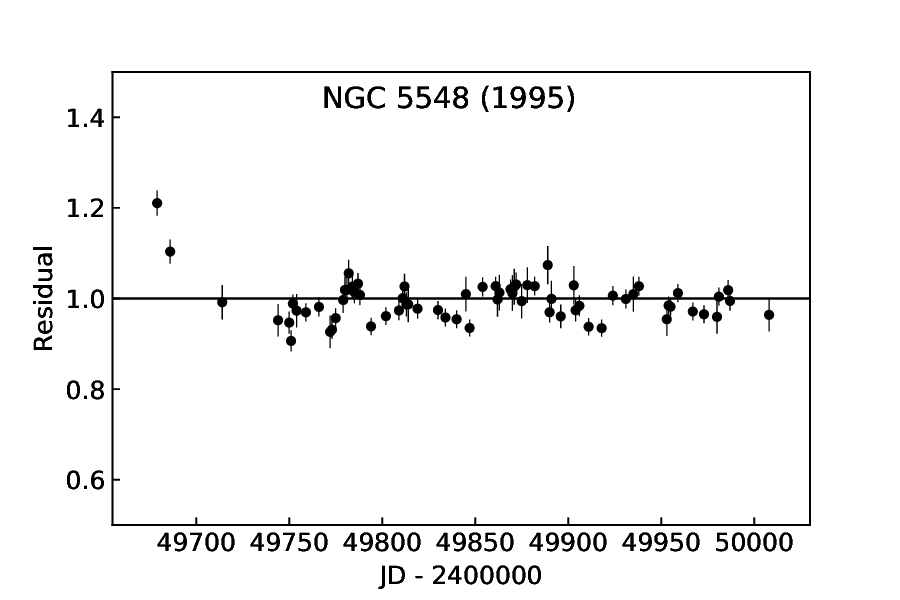}}\quad
   \caption{-- Continued.}
 \end{figure*} 

\newpage
\clearpage
\begin{figure*}  % Figure 2 - Part 3
   \setcounter{figure}{1}
   \centering
   \subfloat{\includegraphics[width=.45\textwidth]{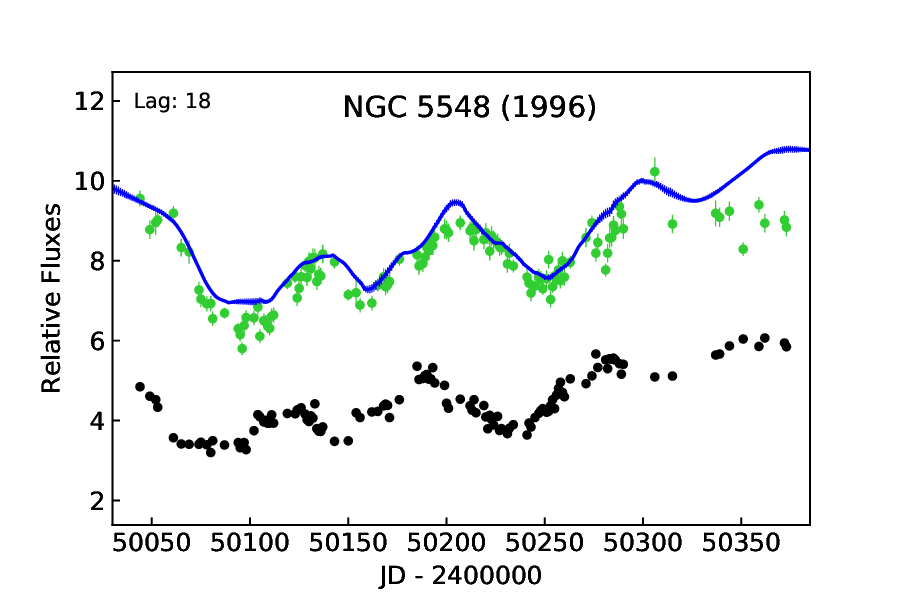}}\quad
   \subfloat{\includegraphics[width=.45\textwidth]{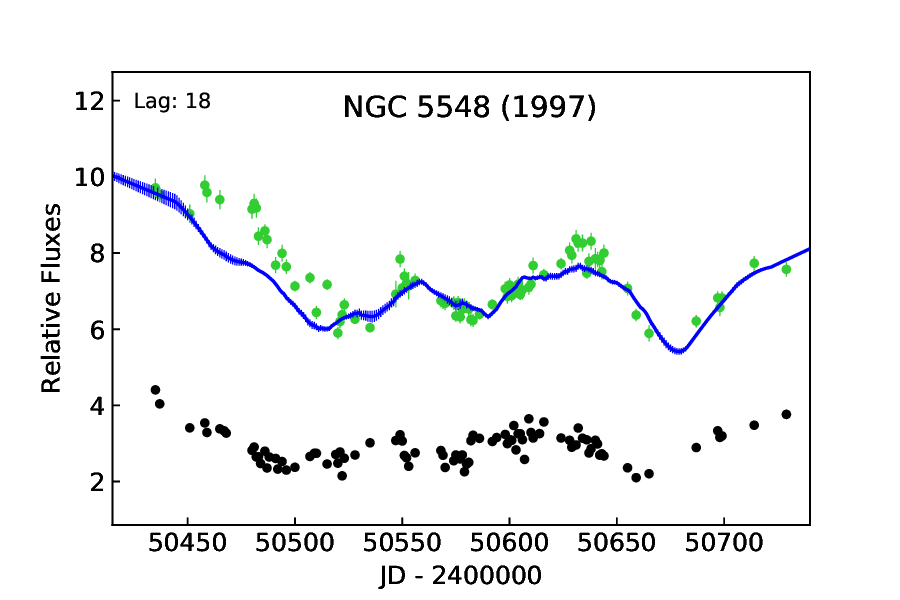}}\quad
   \subfloat{\includegraphics[width=.45\textwidth]{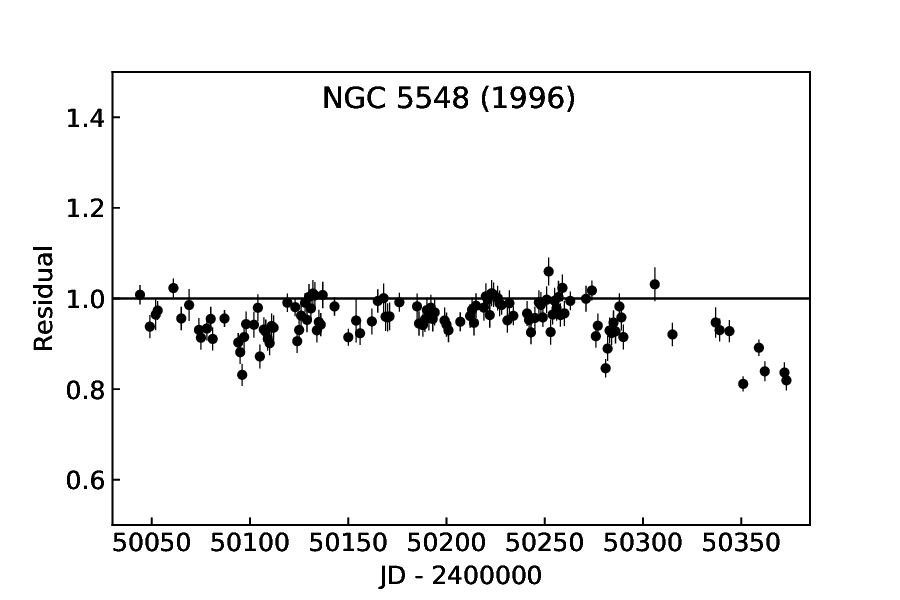}}\quad \subfloat{\includegraphics[width=.45\textwidth]{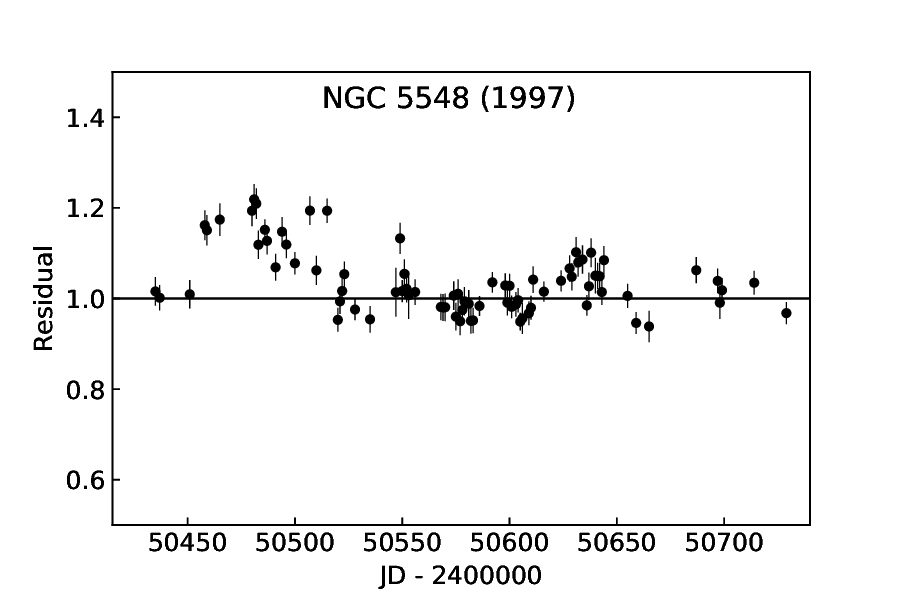}}\quad
   \subfloat{\includegraphics[width=.45\textwidth]{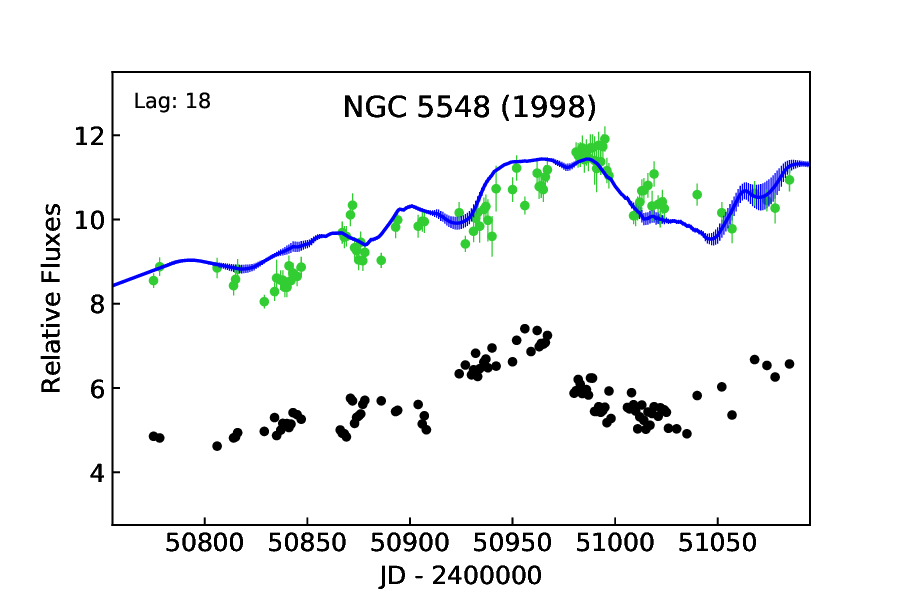}}\quad
   \subfloat{\includegraphics[width=.45\textwidth]{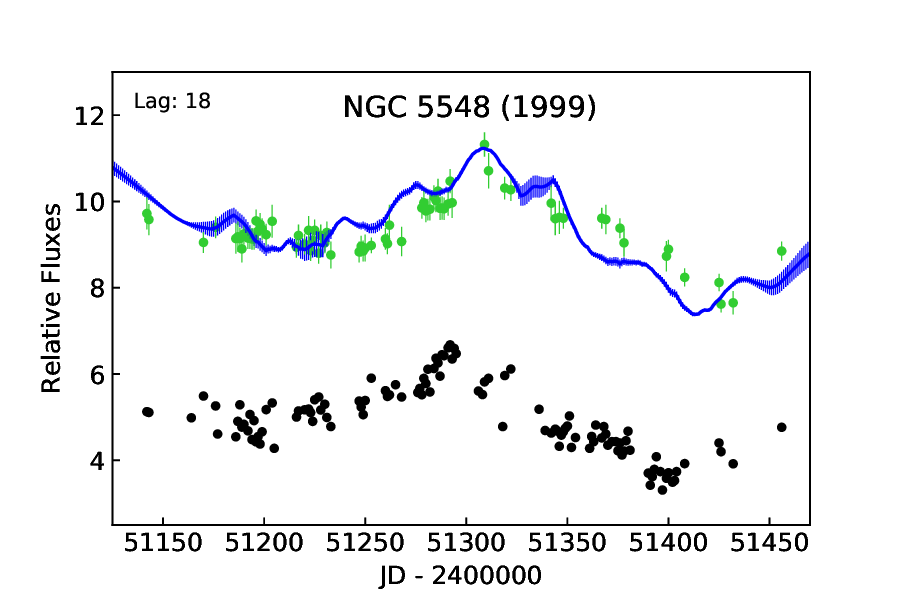}}\quad
   \subfloat{\includegraphics[width=.45\textwidth]{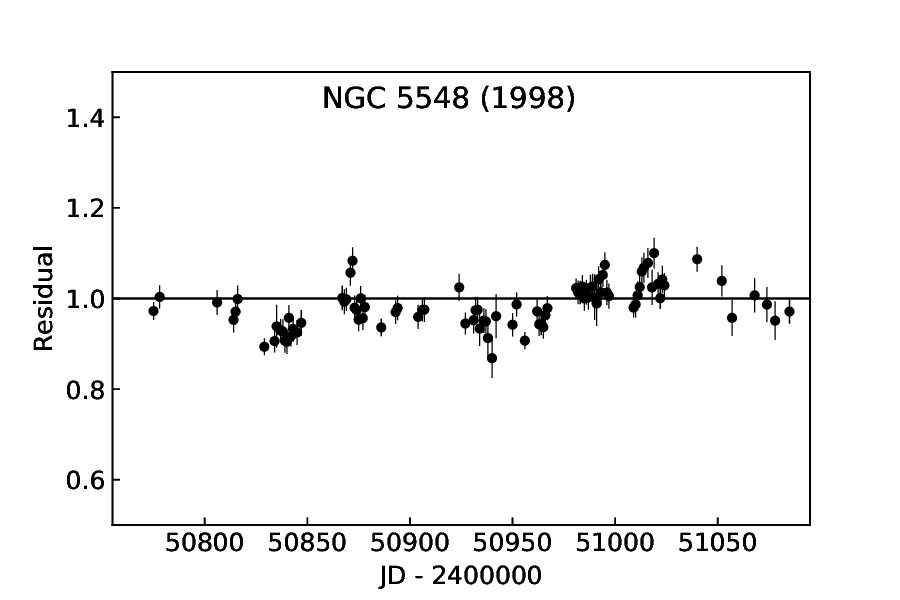}}\quad
   \subfloat{\includegraphics[width=.45\textwidth]{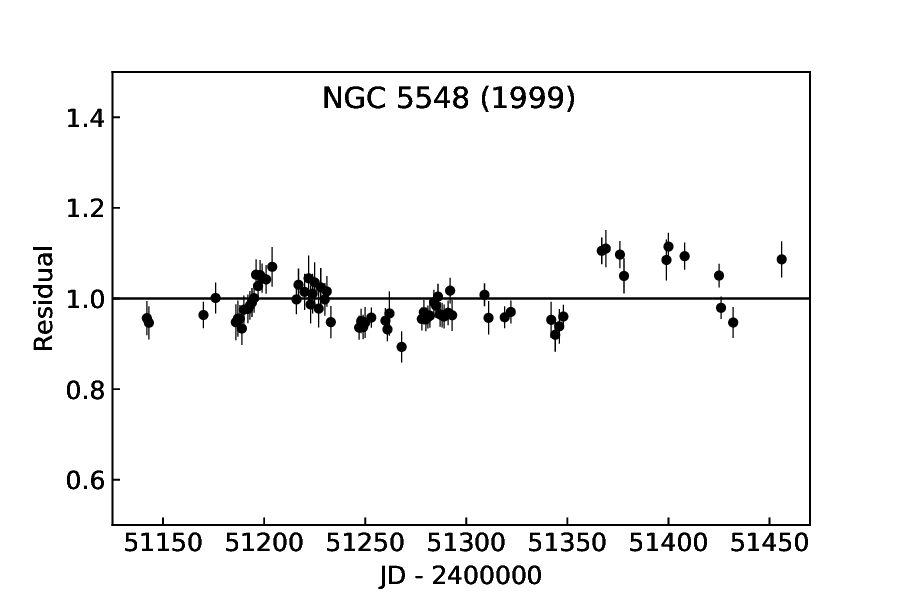}}\quad
   \caption{-- Continued.}
 \end{figure*} 

\newpage
\clearpage
\begin{figure*}   % Figure 2 - Part 4
   \setcounter{figure}{1}
   \centering
   \subfloat{\includegraphics[width=.45\textwidth]{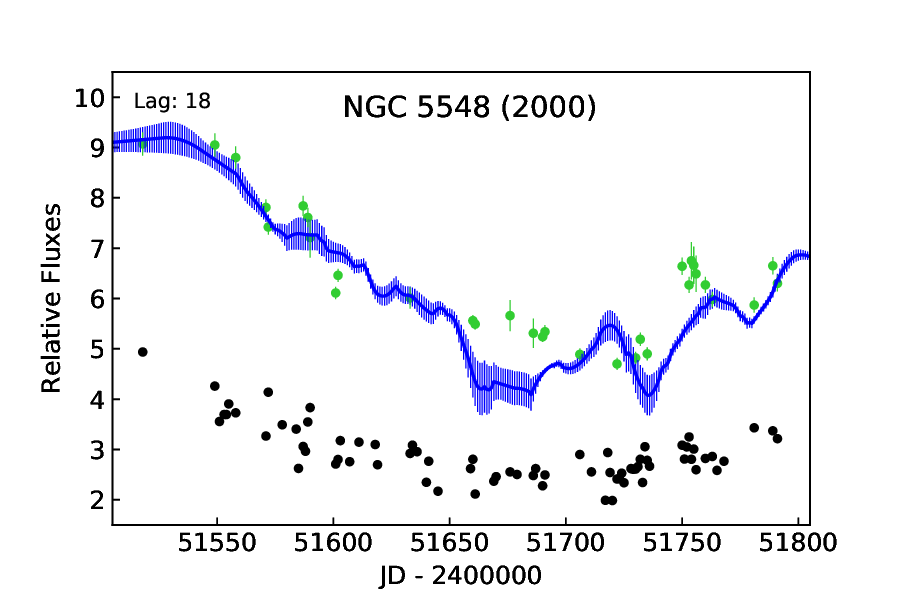}}\quad
   \subfloat{\includegraphics[width=.45\textwidth]{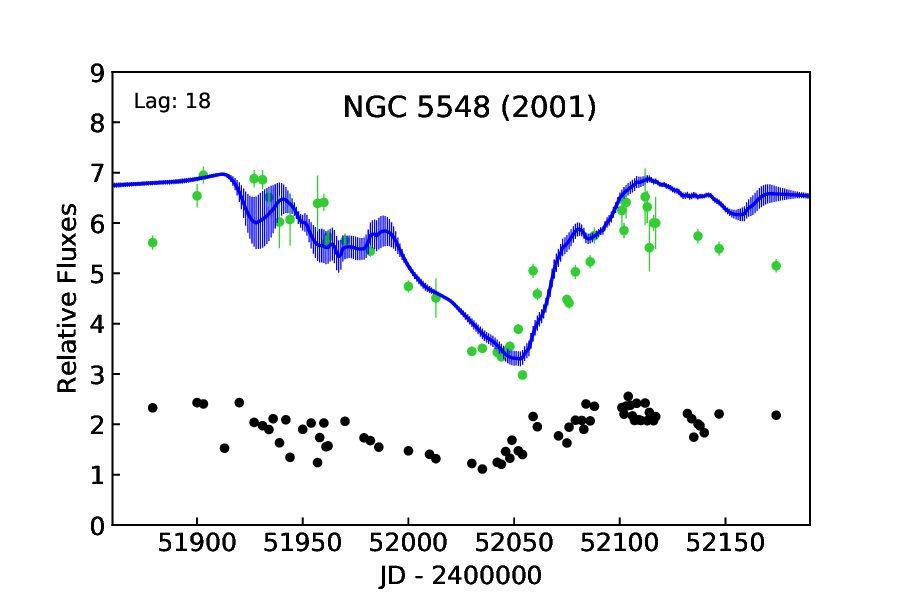}}\quad
   \subfloat{\includegraphics[width=.45\textwidth]{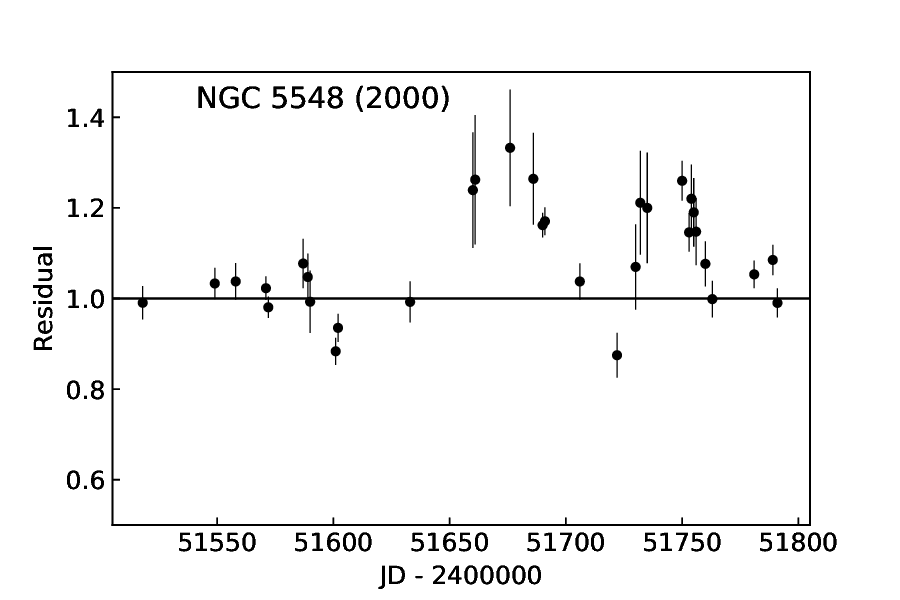}}\quad
   \subfloat{\includegraphics[width=.45\textwidth]{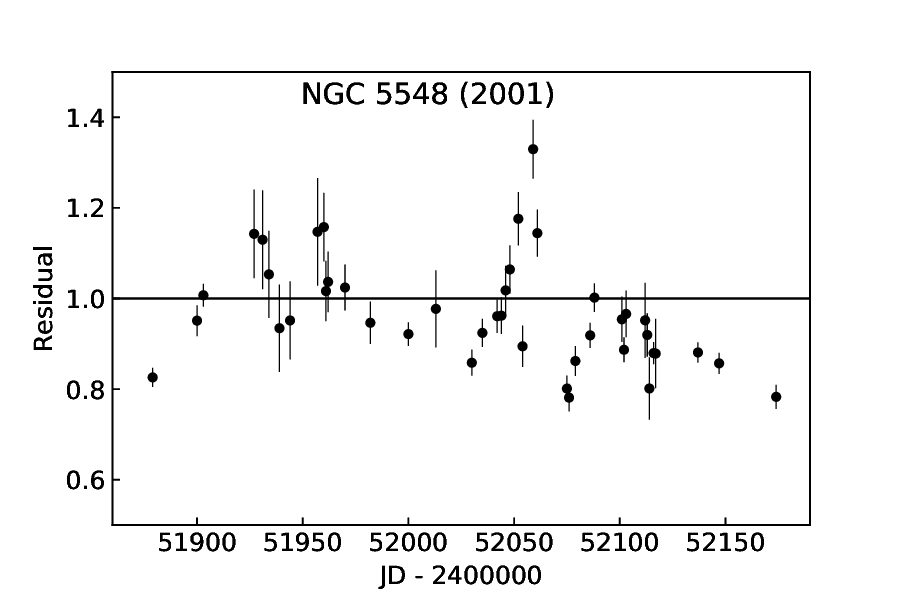}}\quad
   \caption{-- Concluded.}
 \end{figure*} 

\newpage
\clearpage    
\begin{figure*} % Figure 3
   \centering
   \includegraphics[width=.5\textwidth]{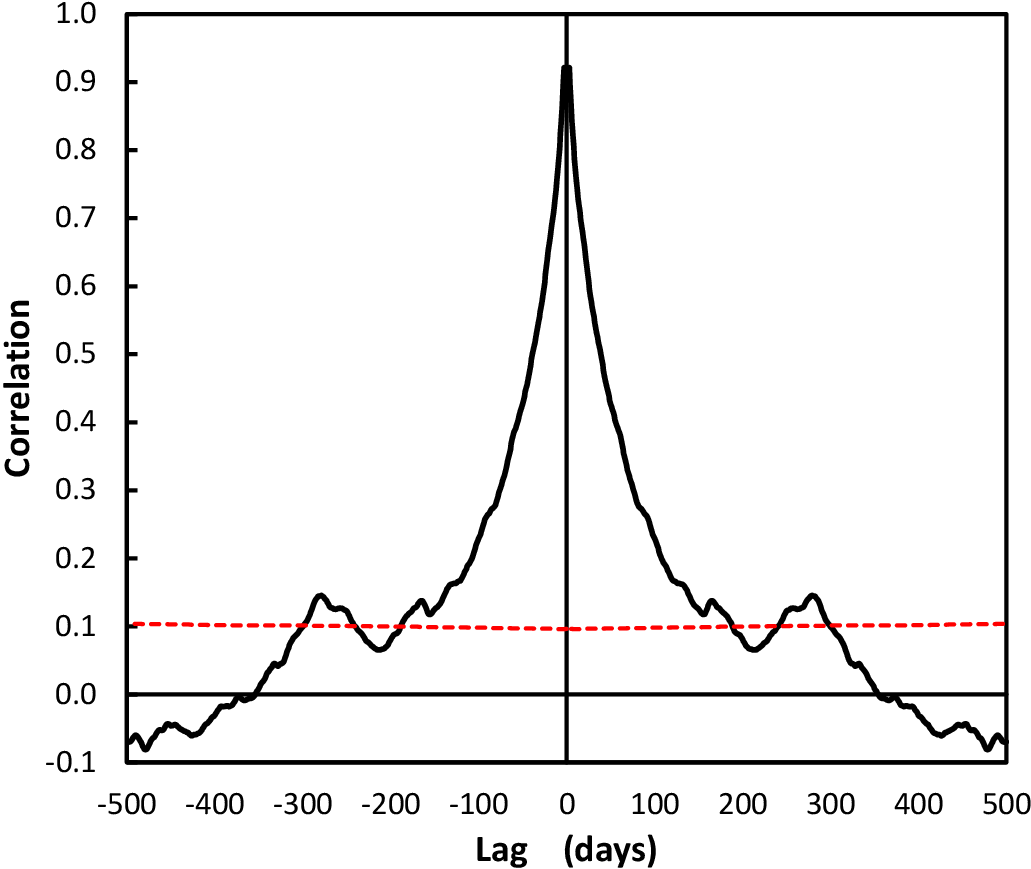}
   \caption{{Interpolated autocorrelation function of the ratios of observed H$\upbeta$ fluxes in NGC~5548 to the predictions from the optical continuum.  The red dashed curve shows the 90\% confidence interval for no correlation.  See text for details.}}
\end{figure*}  
% This figure is from Breathing_analysis/IntRatio

%\newpage
%\clearpage    
\begin{figure*} % Figure 4
   \centering
   \includegraphics[width=.5\textwidth]{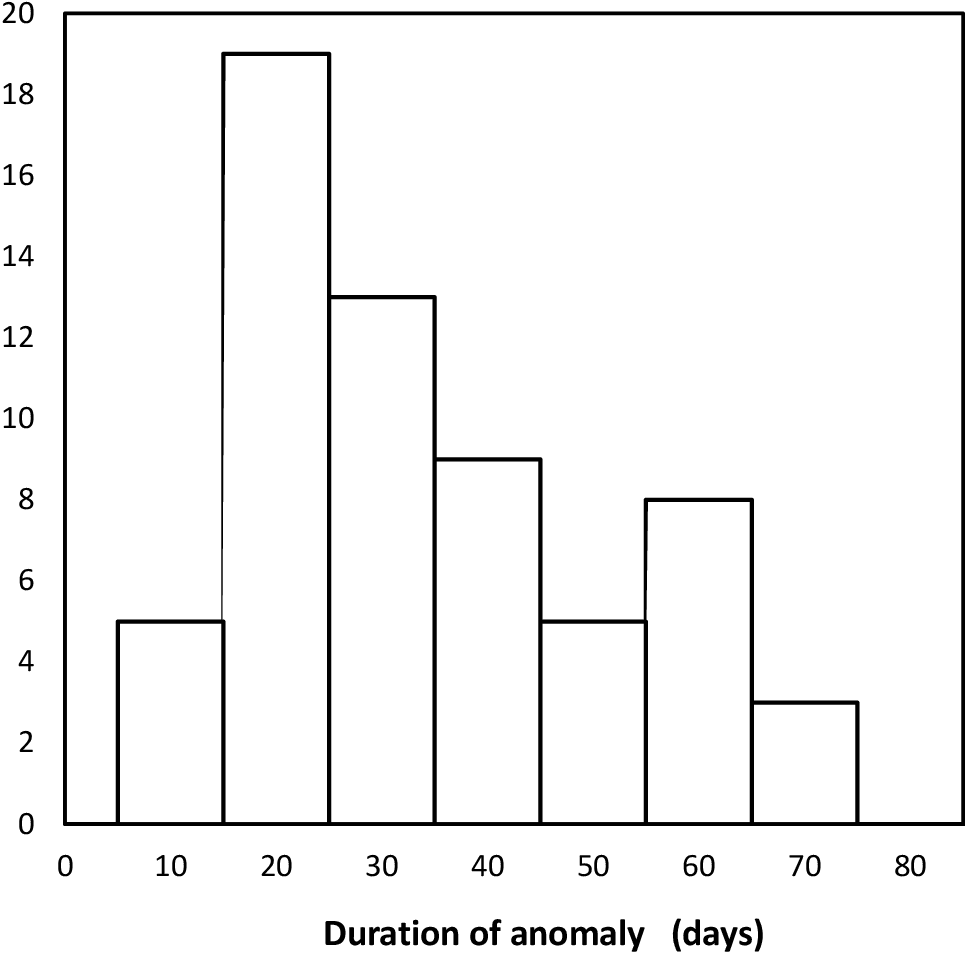}
   \caption{{The distribution of the durations of anomalous H$\upbeta$ events in NGC~5548.  See text for explanation.}}
\end{figure*}  
% This figure is from Timescale+amplitudes/Durations

\begin{figure*} % Figure 5
   \centering
   \includegraphics[width=.5\textwidth]{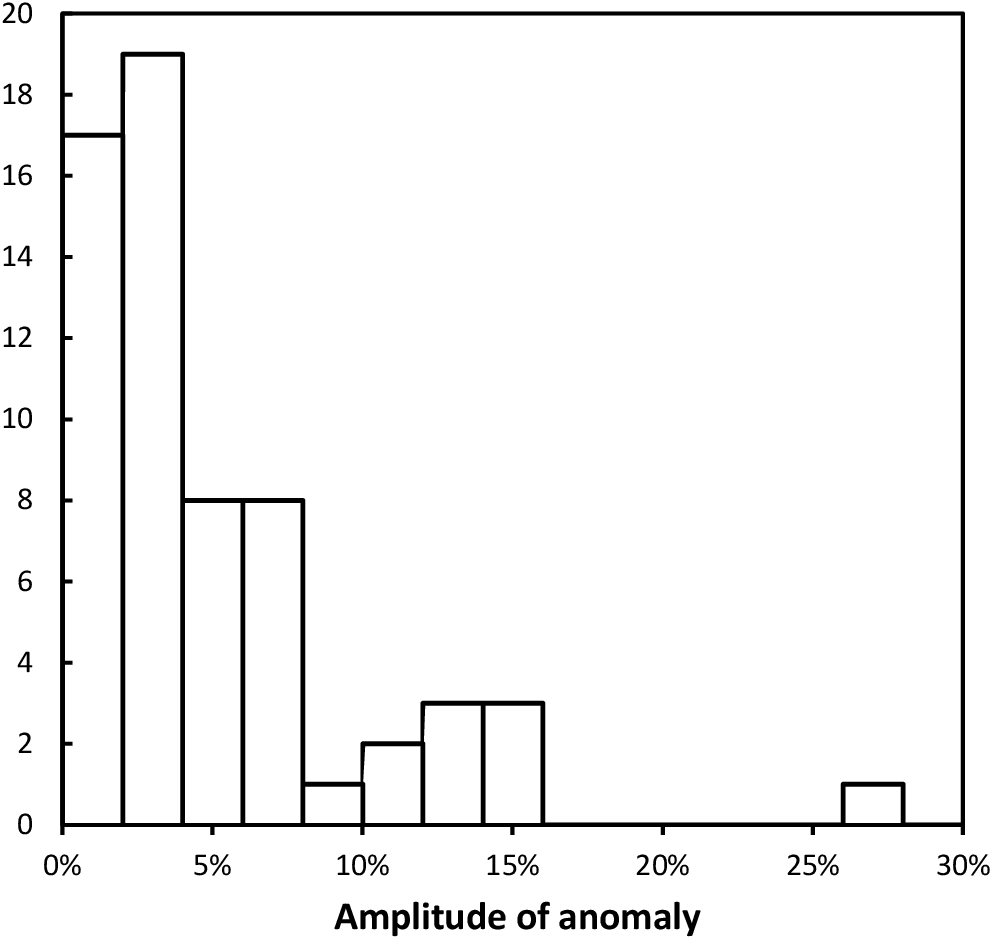}
   \caption{{The distribution of the deviations of peak (or trough) amplitudes of anomalous H$\upbeta$ events in NGC~5548 relative to predictions.  See text for details.}}
\end{figure*}  
% This figure is from Timescale+amplitudes/Durations

%\newpage
%\clearpage    
\begin{figure*} % Figure 6
   \centering
   \includegraphics[width=.7\textwidth]{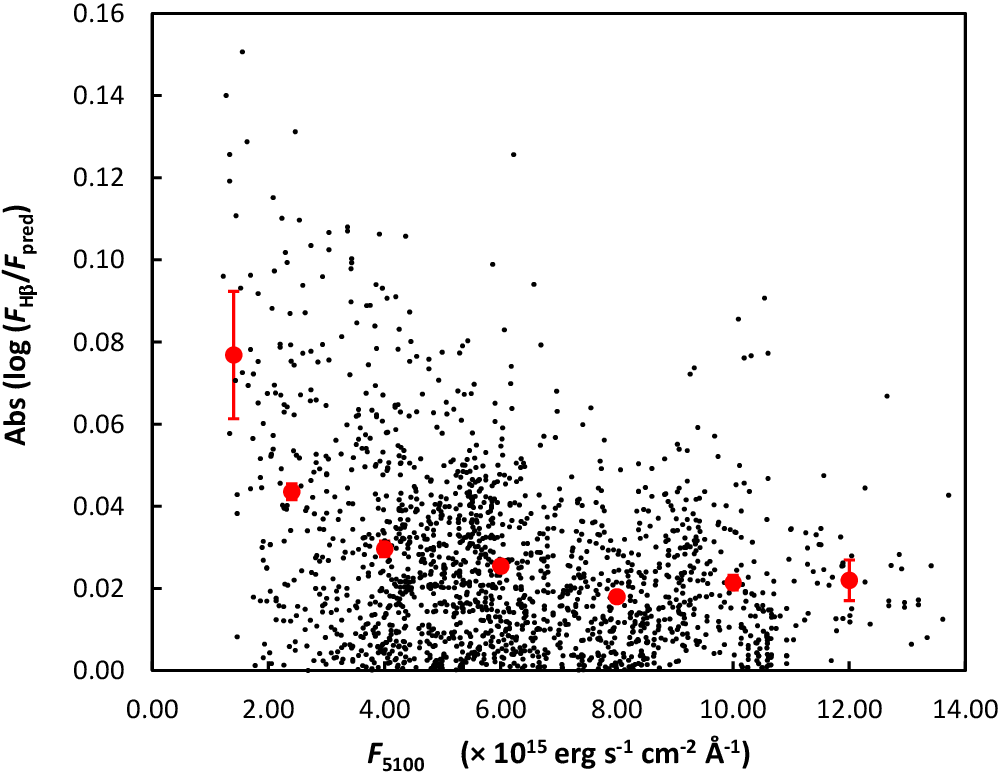}
   \caption{The absolute value of the logarithm of the residuals of the NGC~5548 H$\upbeta$ fluxes from the fluxes predicted from the continuum variability versus the $\lambda$5100 continuum flux.  The red circles show the means.}
\end{figure*}  
% This figure is from

\begin{figure*} % Figure 7 - Part 1
   \centering
   \subfloat{\includegraphics[width=.45\textwidth]{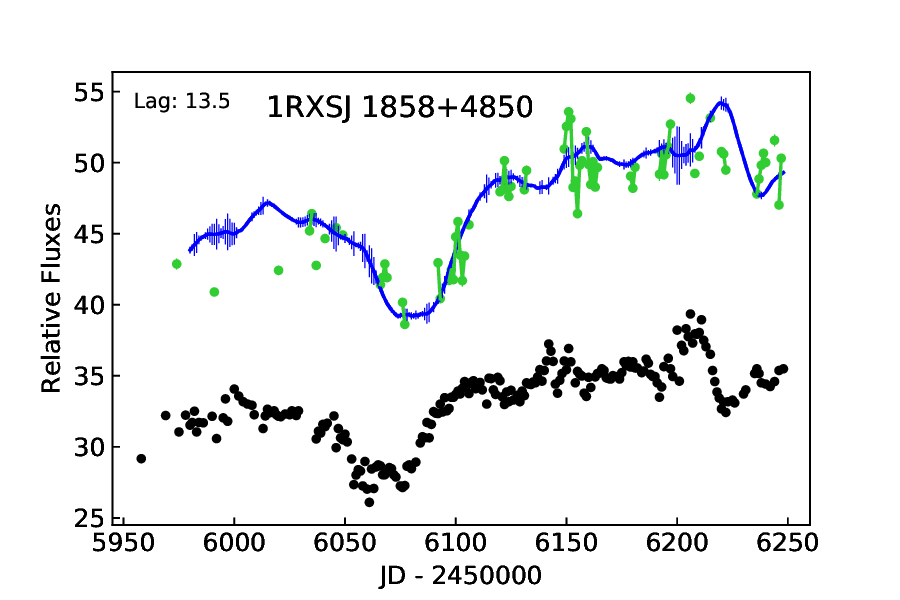}}\quad
   \subfloat{\includegraphics[width=.45\textwidth]{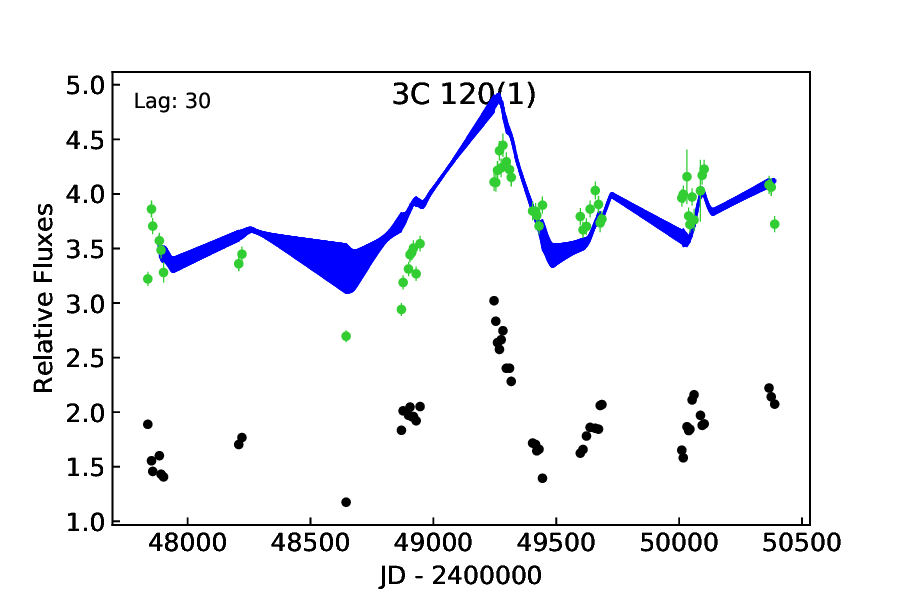}}\quad
   \subfloat{\includegraphics[width=.45\textwidth]{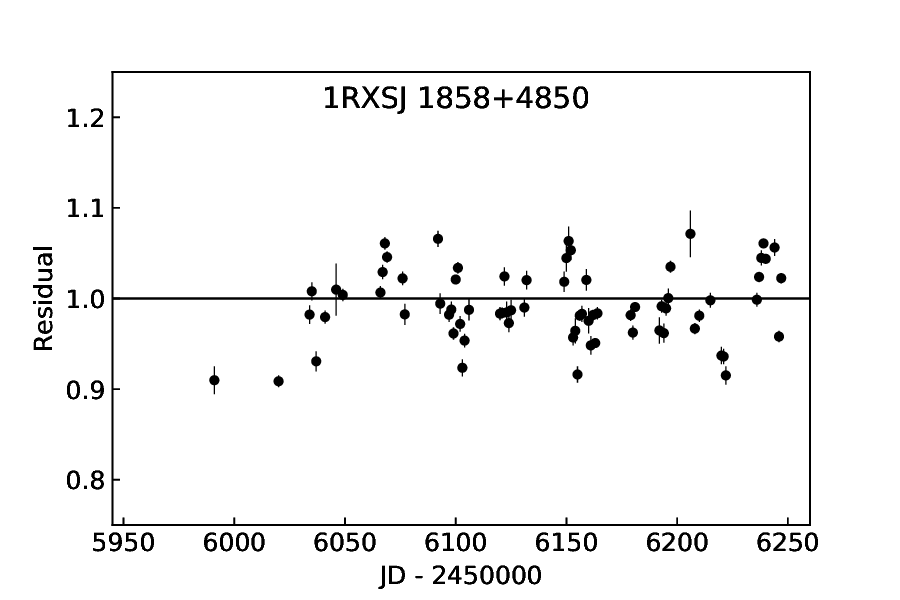}}\quad
   \subfloat{\includegraphics[width=.45\textwidth]{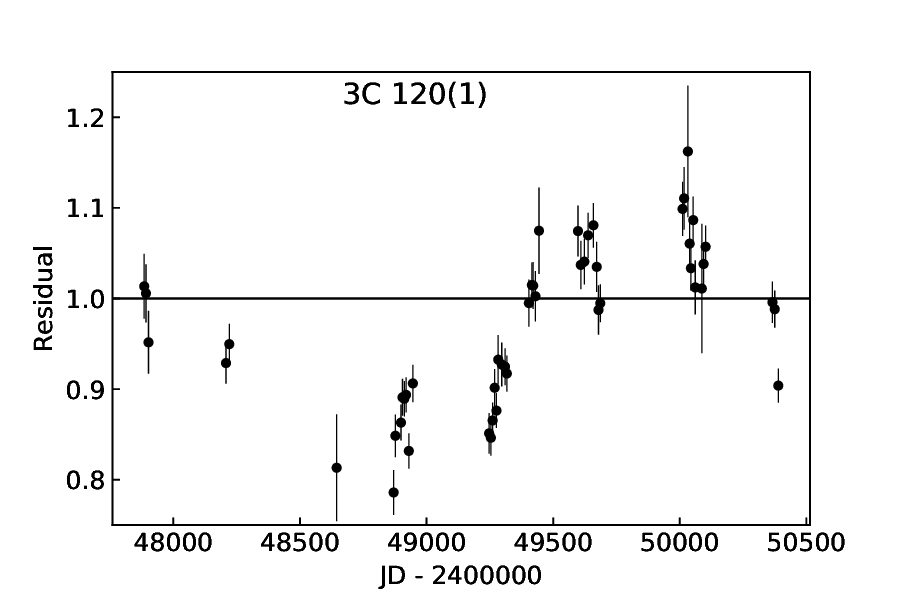}}\quad
   \subfloat{\includegraphics[width=.45\textwidth]{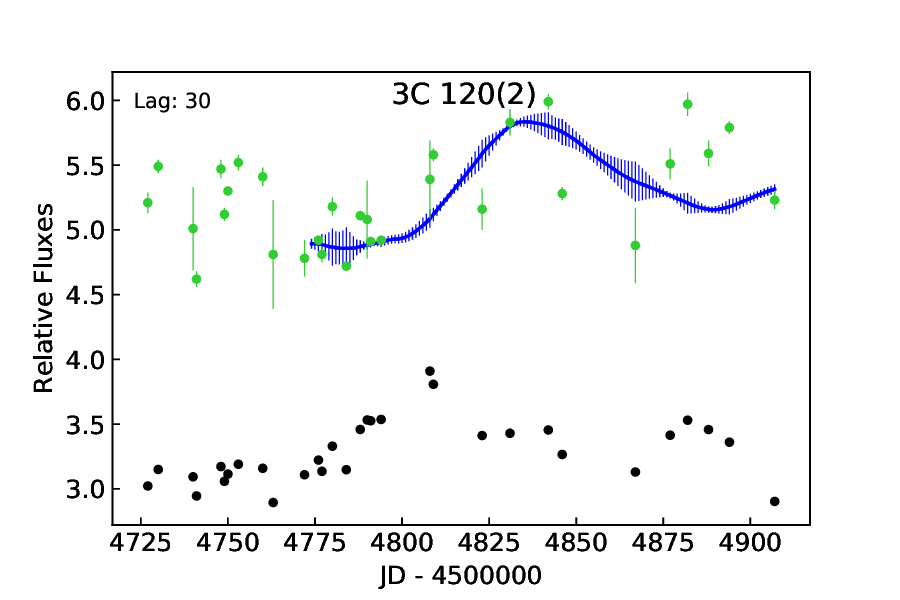}}\quad
   \subfloat{\includegraphics[width=.45\textwidth]{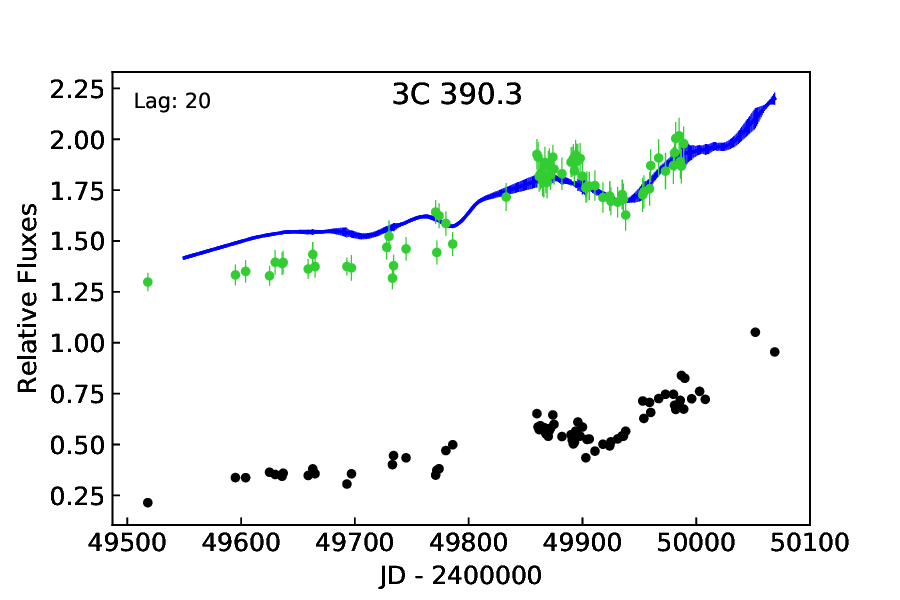}}\quad
   \subfloat{\includegraphics[width=.45\textwidth]{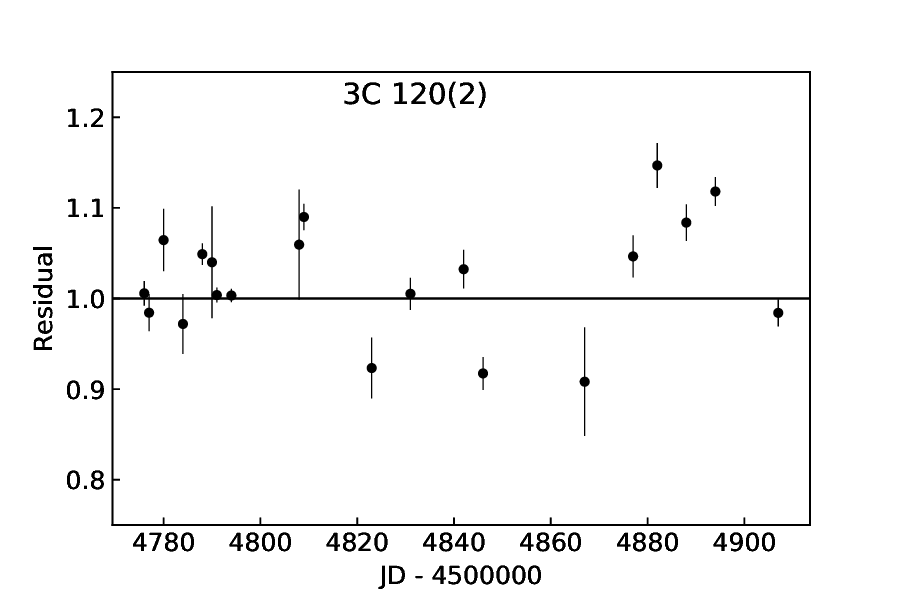}}\quad
   \subfloat{\includegraphics[width=.45\textwidth]{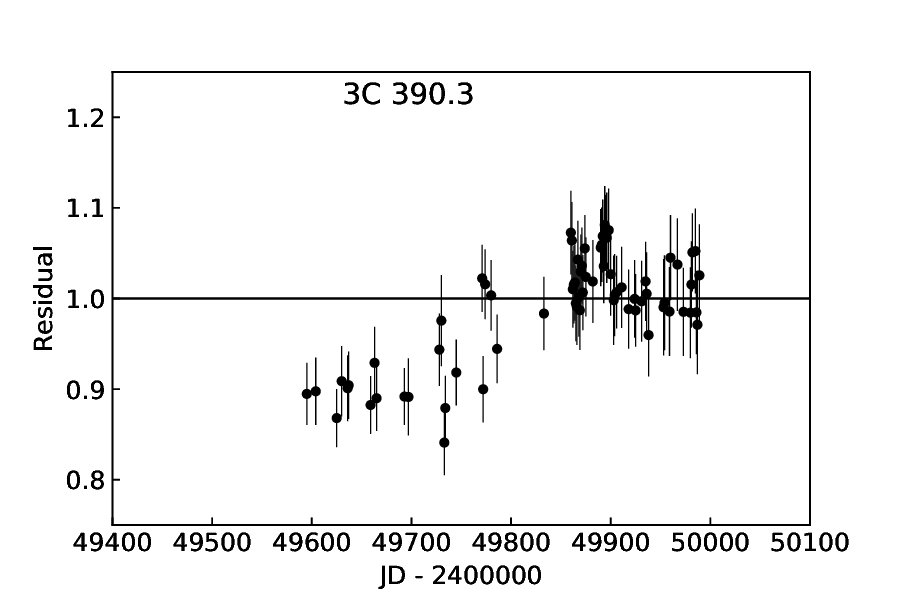}}\quad
   \caption{As in Figure 2, but for other AGNs.  The adopted lags (in days) are indicated.}
\end{figure*}
   
\newpage
\clearpage
\begin{figure*} % Figure 7 - Part 2
   \setcounter{figure}{6}
   \centering
   \subfloat{\includegraphics[width=.45\textwidth]{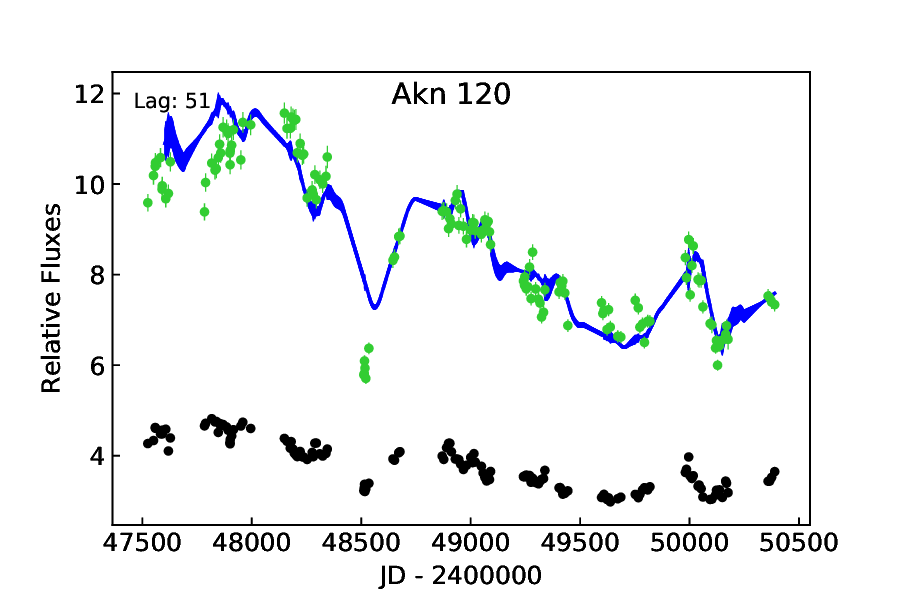}}\quad
   \subfloat{\includegraphics[width=.45\textwidth]{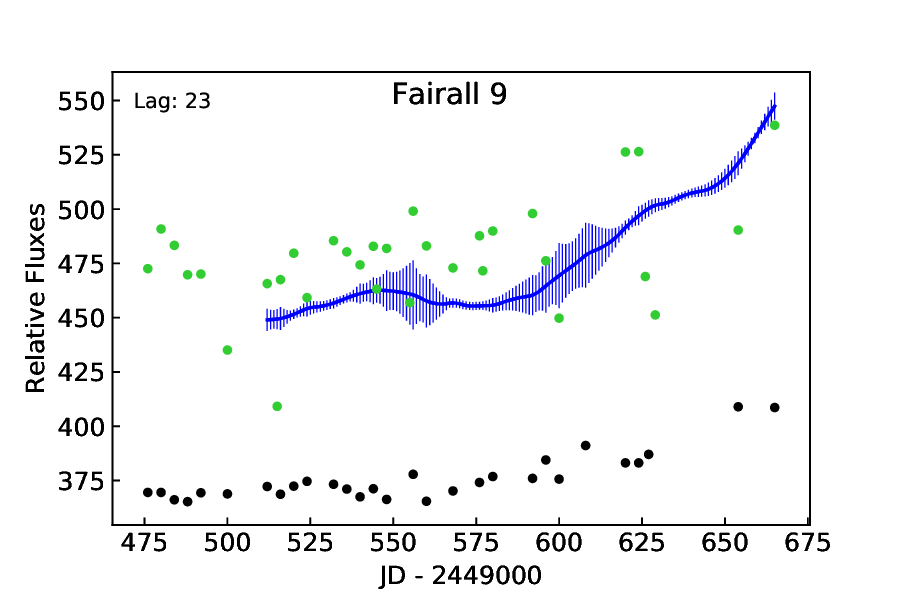}}\quad
   \subfloat{\includegraphics[width=.45\textwidth]{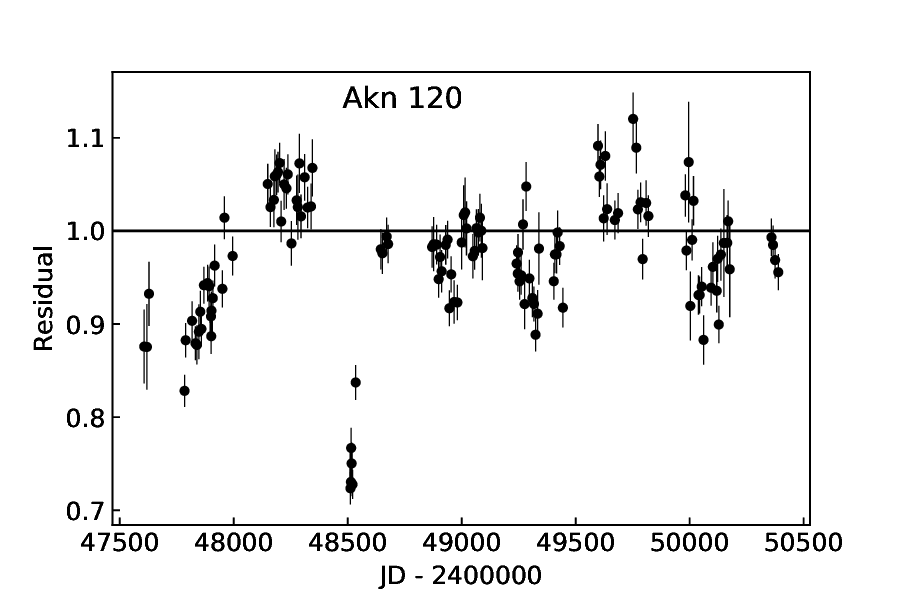}}\quad  
   \subfloat{\includegraphics[width=.45\textwidth]{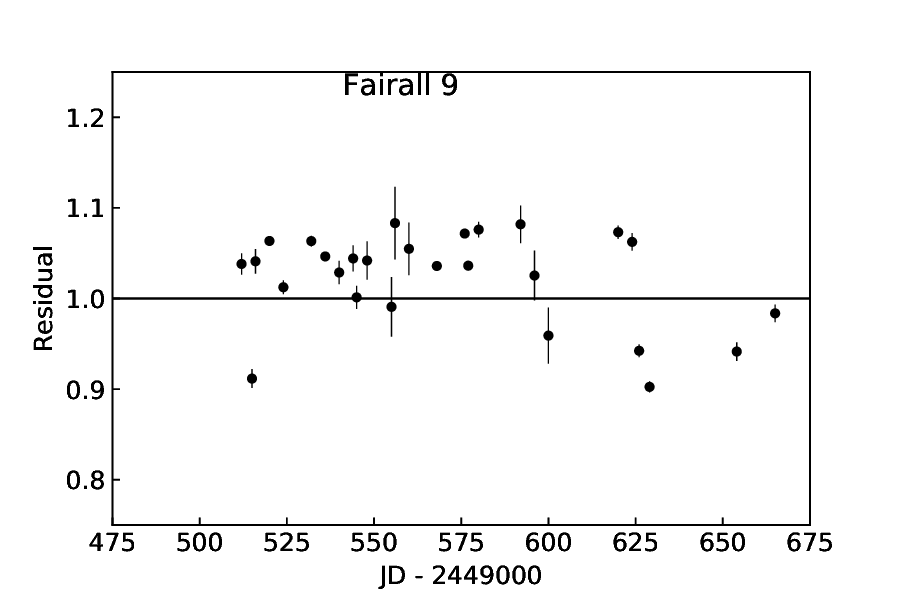}}\quad
   \subfloat{\includegraphics[width=.45\textwidth]{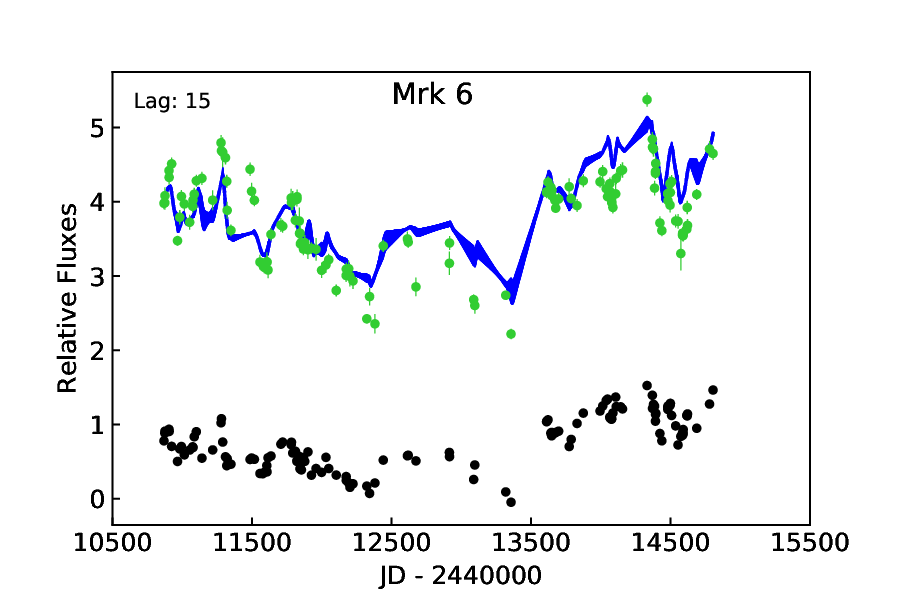}}\quad  
   \subfloat{\includegraphics[width=.45\textwidth]{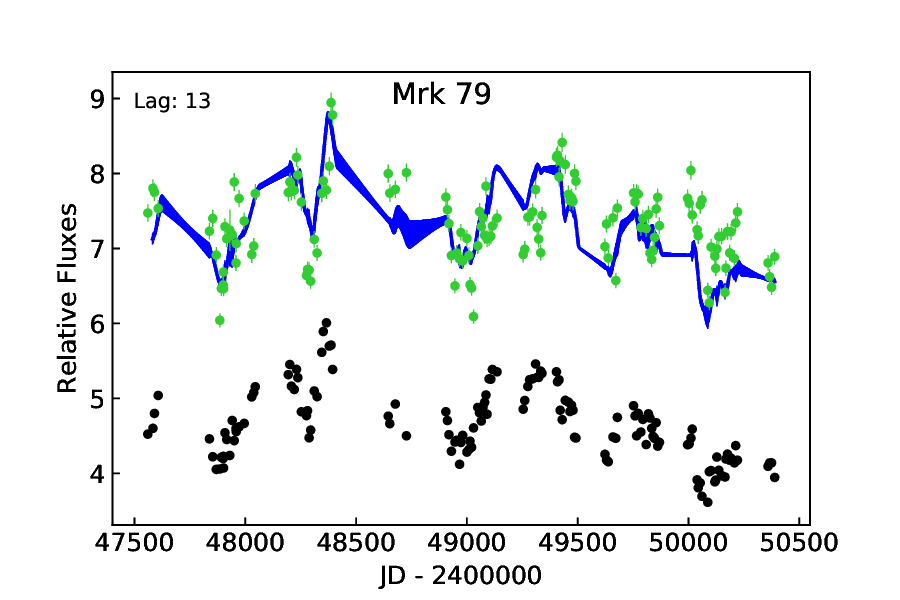}}\quad
   \subfloat{\includegraphics[width=.45\textwidth]{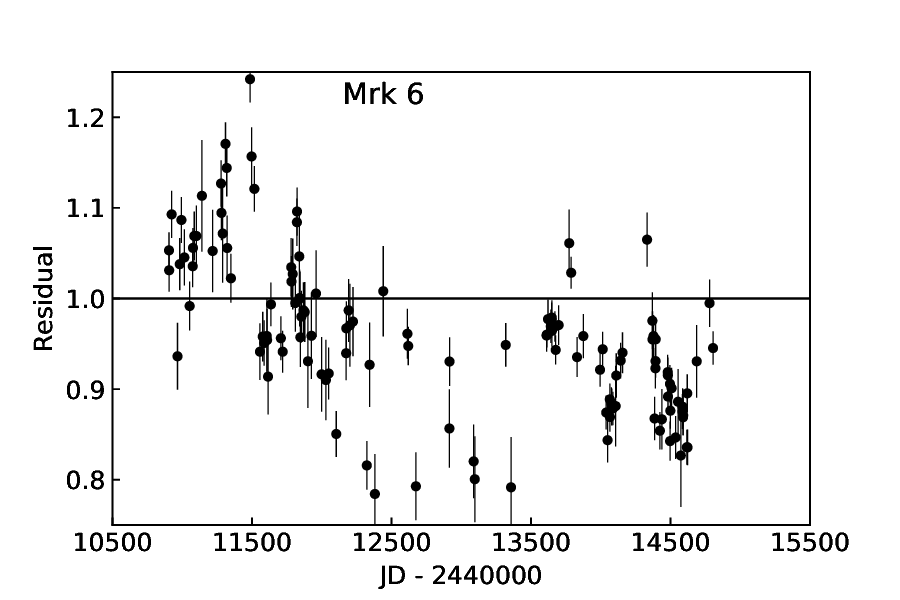}}\quad 
   \subfloat{\includegraphics[width=.45\textwidth]{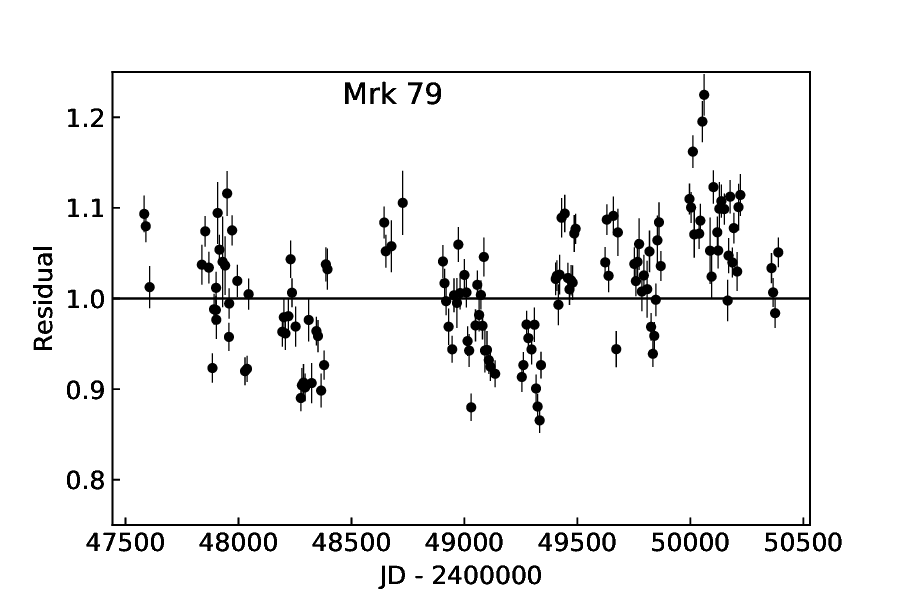}}\quad
   \caption{-- Continued.}
 \end{figure*} 
   
\newpage
\clearpage
\begin{figure*} 
   \setcounter{figure}{6} % Figure 7 - Part 3
   \centering
   \subfloat{\includegraphics[width=.45\textwidth]{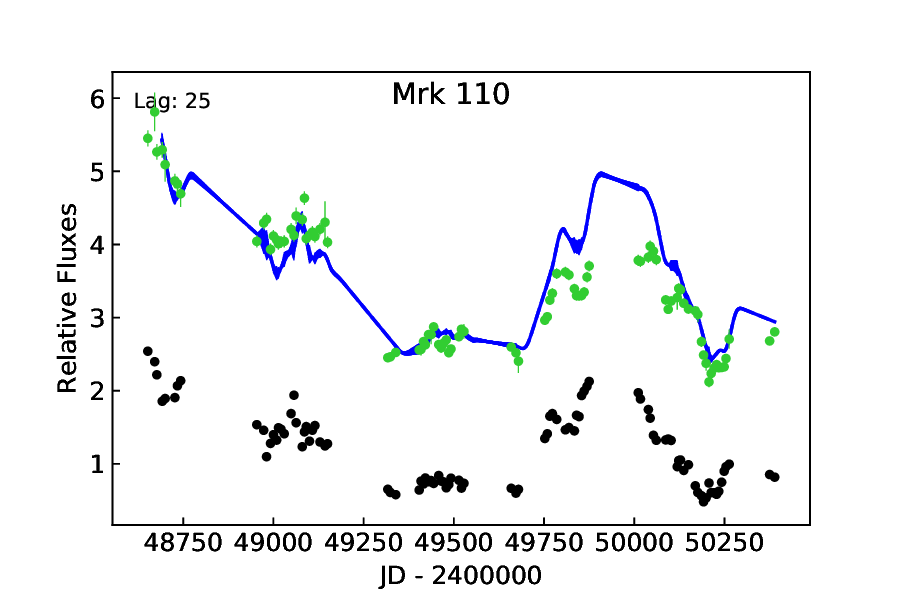}}\quad
   \subfloat{\includegraphics[width=.45\textwidth]{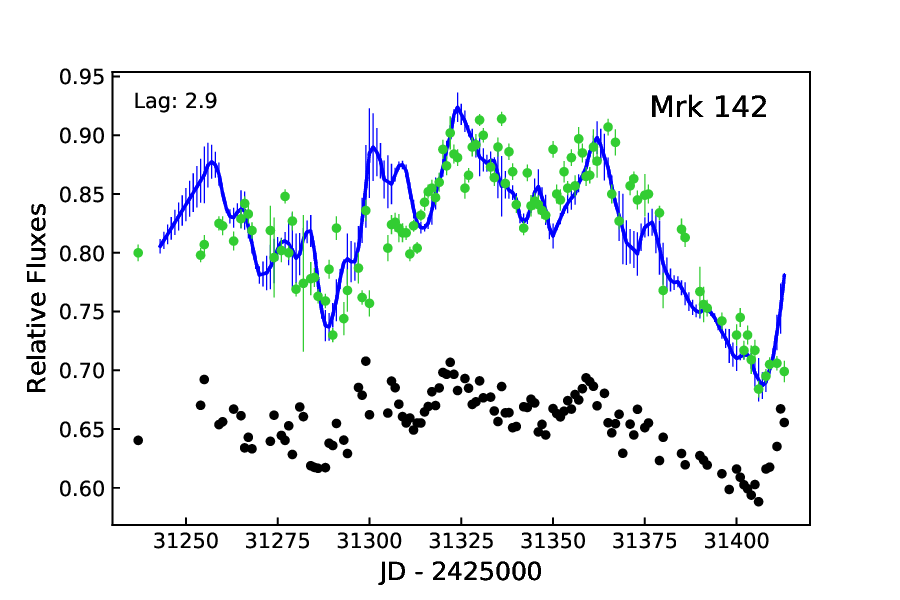}}\quad 
   \subfloat{\includegraphics[width=.45\textwidth]{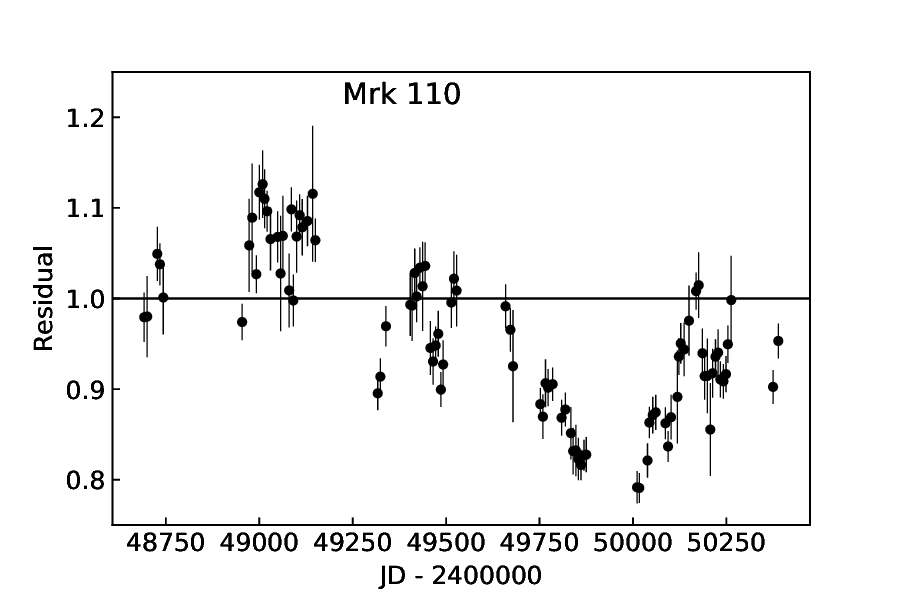}}\quad
   \subfloat{\includegraphics[width=.45\textwidth]{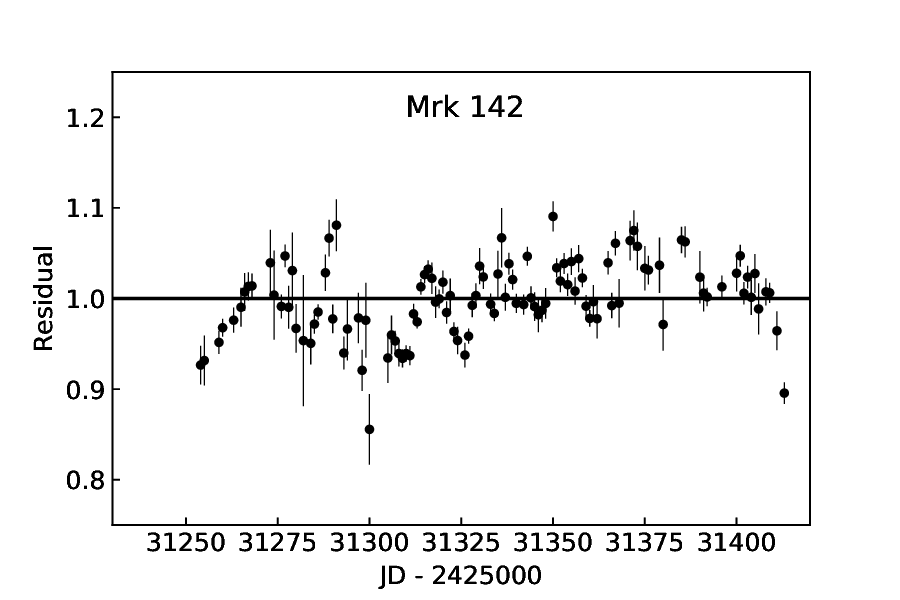}}\quad
   \subfloat{\includegraphics[width=.45\textwidth]{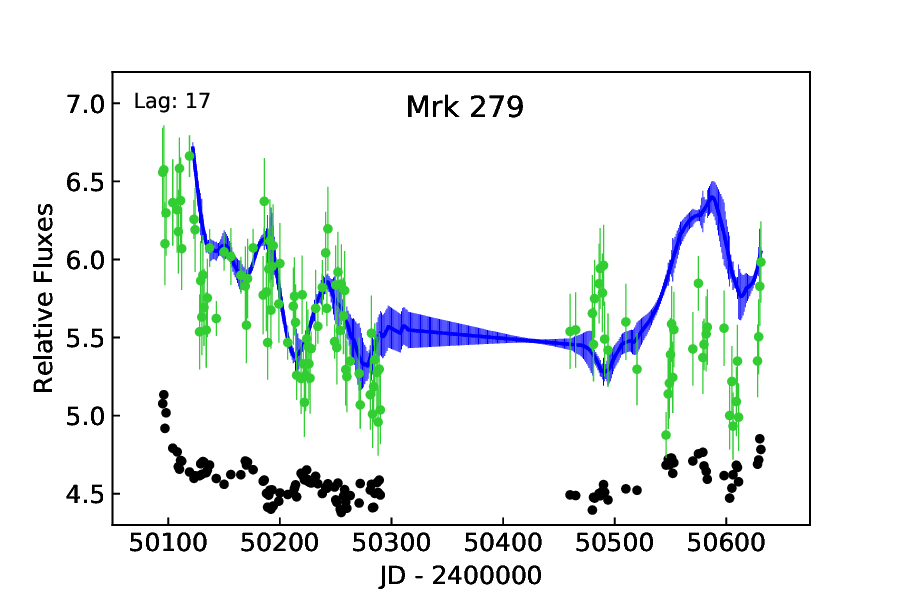}}\quad
   \subfloat{\includegraphics[width=.45\textwidth]{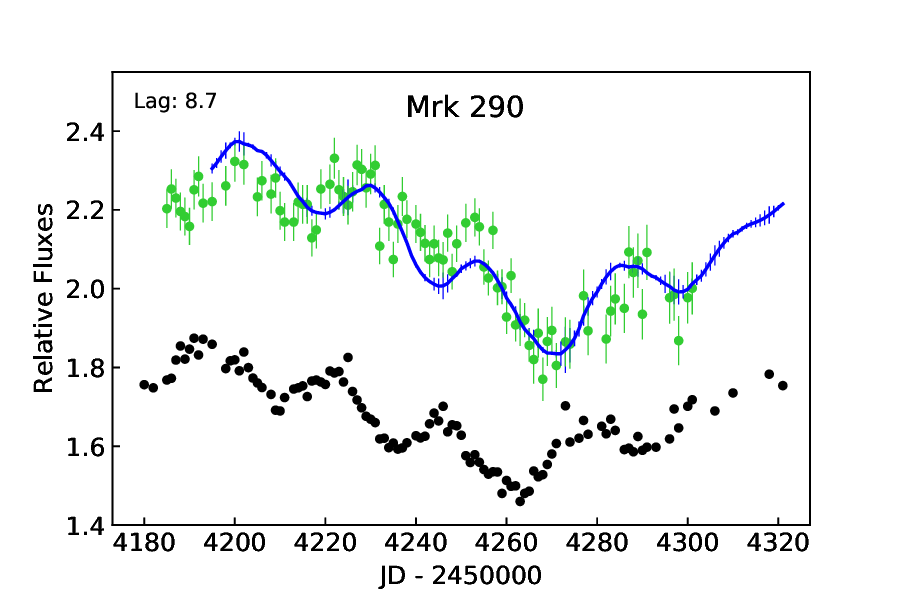}}\quad
   \subfloat{\includegraphics[width=.45\textwidth]{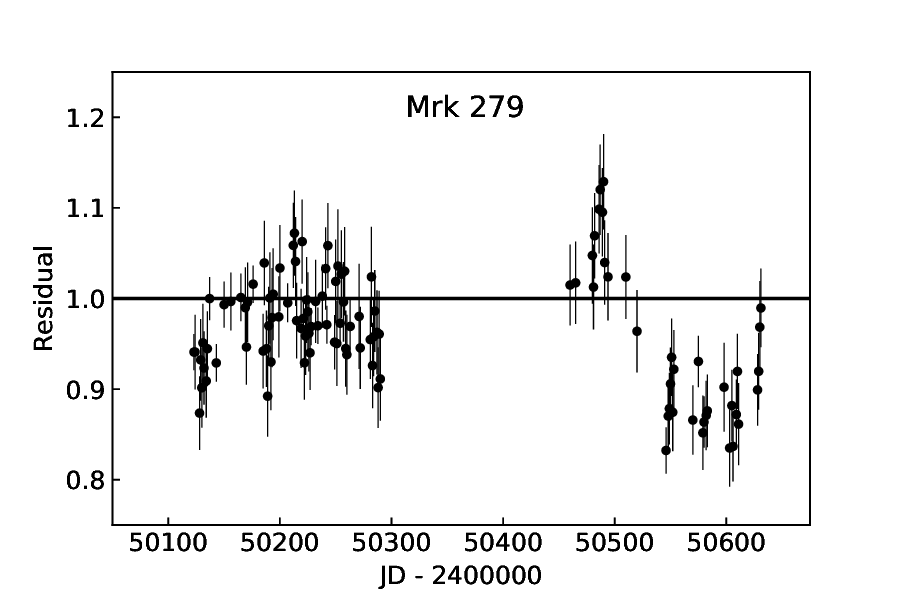}}\quad
   \subfloat{\includegraphics[width=.45\textwidth]{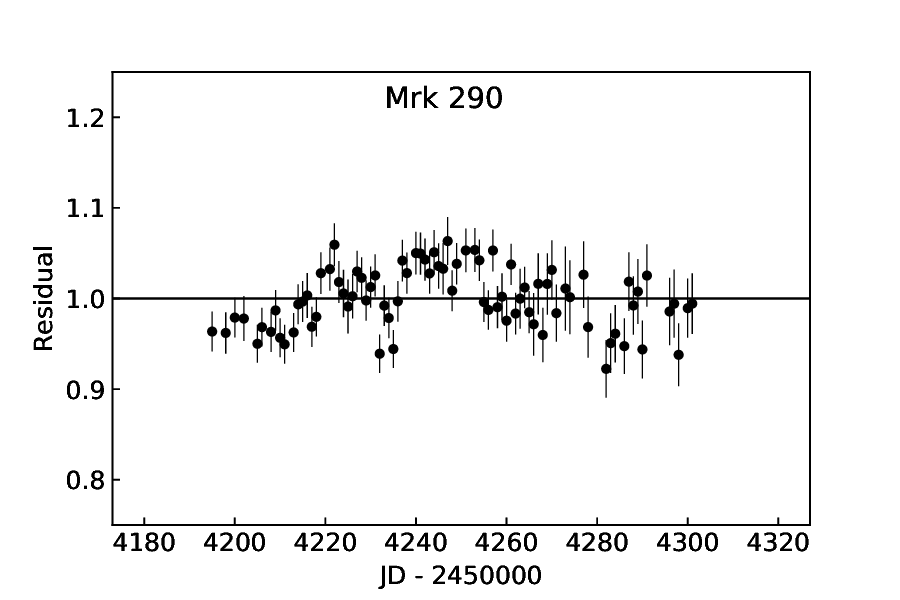}}\quad
     \caption{-- Continued.}
\end{figure*}   

\newpage
\clearpage  
\begin{figure*}  % Figure 7 - Part 4
   \setcounter{figure}{6}
   \centering
   \subfloat{\includegraphics[width=.45\textwidth]{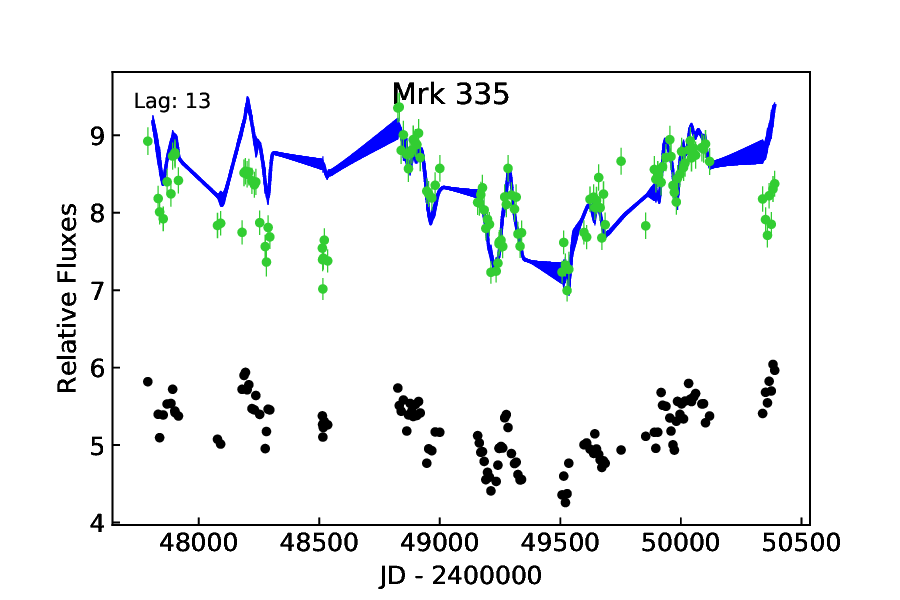}}\quad
   \subfloat{\includegraphics[width=.45\textwidth]{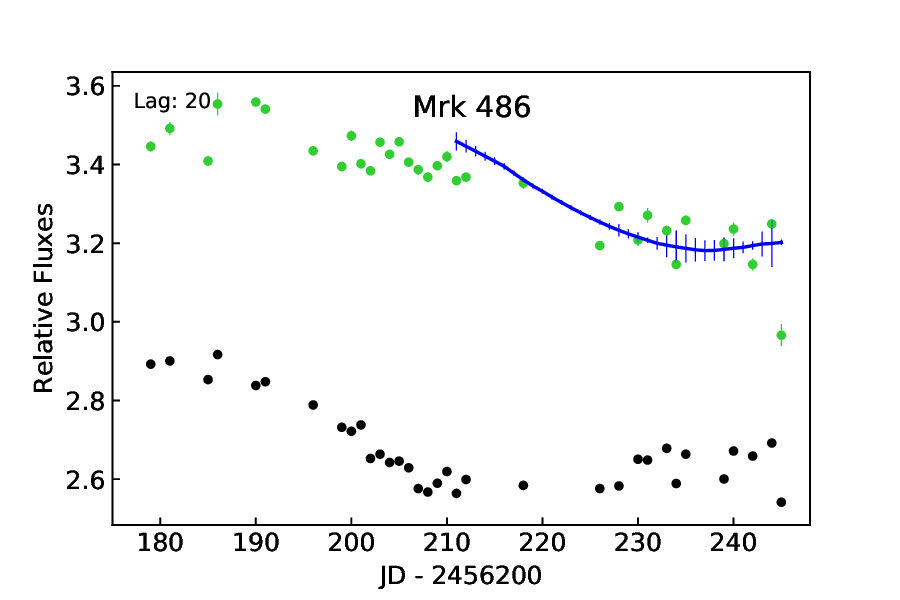}}\quad
   \subfloat{\includegraphics[width=.45\textwidth]{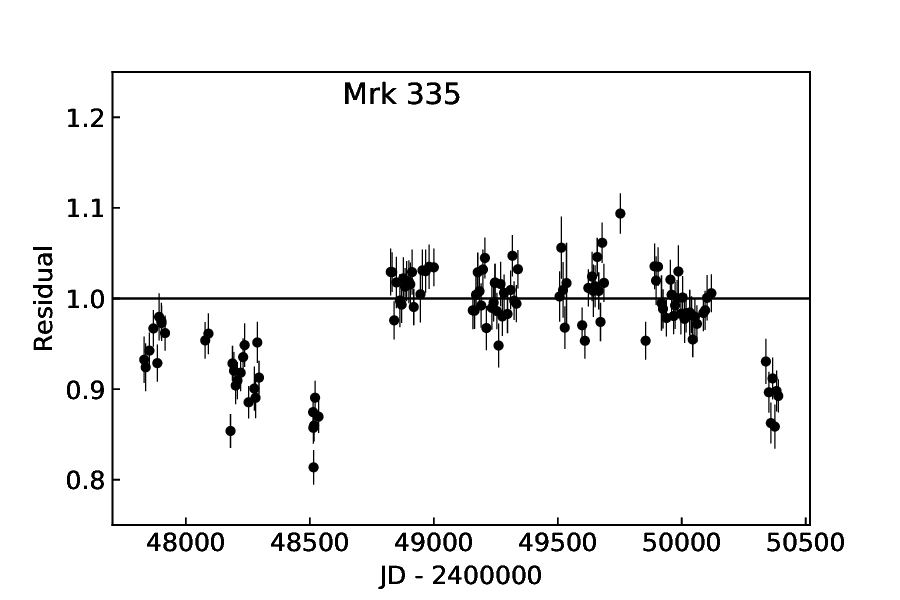}}\quad
   \subfloat{\includegraphics[width=.45\textwidth]{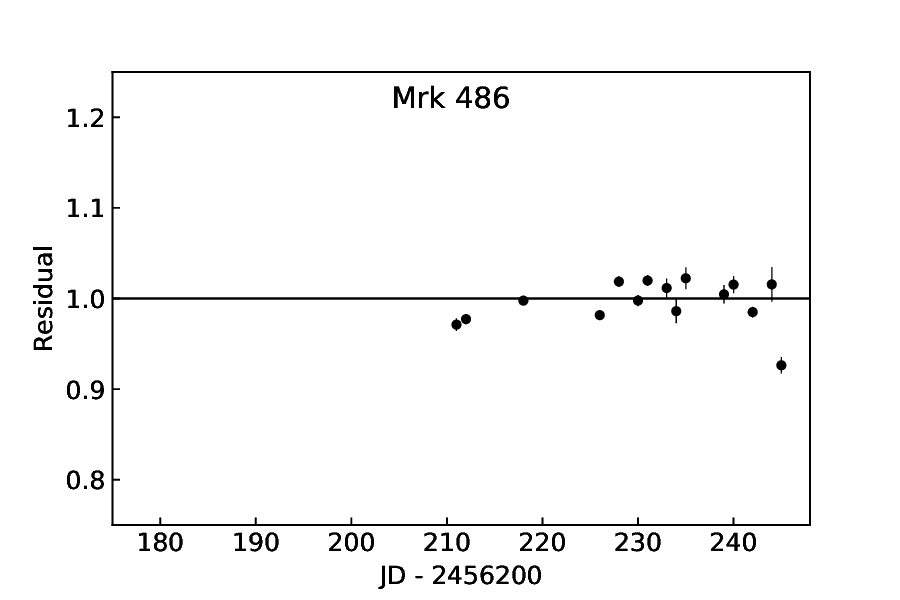}}\quad 
   \subfloat{\includegraphics[width=.45\textwidth]{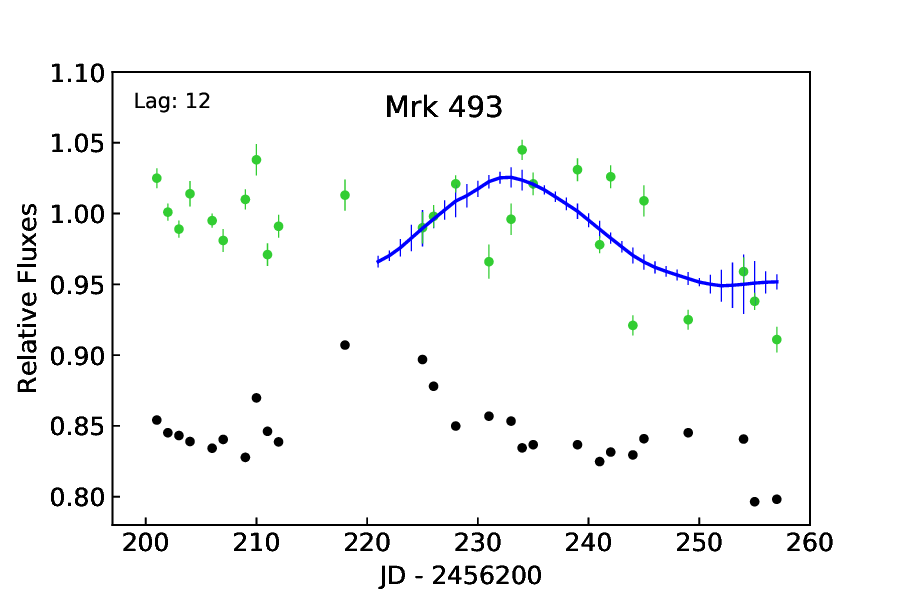}}\quad
   \subfloat{\includegraphics[width=.45\textwidth]{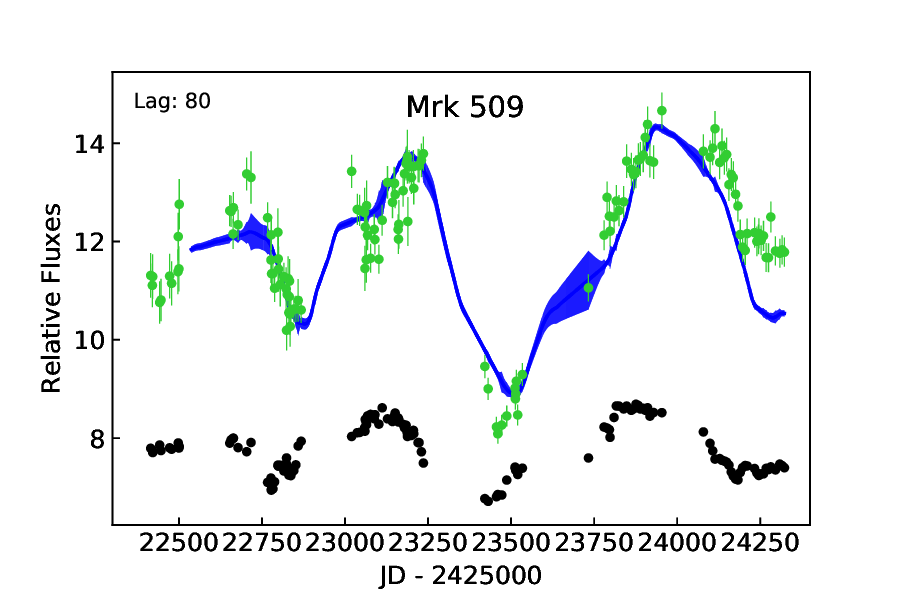}}\quad 
   \subfloat{\includegraphics[width=.45\textwidth]{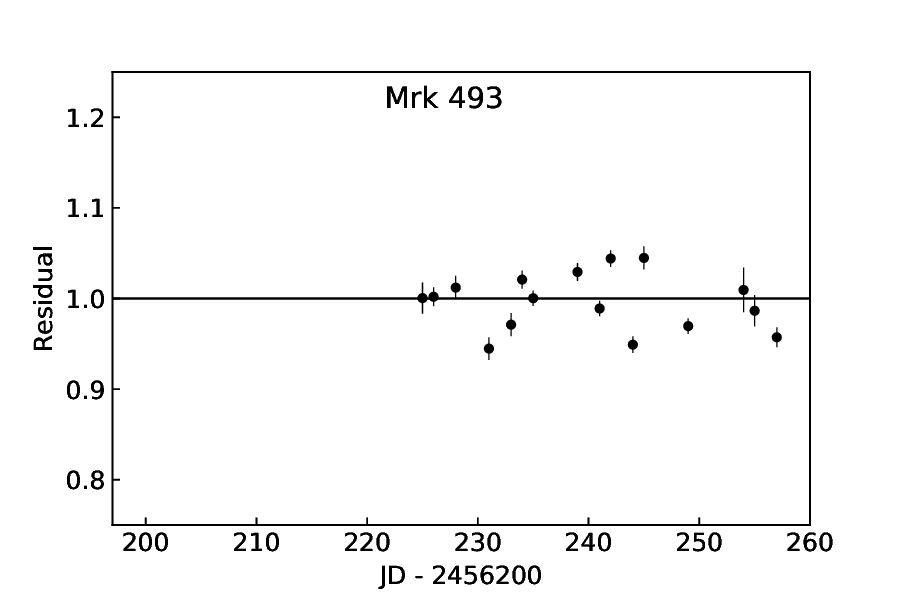}}\quad  
   \subfloat{\includegraphics[width=.45\textwidth]{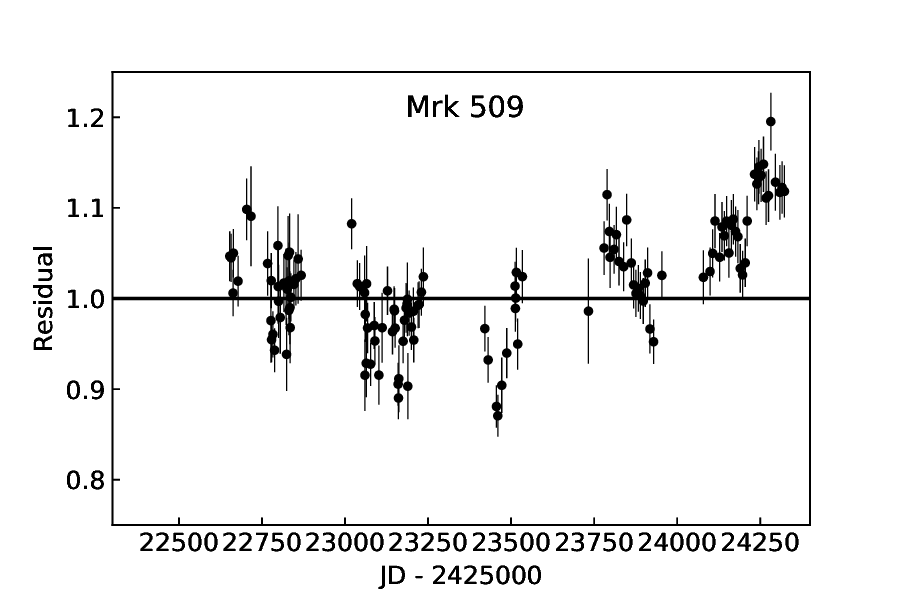}}\quad
     \caption{-- Continued.}
 \end{figure*}
 
\newpage
\clearpage  
\begin{figure*} % Figure 7 - Part 5
   \setcounter{figure}{6}
   \centering
   \subfloat{\includegraphics[width=.45\textwidth]{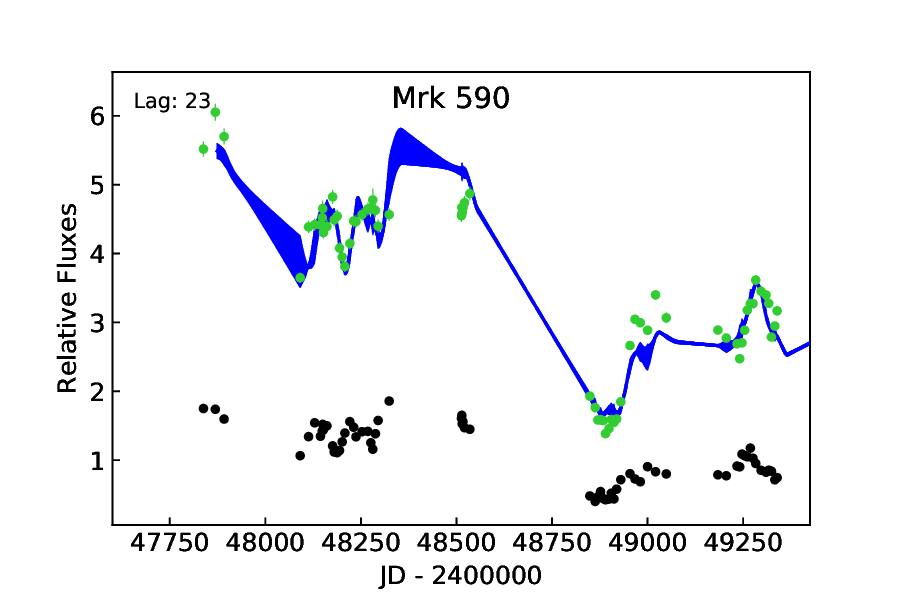}}\quad 
   \subfloat{\includegraphics[width=.45\textwidth]{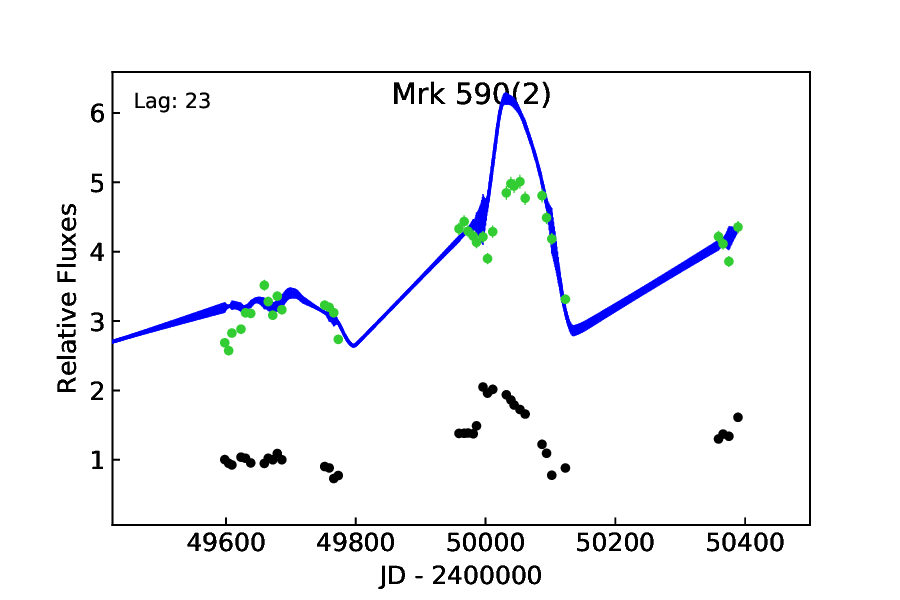}}\quad  
   \subfloat{\includegraphics[width=.45\textwidth]{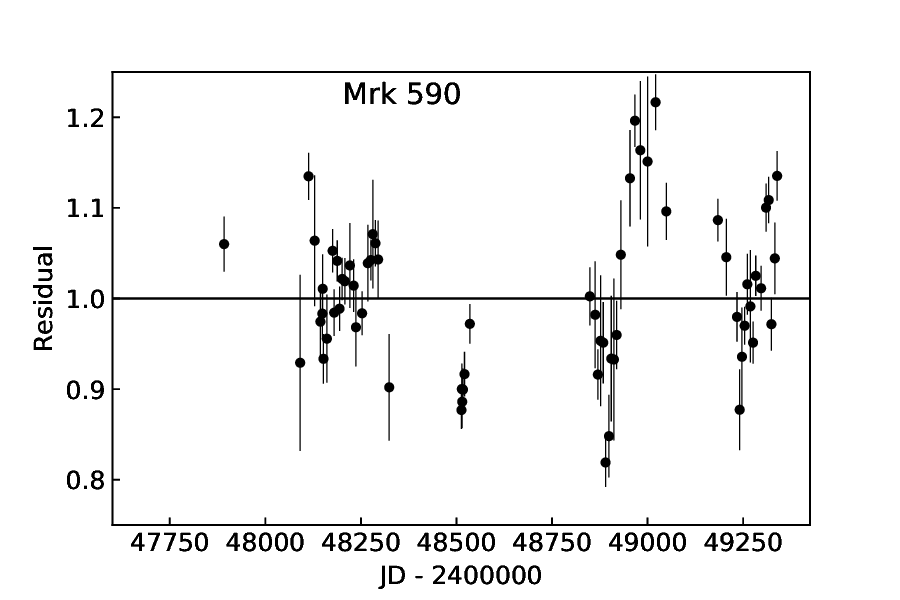}}\quad
   \subfloat{\includegraphics[width=.45\textwidth]{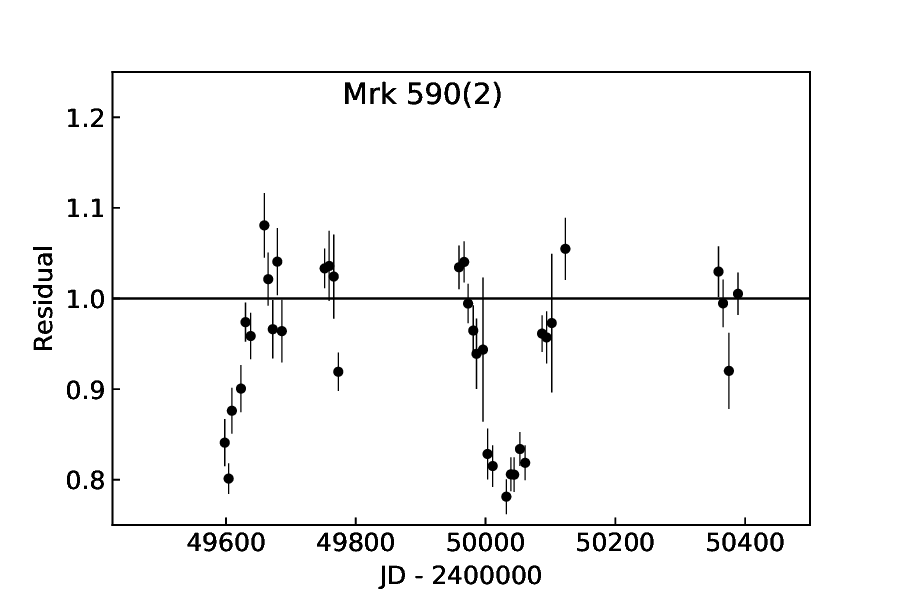}}\quad
   \subfloat{\includegraphics[width=.45\textwidth]{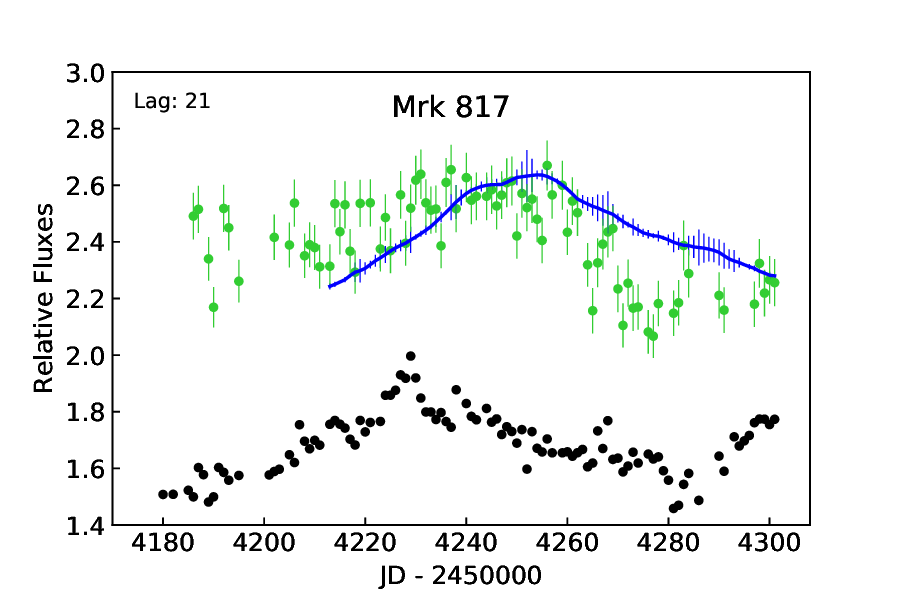}}\quad   
   \subfloat{\includegraphics[width=.45\textwidth]{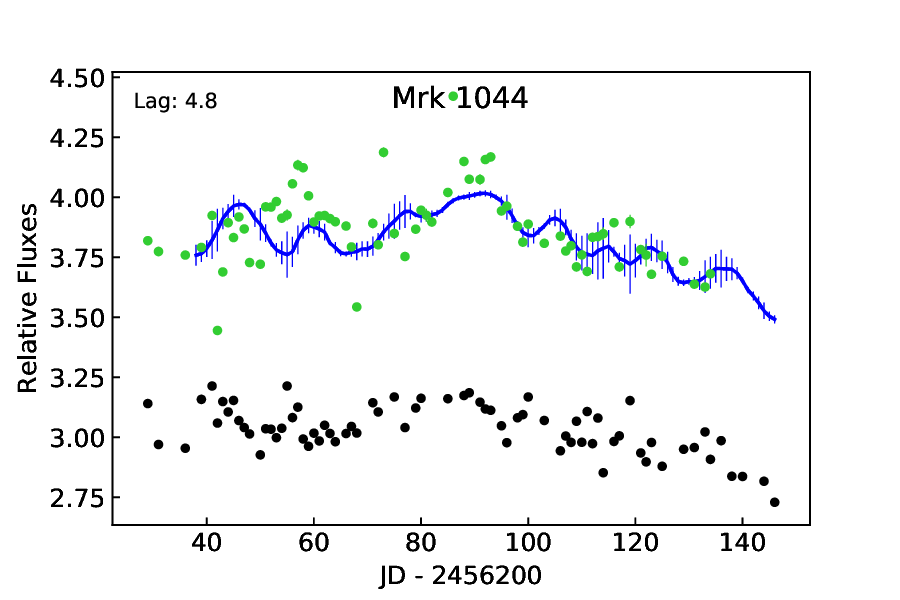}}\quad
   \subfloat{\includegraphics[width=.45\textwidth]{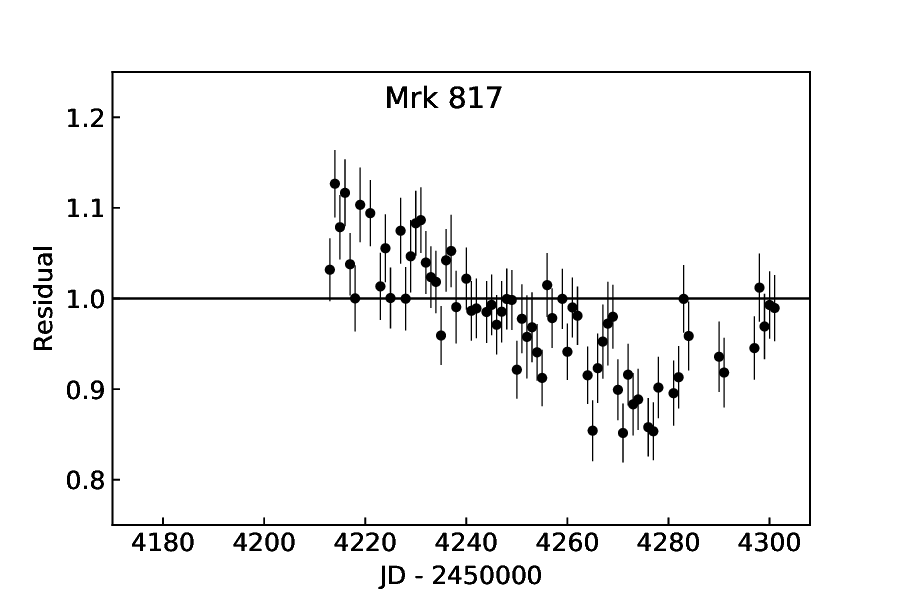}}\quad
   \subfloat{\includegraphics[width=.45\textwidth]{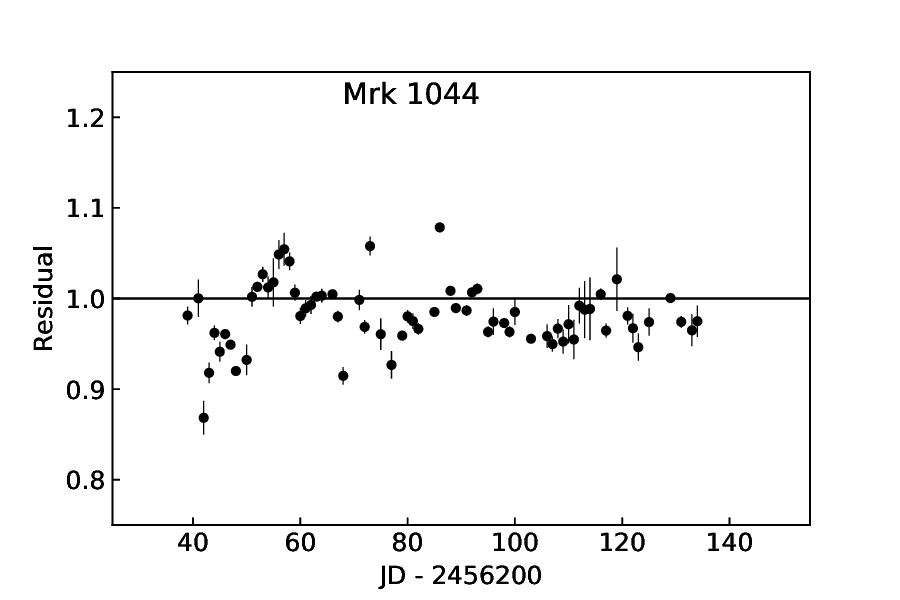}}\quad
   \caption{-- Continued.}
 \end{figure*}
 
\newpage
\clearpage    
\begin{figure*} % Figure 7 - Part 6
   \setcounter{figure}{6} 
   \centering
   \subfloat{\includegraphics[width=.45\textwidth]{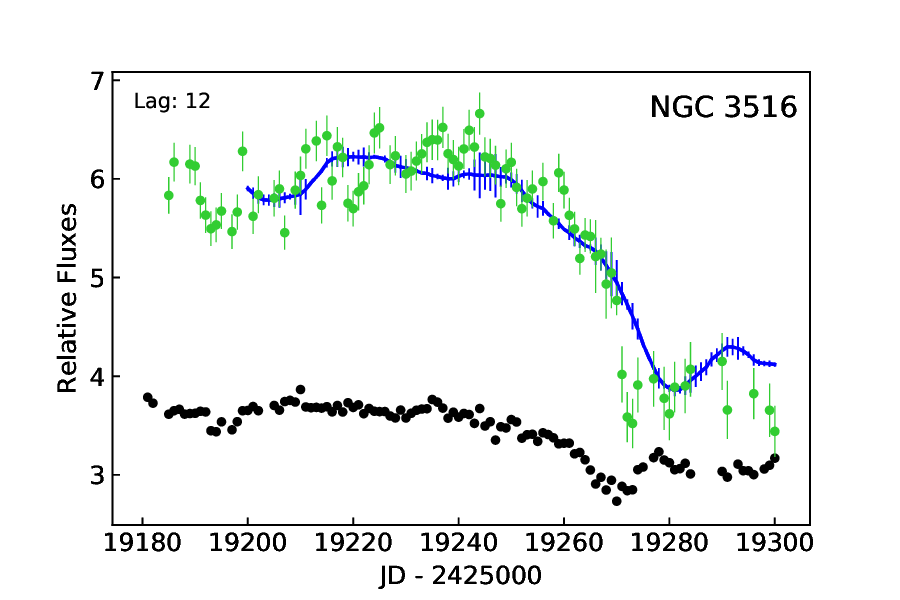}}\quad
   \subfloat{\includegraphics[width=.45\textwidth]{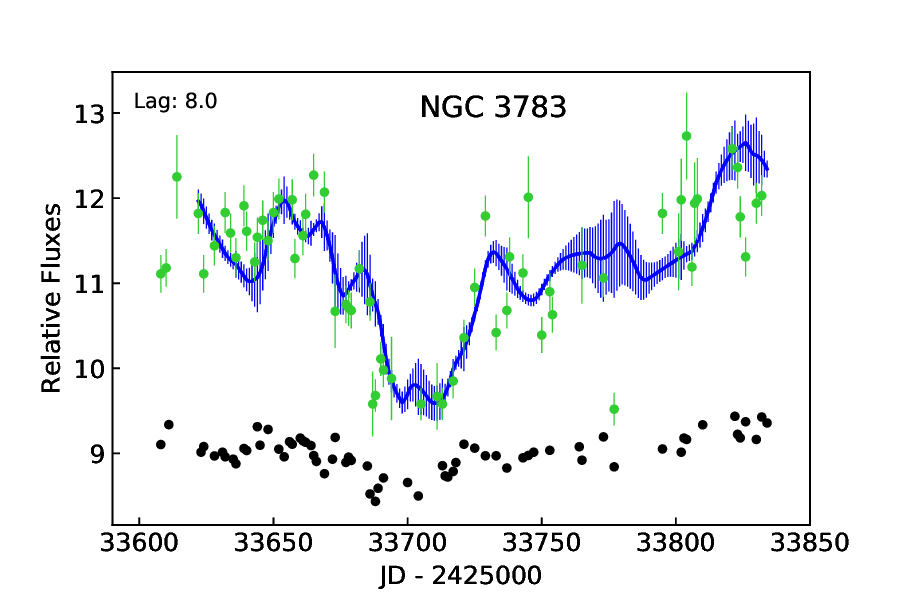}}\quad
   \subfloat{\includegraphics[width=.45\textwidth]{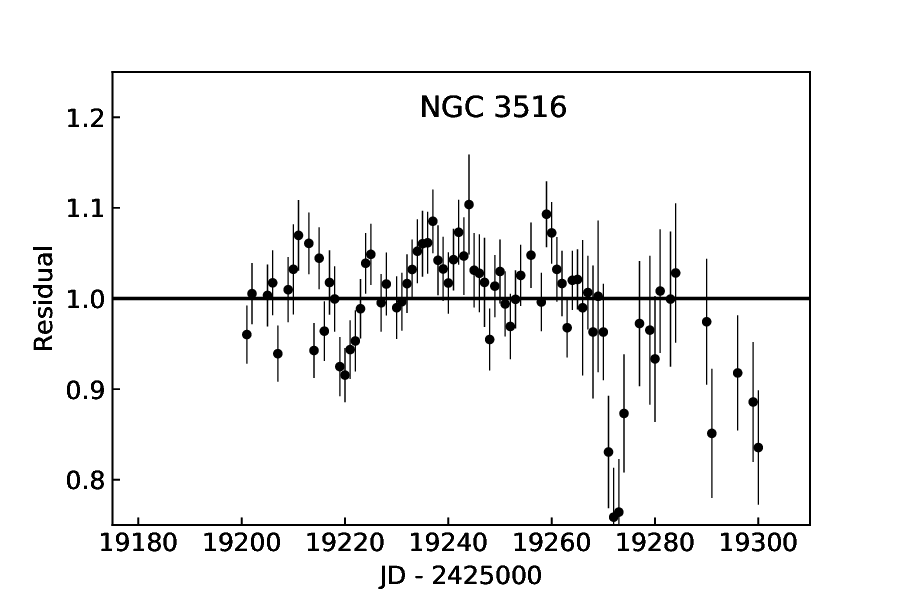}}\quad   
   \subfloat{\includegraphics[width=.45\textwidth]{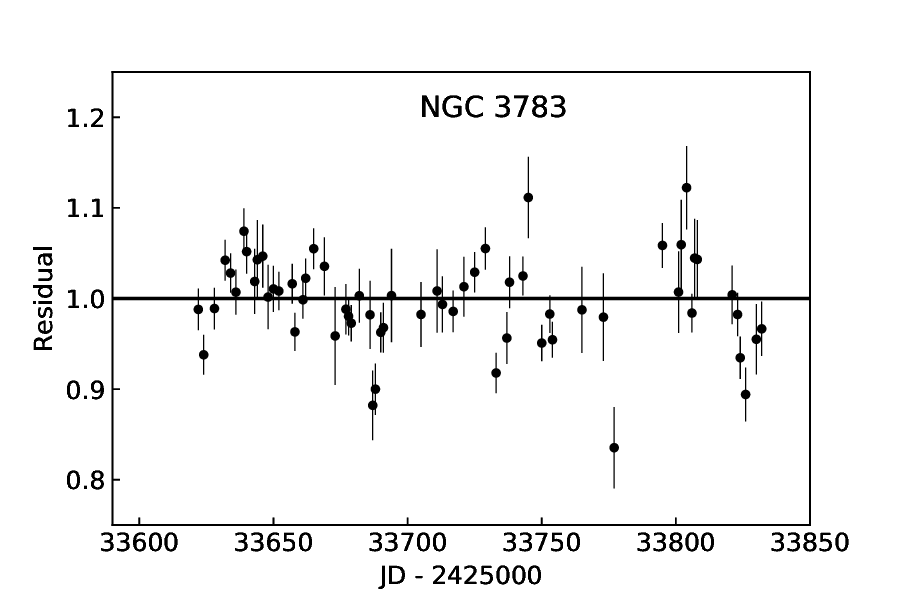}}\quad
   \subfloat{\includegraphics[width=.45\textwidth]{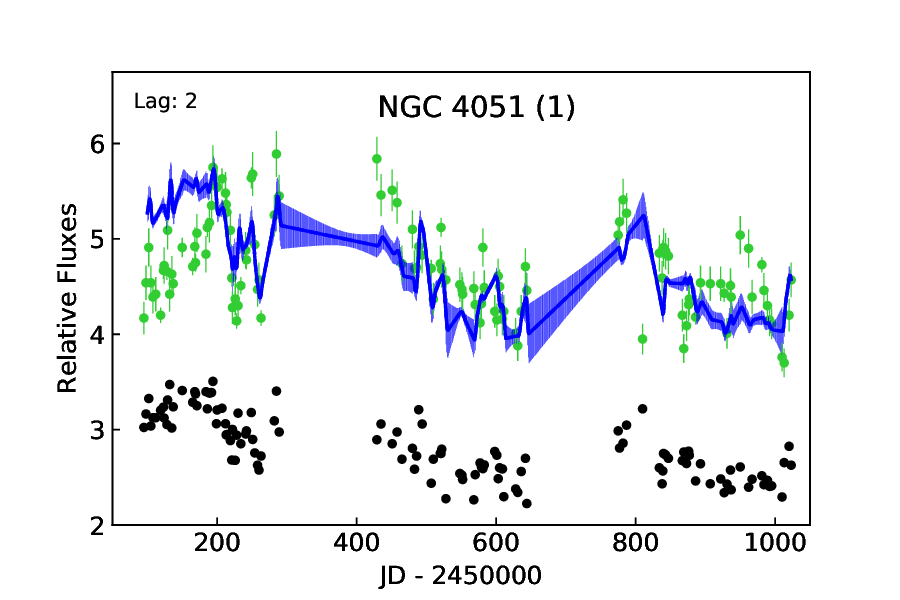}}\quad
   \subfloat{\includegraphics[width=.45\textwidth]{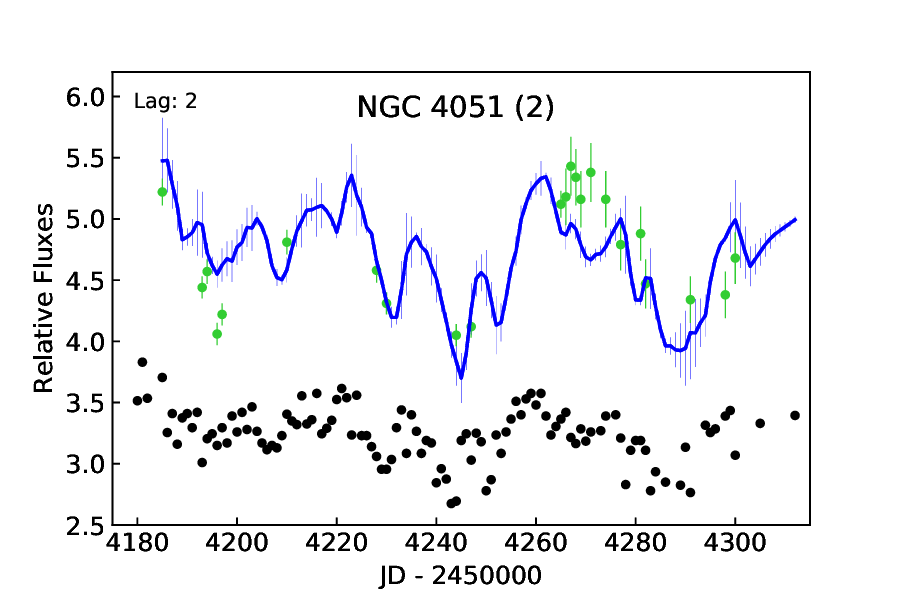}}\quad   
   \subfloat{\includegraphics[width=.45\textwidth]{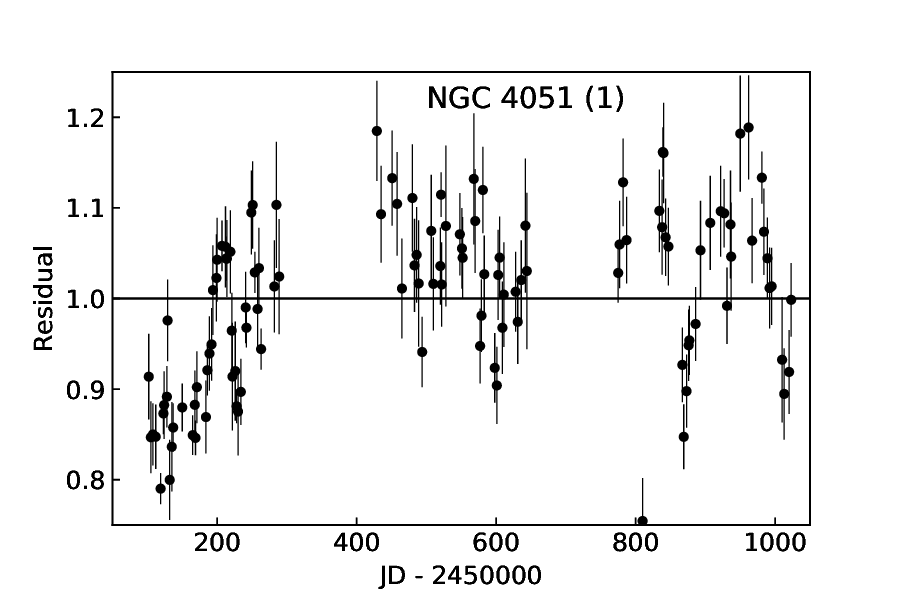}}\quad
   \subfloat{\includegraphics[width=.45\textwidth]{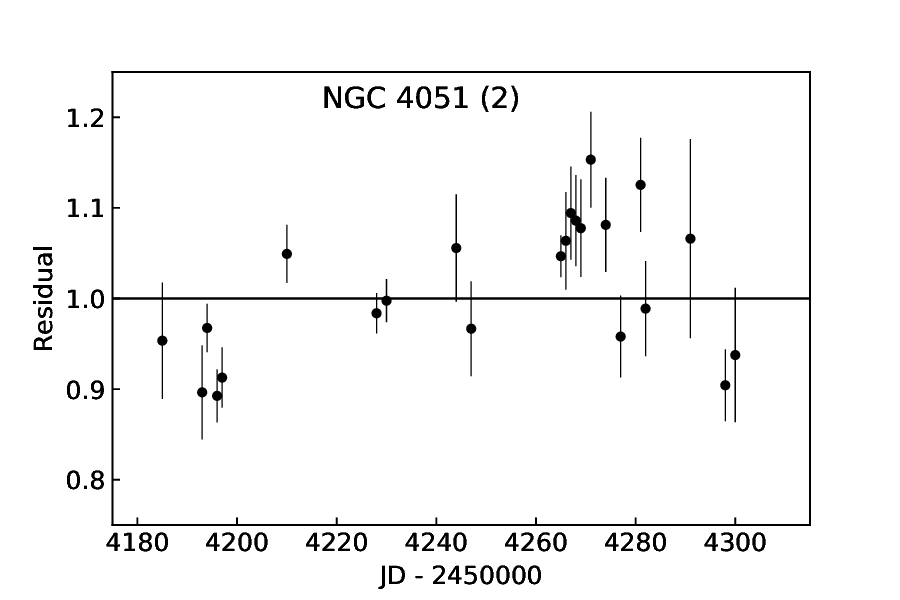}}\quad   
   \caption{-- Continued.}
\end{figure*}

\newpage
\clearpage  
\begin{figure*} % Figure 7 - Part 7
   \setcounter{figure}{6}
   \centering
   \subfloat{\includegraphics[width=.45\textwidth]{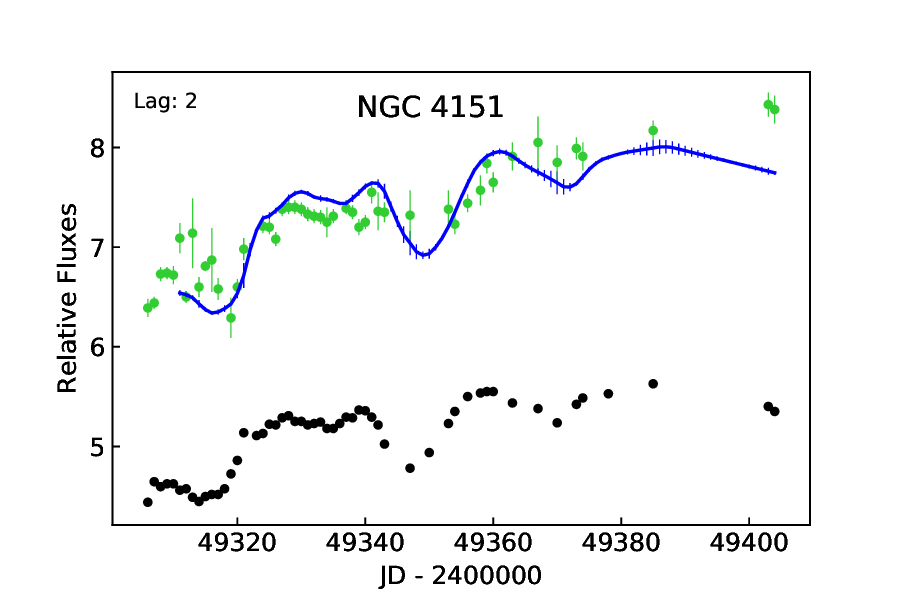}}\quad  
   \subfloat{\includegraphics[width=.45\textwidth]{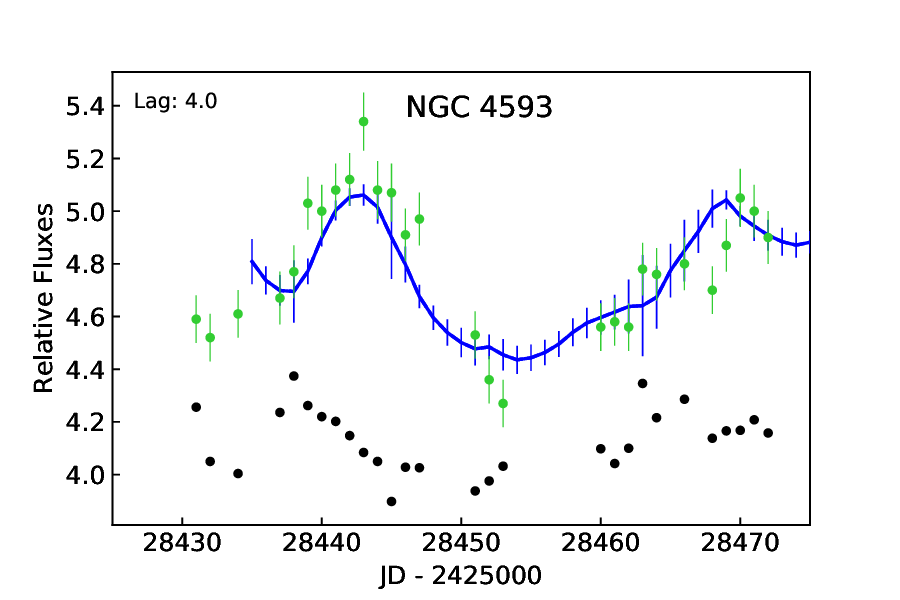}}\quad 
   \subfloat{\includegraphics[width=.45\textwidth]{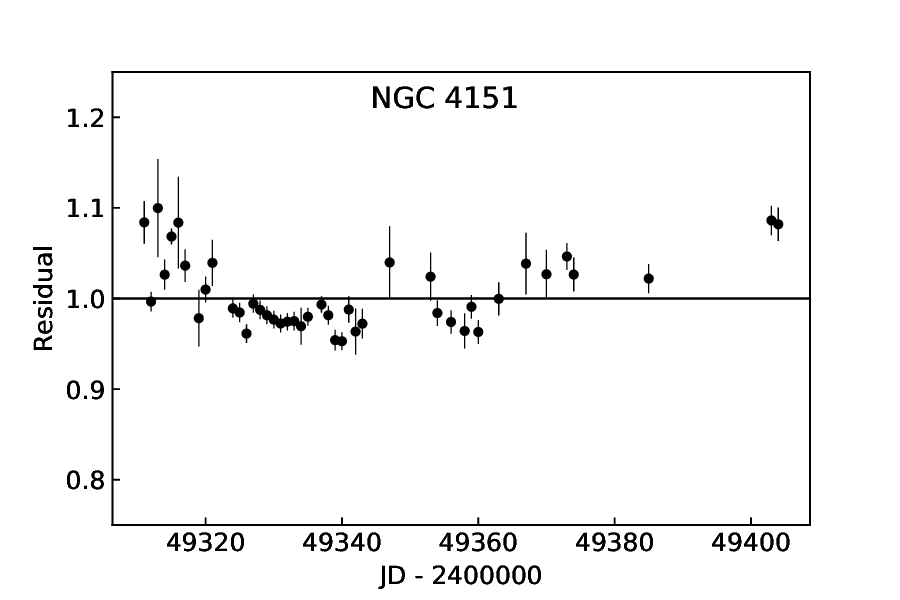}}\quad
   \subfloat{\includegraphics[width=.45\textwidth]{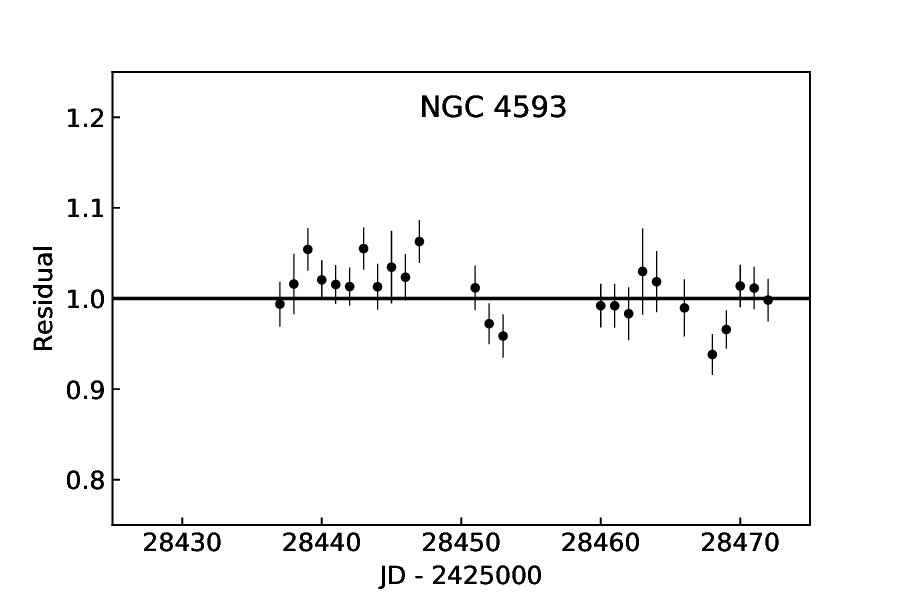}}\quad
   \subfloat{\includegraphics[width=.45\textwidth]{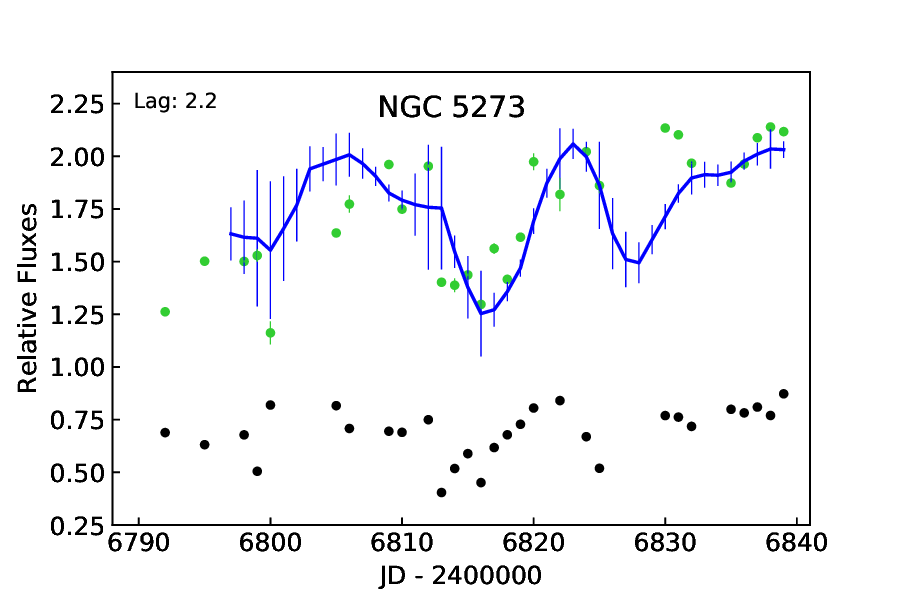}}\quad 
   \subfloat{\includegraphics[width=.45\textwidth]{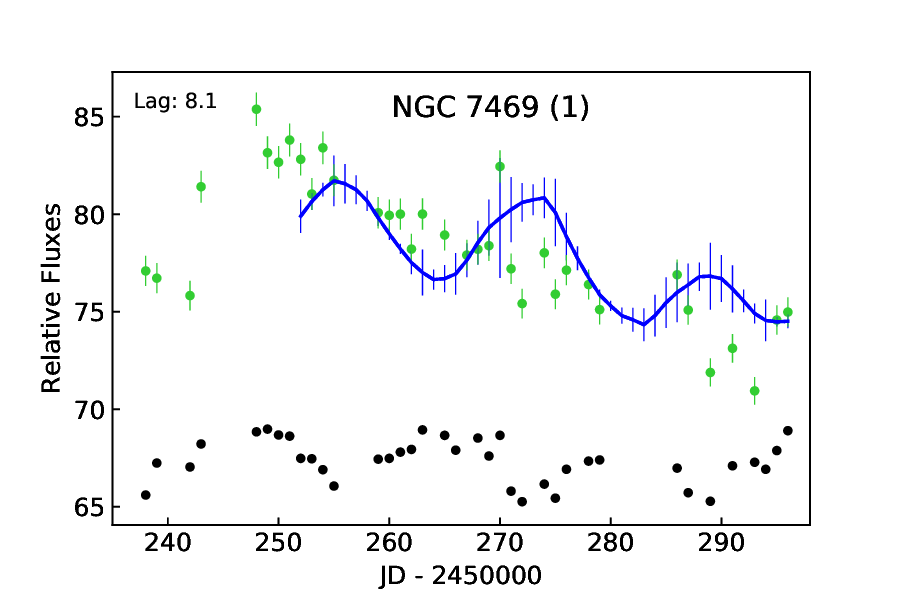}}\quad   
   \subfloat{\includegraphics[width=.45\textwidth]{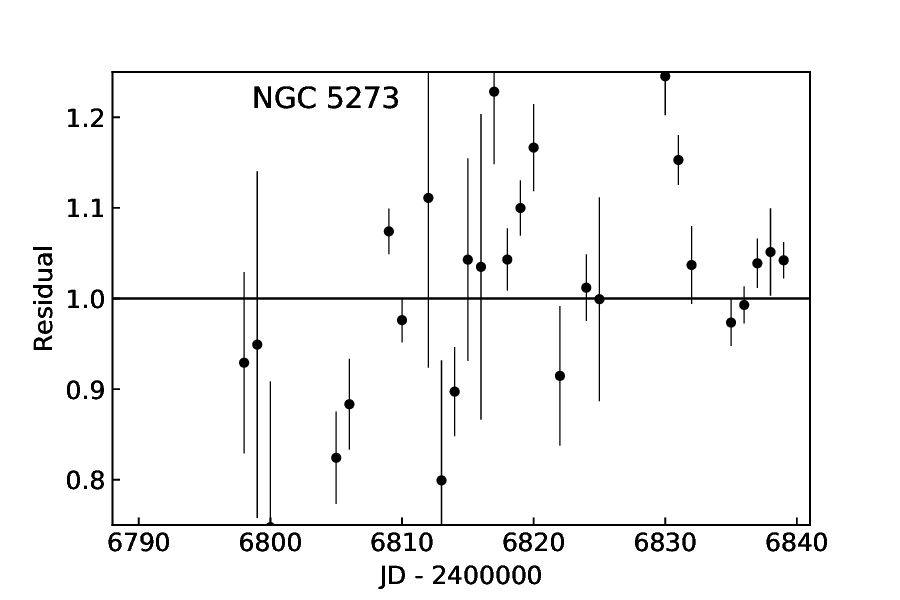}}\quad
   \subfloat{\includegraphics[width=.45\textwidth]{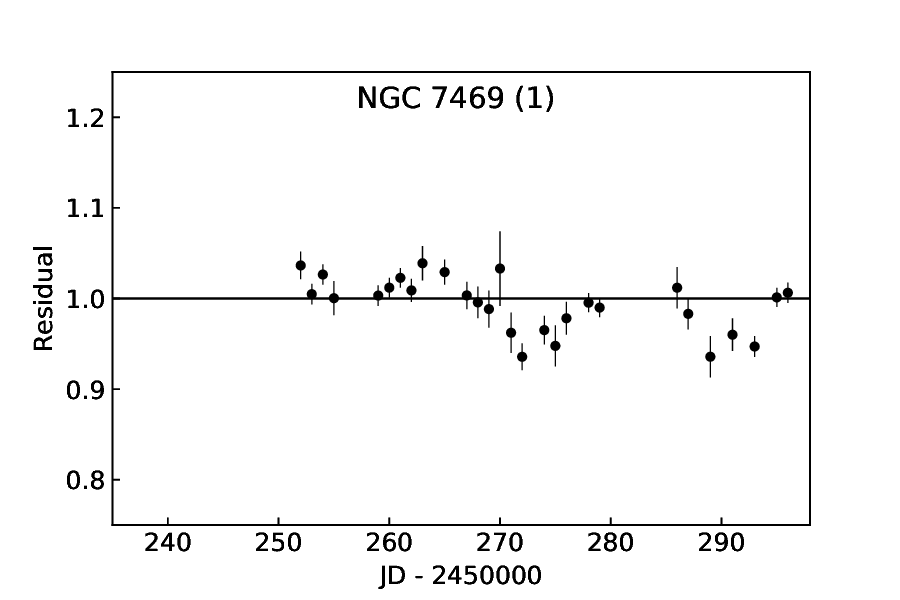}}\quad
\caption{-- Continued.}
\end{figure*}  

\newpage
\clearpage   
\begin{figure*} % FIgure 7 - Part 8
   \setcounter{figure}{6}
   \centering
   \subfloat{\includegraphics[width=.45\textwidth]{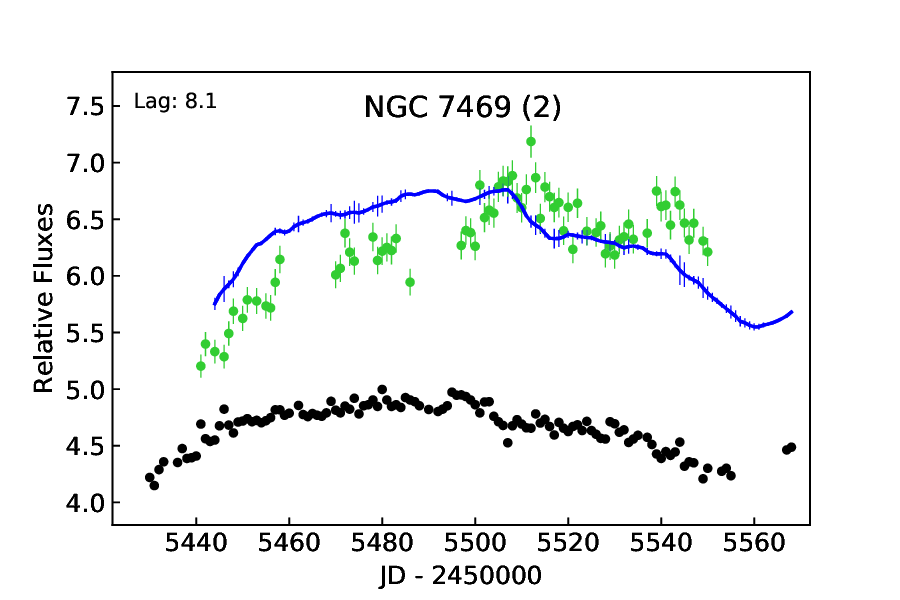}}\quad
   \subfloat{\includegraphics[width=.45\textwidth]{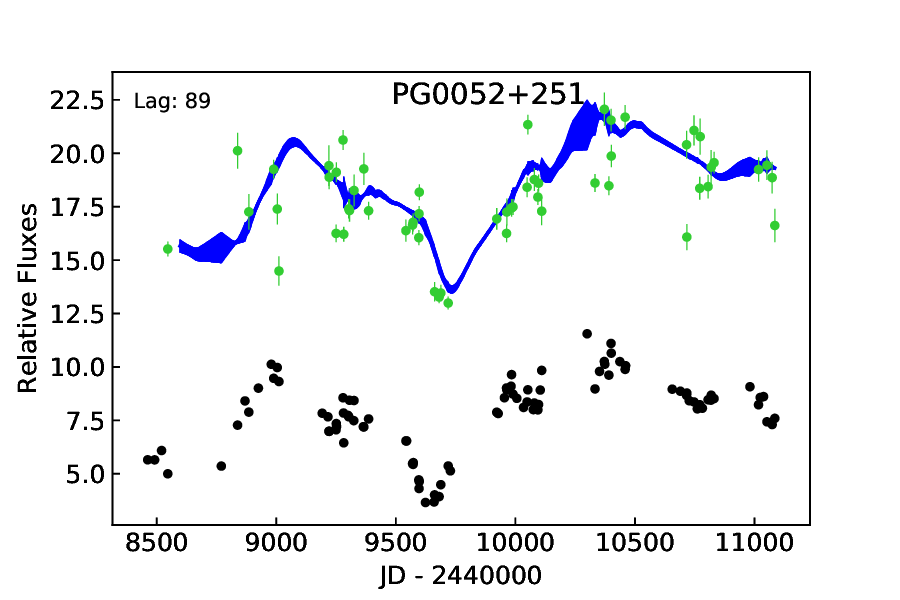}}\quad
   \subfloat{\includegraphics[width=.45\textwidth]{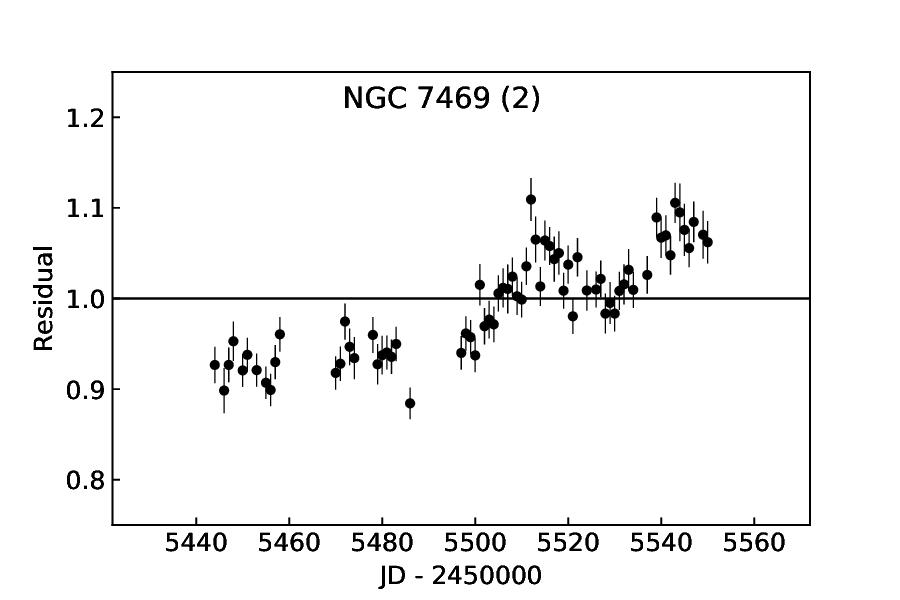}}\quad
   \subfloat{\includegraphics[width=.45\textwidth]{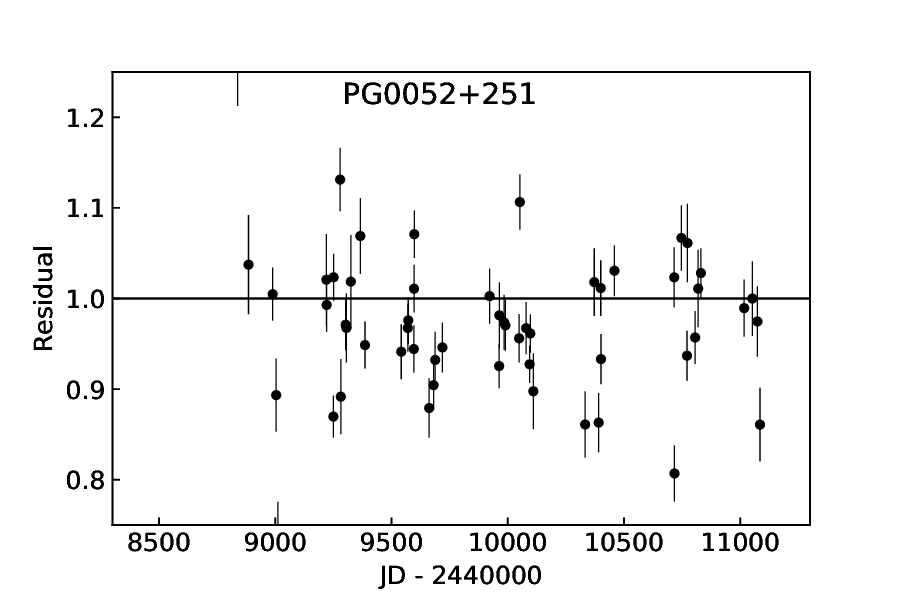}}\quad
   \subfloat{\includegraphics[width=.45\textwidth]{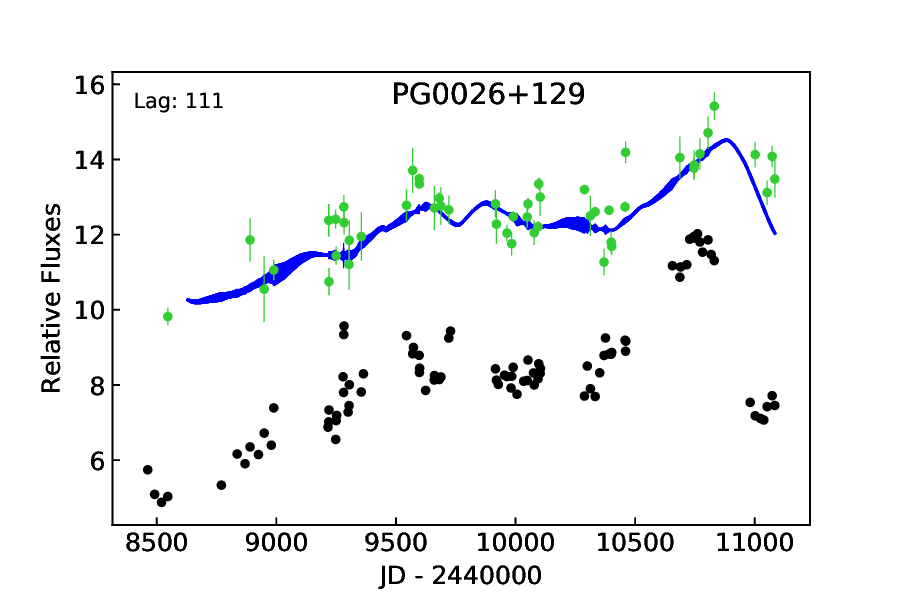}}\quad
   \subfloat{\includegraphics[width=.45\textwidth]{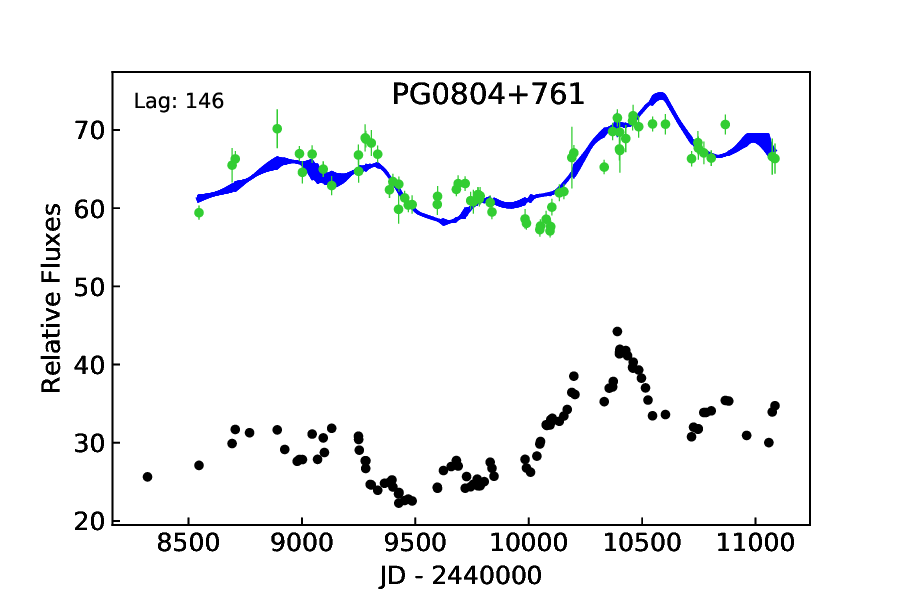}}\quad
   \subfloat{\includegraphics[width=.45\textwidth]{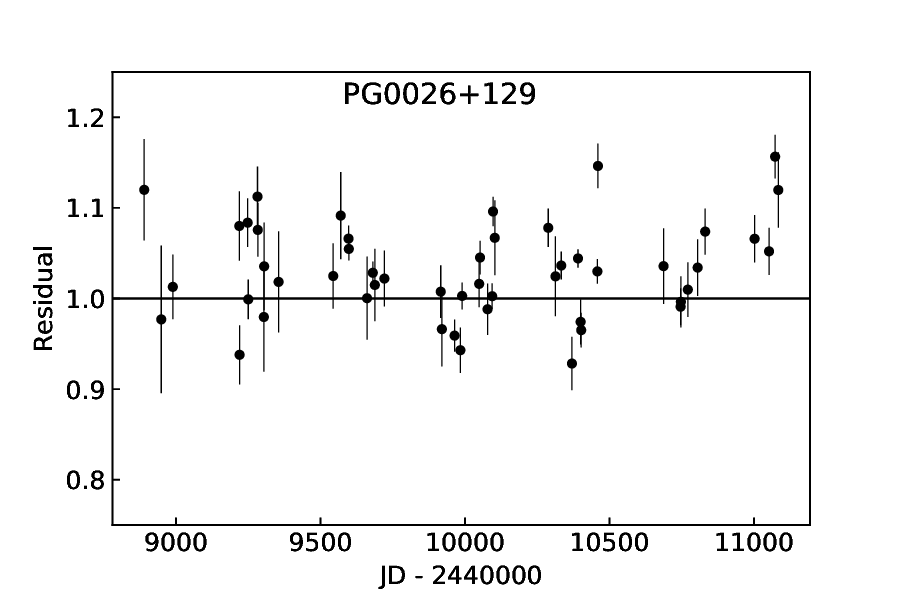}}\quad
   \subfloat{\includegraphics[width=.45\textwidth]{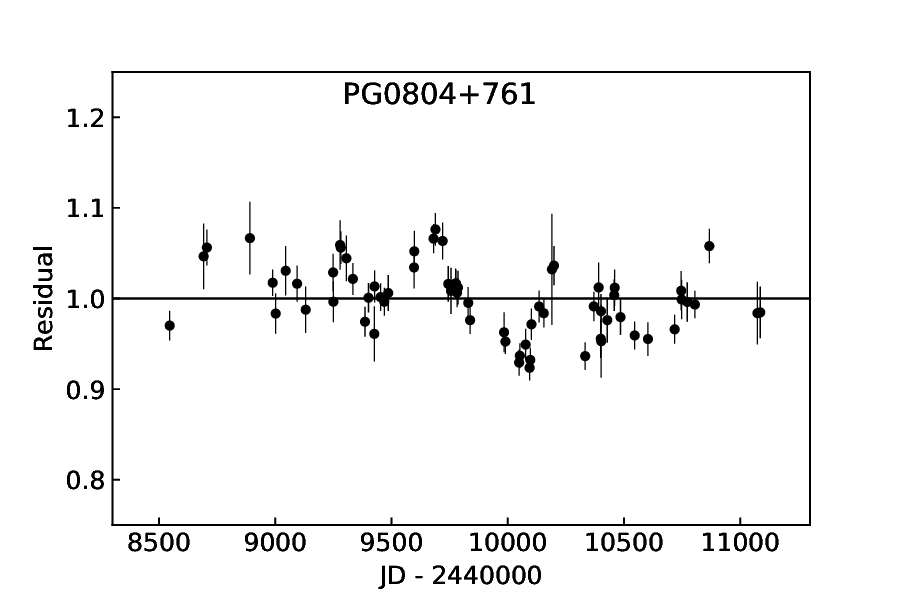}}\quad
   \caption{-- Continued.}
\end{figure*}  
  
\newpage
\clearpage   
\begin{figure*} % Figure 7 - Part 9
   \setcounter{figure}{6}
   \centering
   \subfloat{\includegraphics[width=.45\textwidth]{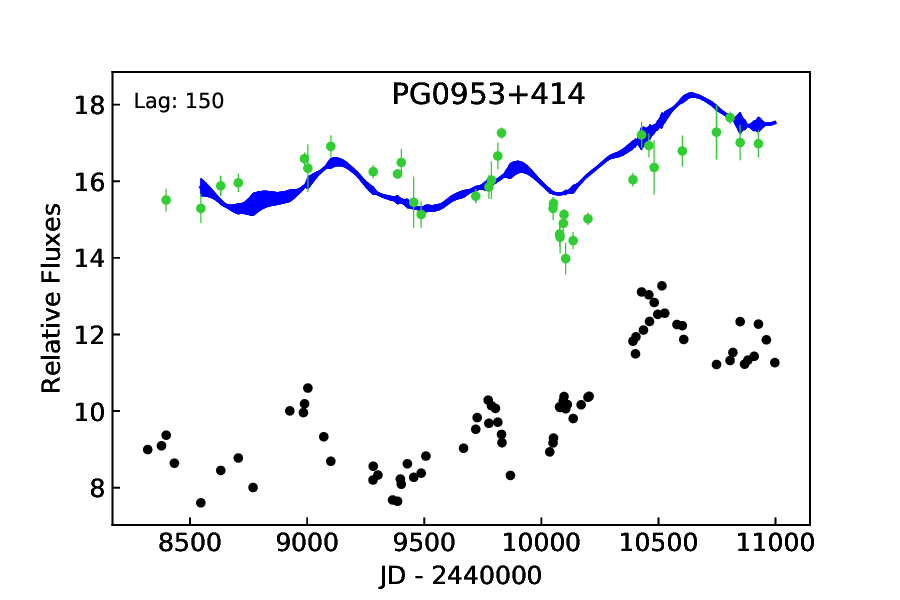}}\quad
   \subfloat{\includegraphics[width=.45\textwidth]{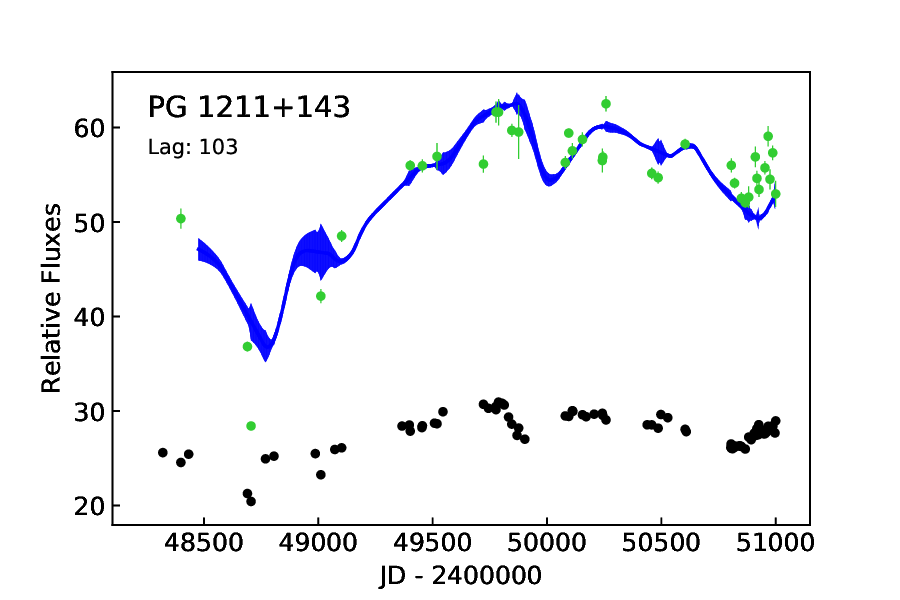}}\quad   
   \subfloat{\includegraphics[width=.45\textwidth]{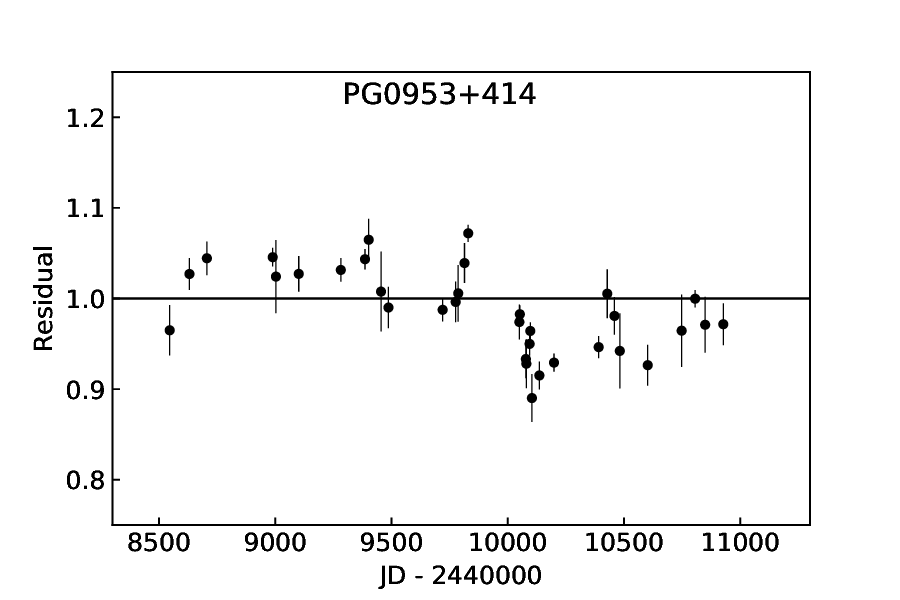}}\quad
   \subfloat{\includegraphics[width=.45\textwidth]{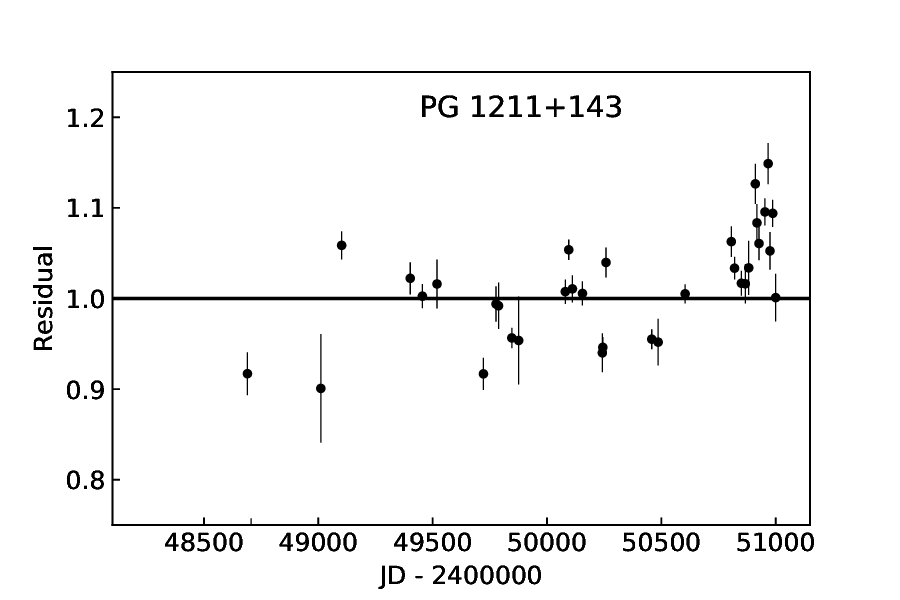}}\quad
   \subfloat{\includegraphics[width=.45\textwidth]{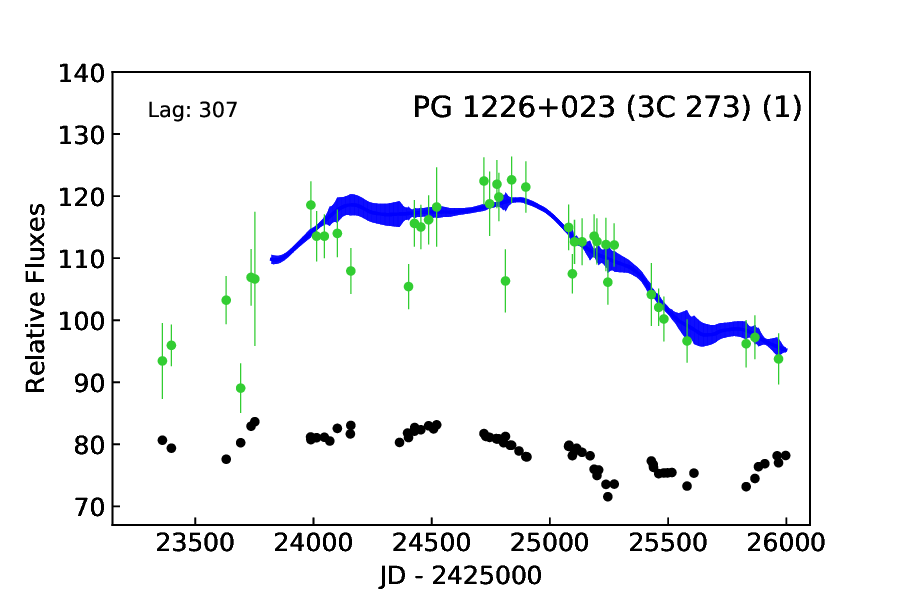}}\quad
   \subfloat{\includegraphics[width=.45\textwidth]{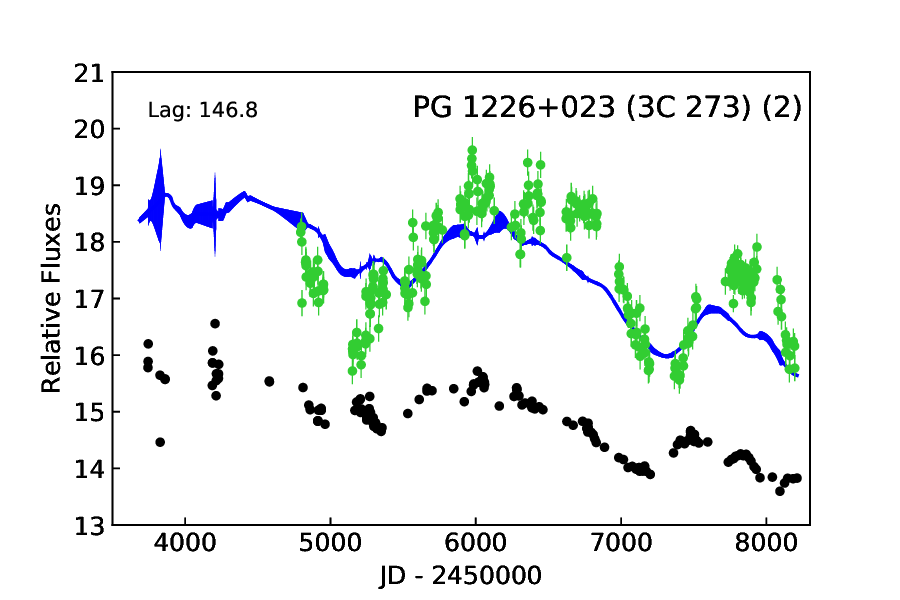}}\quad
   \subfloat{\includegraphics[width=.45\textwidth]{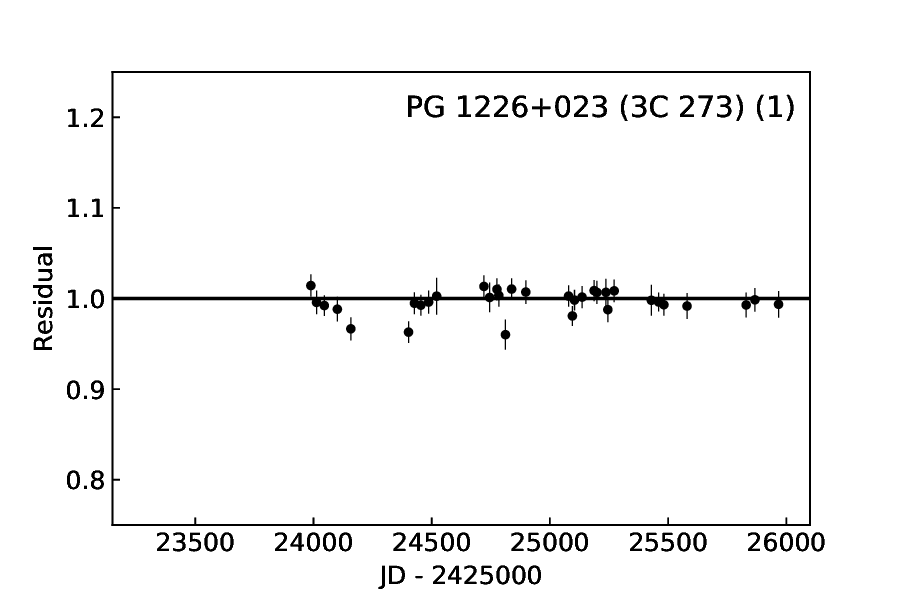}}\quad
   \subfloat{\includegraphics[width=.45\textwidth]{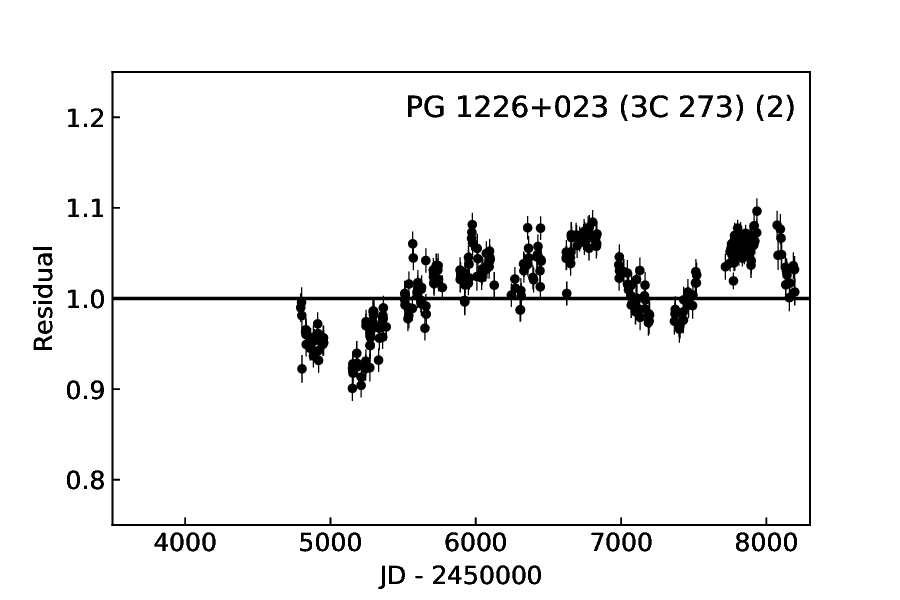}}\quad   
   \caption{-- Continued.}
\end{figure*}   
  
\newpage
\clearpage
\begin{figure*} % Figure 7 - Part 10
   \setcounter{figure}{6}
   \centering
   \subfloat{\includegraphics[width=.45\textwidth]{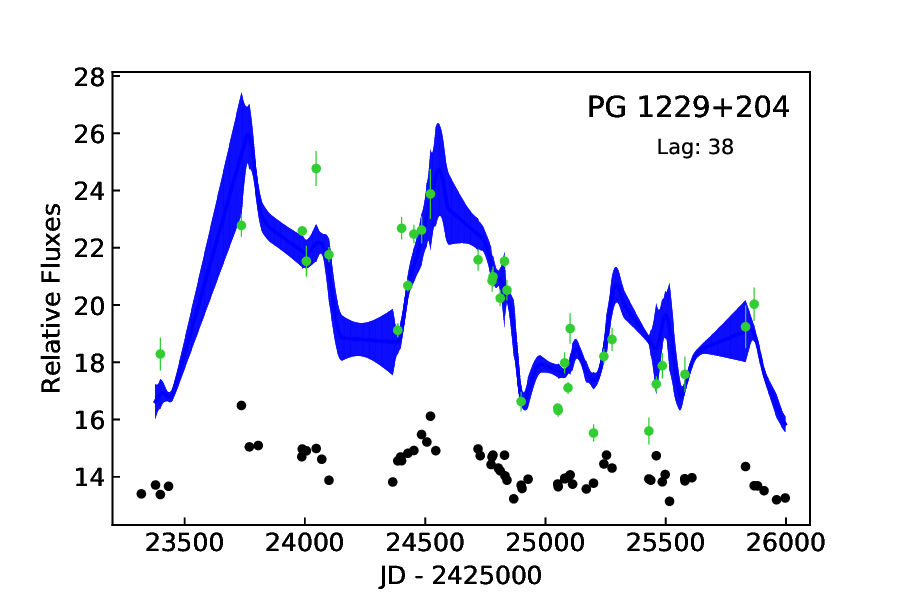}}\quad
   \subfloat{\includegraphics[width=.45\textwidth]{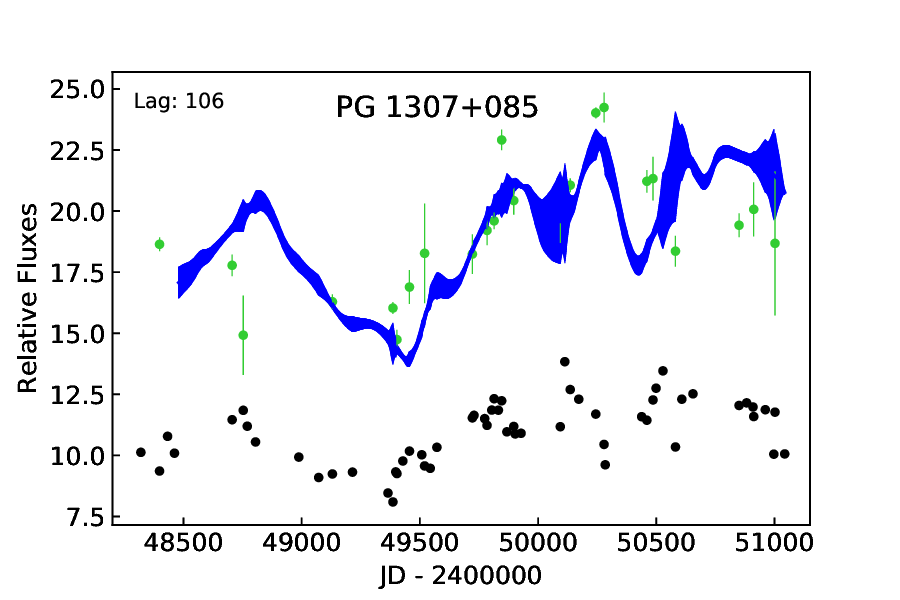}}\quad   
   \subfloat{\includegraphics[width=.45\textwidth]{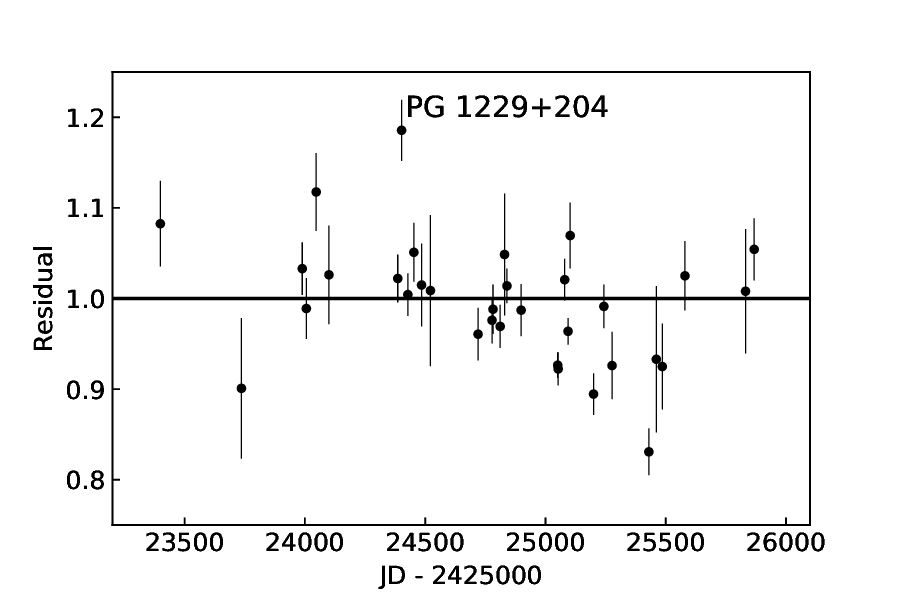}}\quad 
   \subfloat{\includegraphics[width=.45\textwidth]{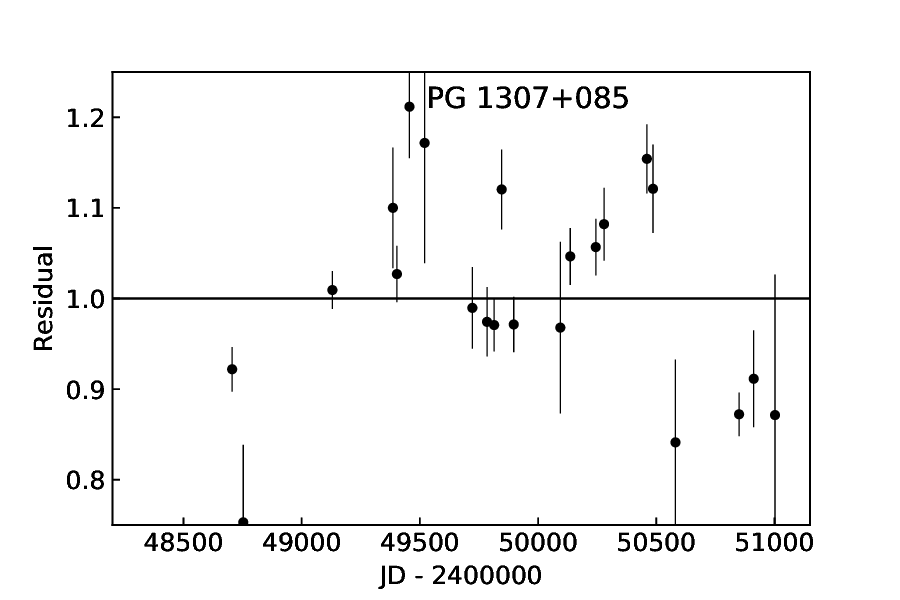}}\quad
   \subfloat{\includegraphics[width=.45\textwidth]{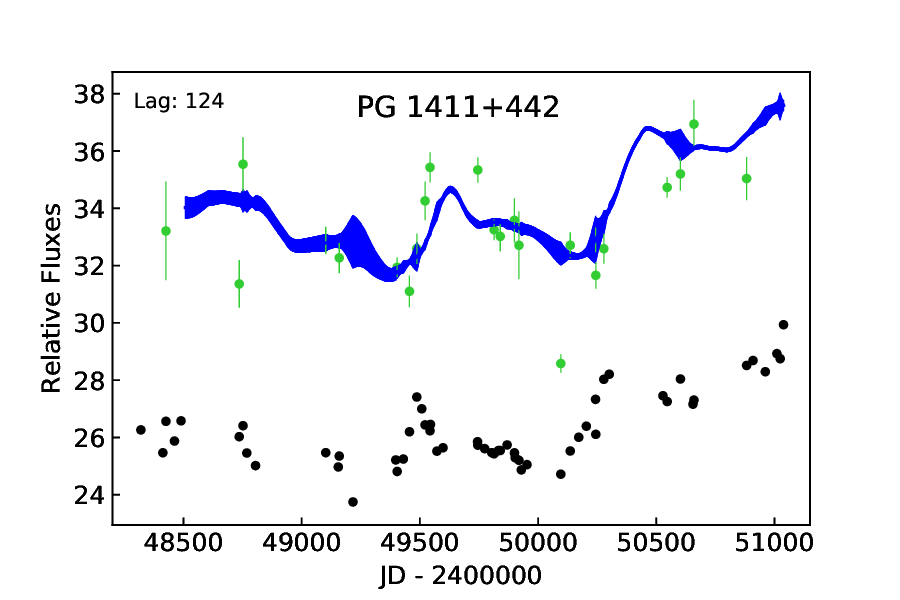}}\quad
   \subfloat{\includegraphics[width=.45\textwidth]{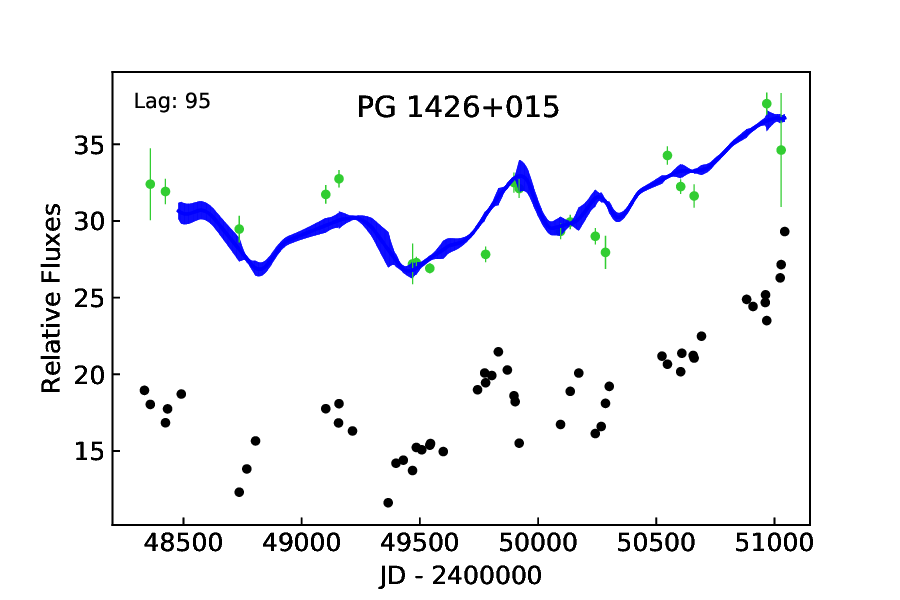}}\quad
   \subfloat{\includegraphics[width=.45\textwidth]{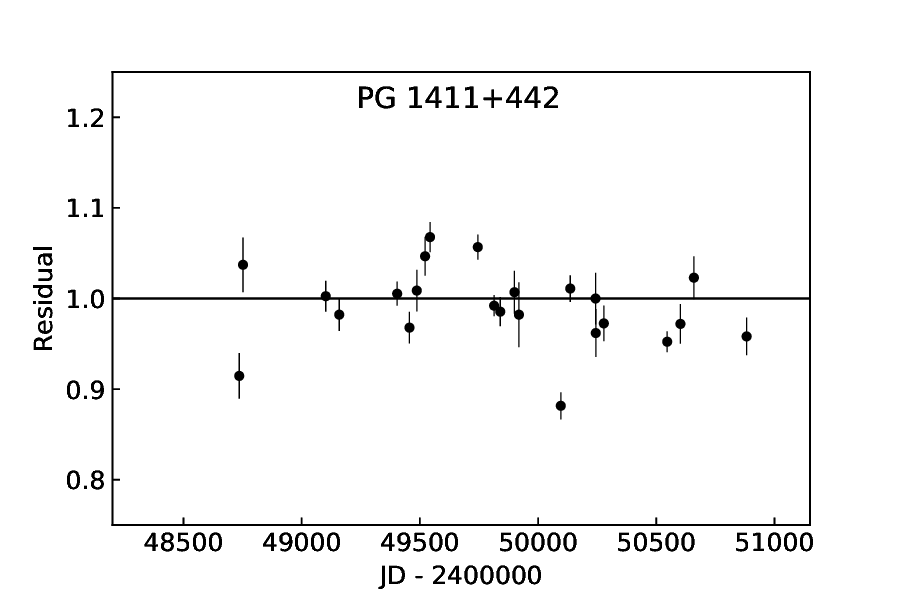}}\quad
   \subfloat{\includegraphics[width=.45\textwidth]{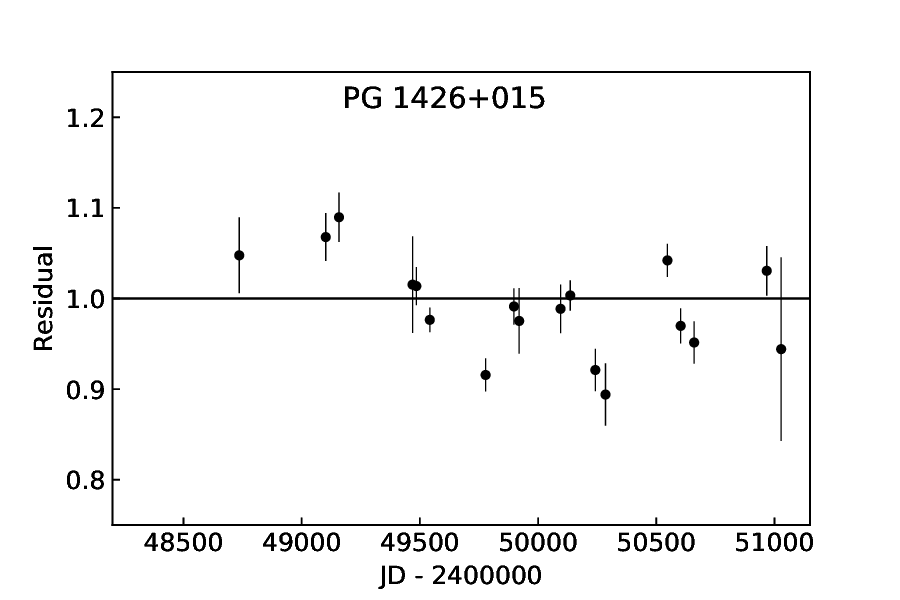}}\quad
   \caption{-- Continued.}
\end{figure*}

\newpage
\clearpage 
\begin{figure*}  % Figure 7 - Part 11
   \setcounter{figure}{6}
   \centering
   \subfloat{\includegraphics[width=.45\textwidth]{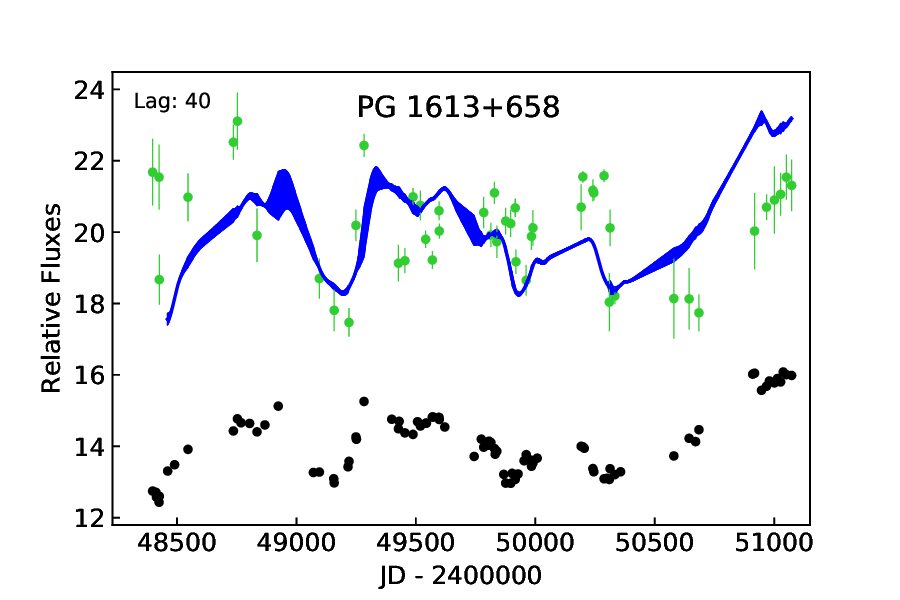}}\quad
   \subfloat{\includegraphics[width=.45\textwidth]{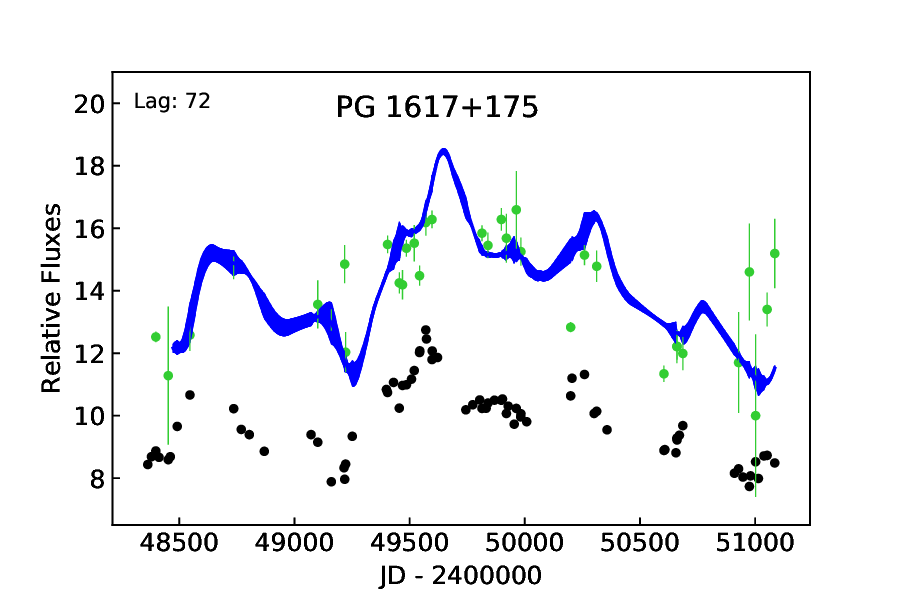}}\quad   
   \subfloat{\includegraphics[width=.45\textwidth]{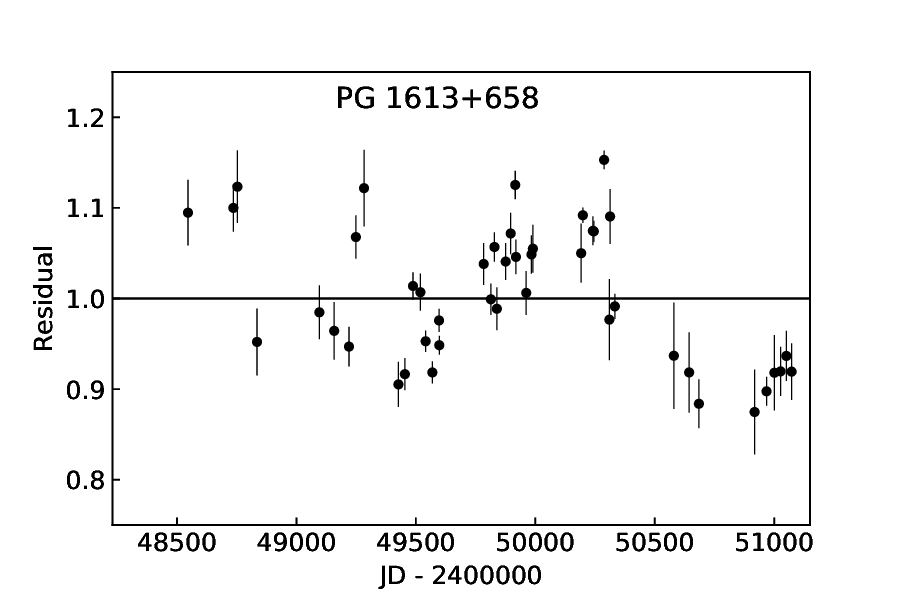}}\quad
   \subfloat{\includegraphics[width=.45\textwidth]{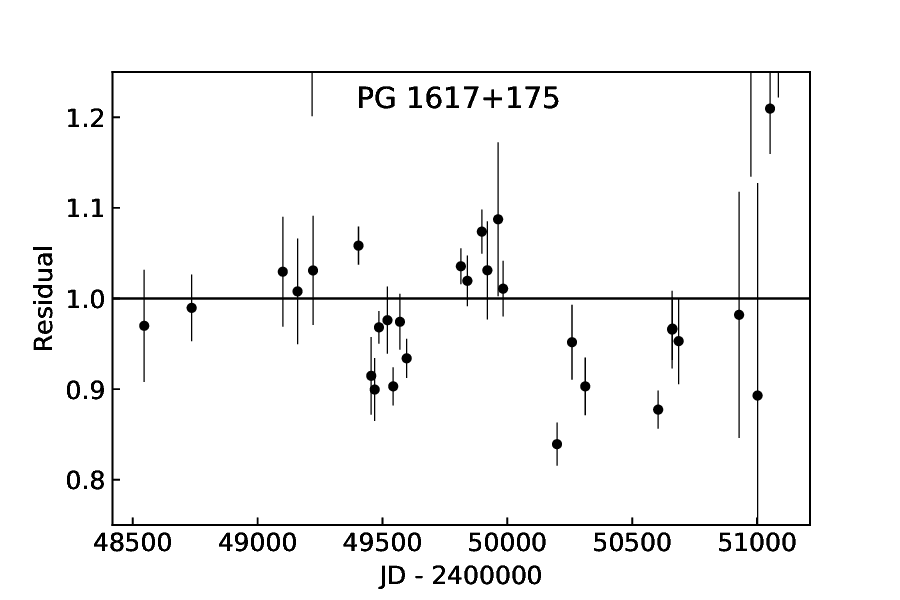}}\quad
   \subfloat{\includegraphics[width=.45\textwidth]{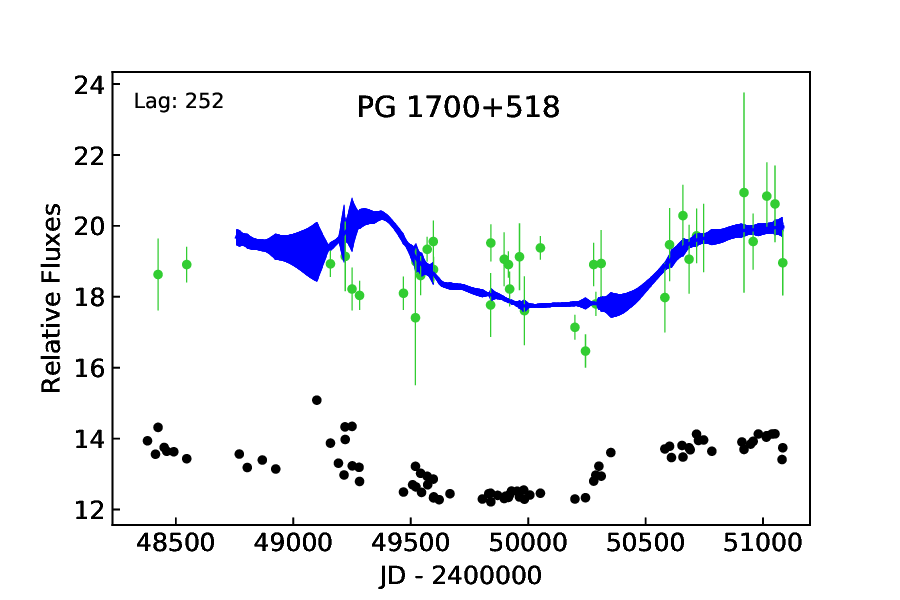}}\quad  
   \subfloat{\includegraphics[width=.45\textwidth]{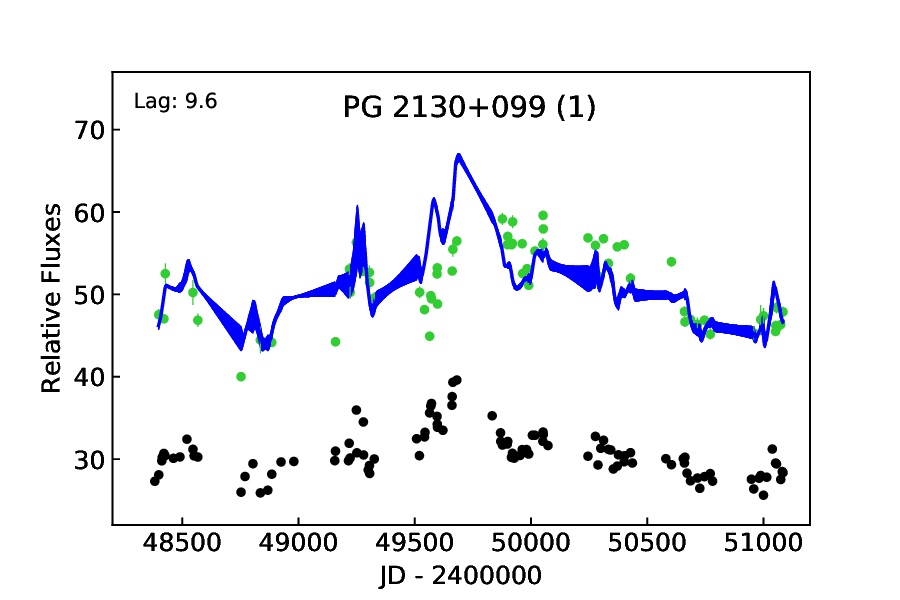}}\quad   
   \subfloat{\includegraphics[width=.45\textwidth]{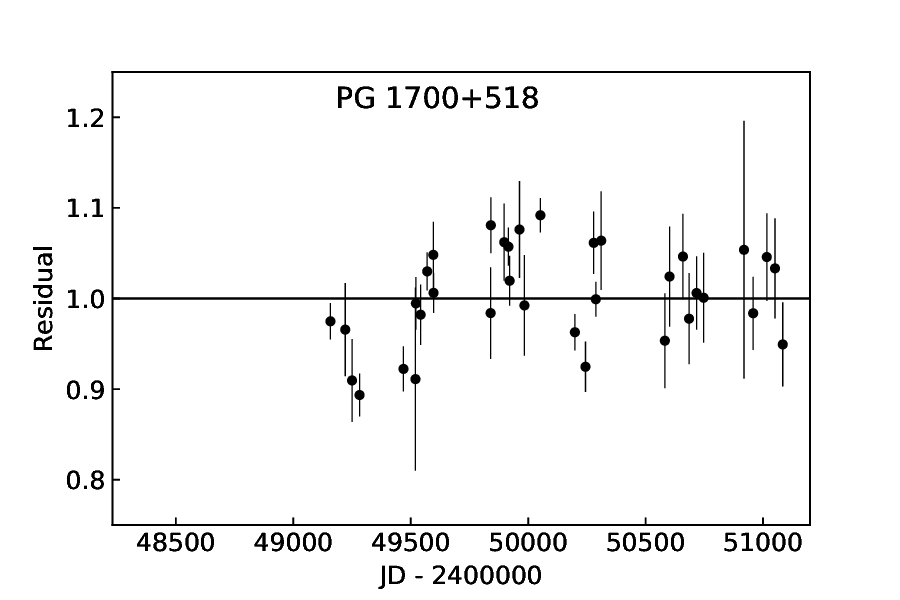}}\quad
   \subfloat{\includegraphics[width=.45\textwidth]{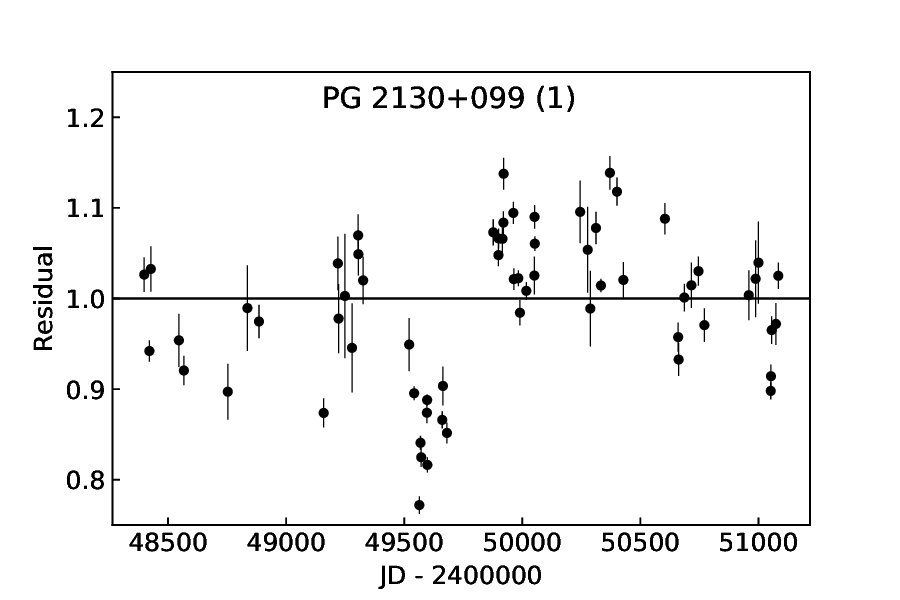}}\quad 
   \caption{-- Continued.}
\end{figure*}

\newpage
\clearpage    
\begin{figure*}  % Figure 7 - Part 12
   \setcounter{figure}{6}
   \centering
   \subfloat{\includegraphics[width=.45\textwidth]{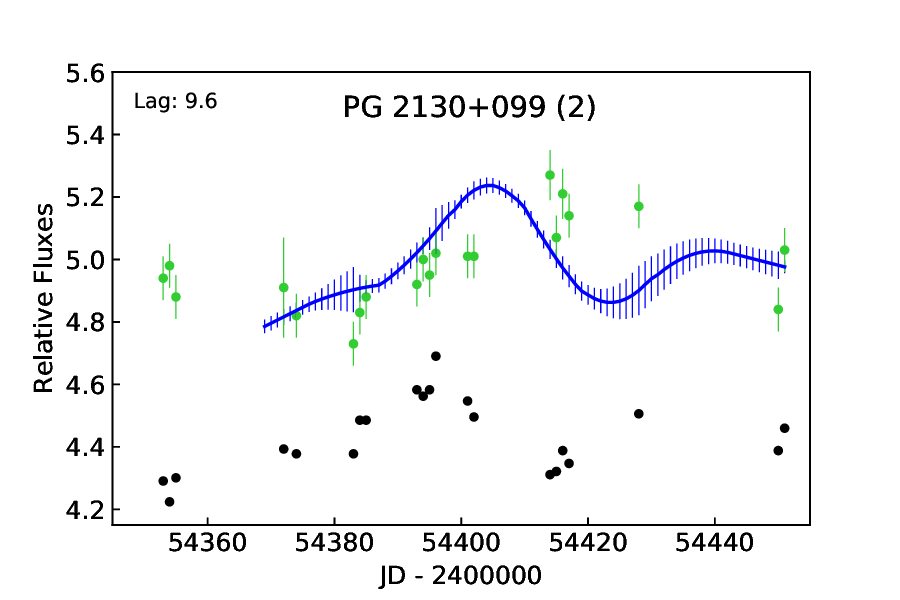}}\quad
   \subfloat{\includegraphics[width=.45\textwidth]{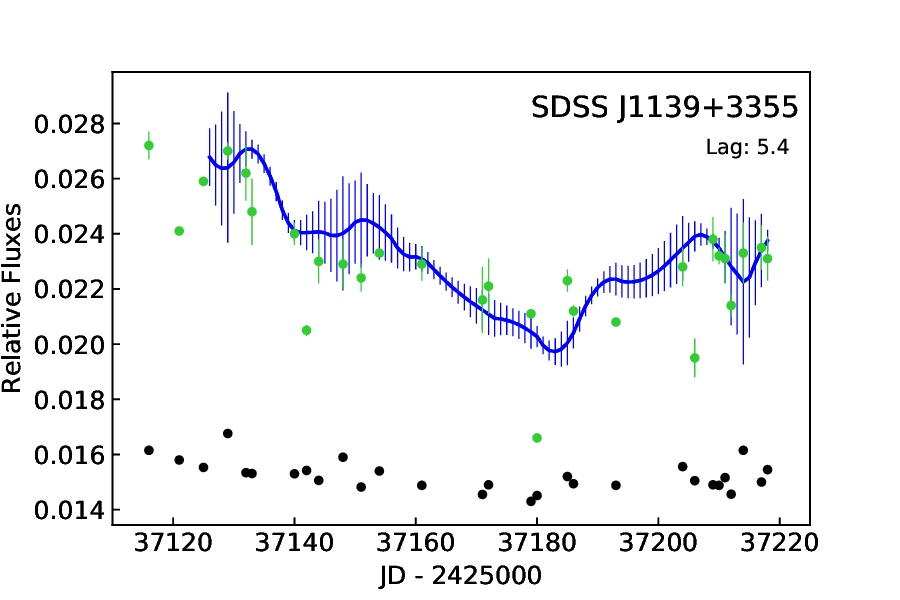}}\quad   
   \subfloat{\includegraphics[width=.45\textwidth]{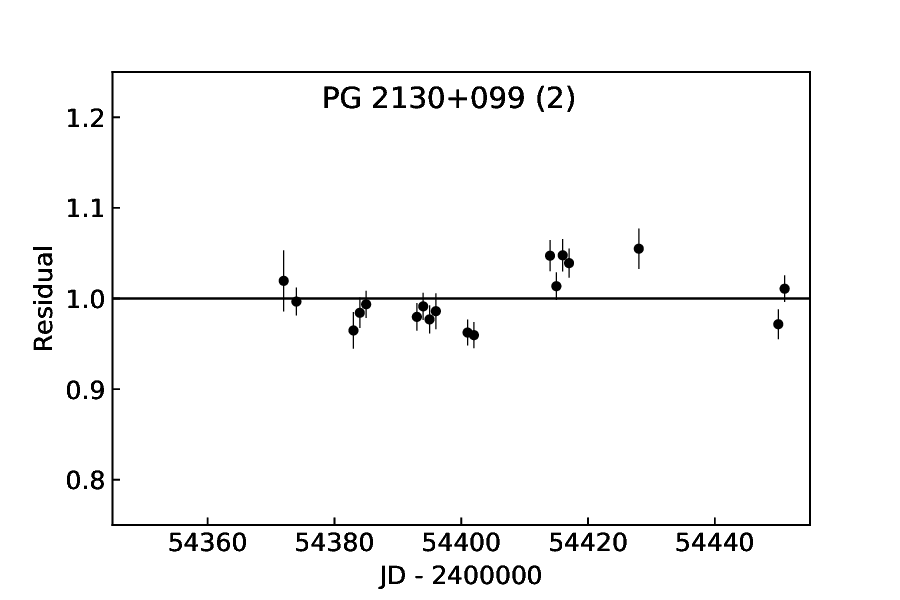}}\quad
   \subfloat{\includegraphics[width=.45\textwidth]{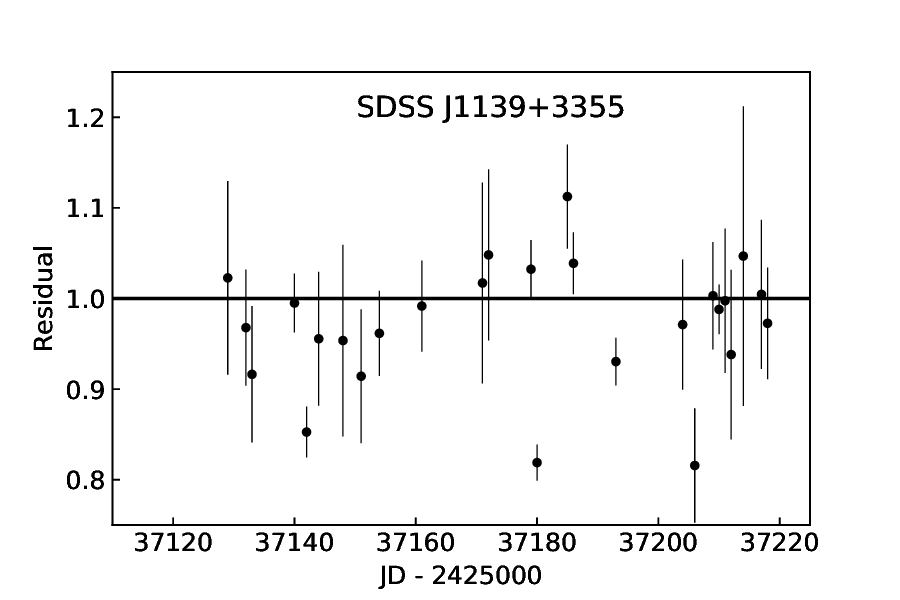}}\quad   
   \subfloat{\includegraphics[width=.45\textwidth]{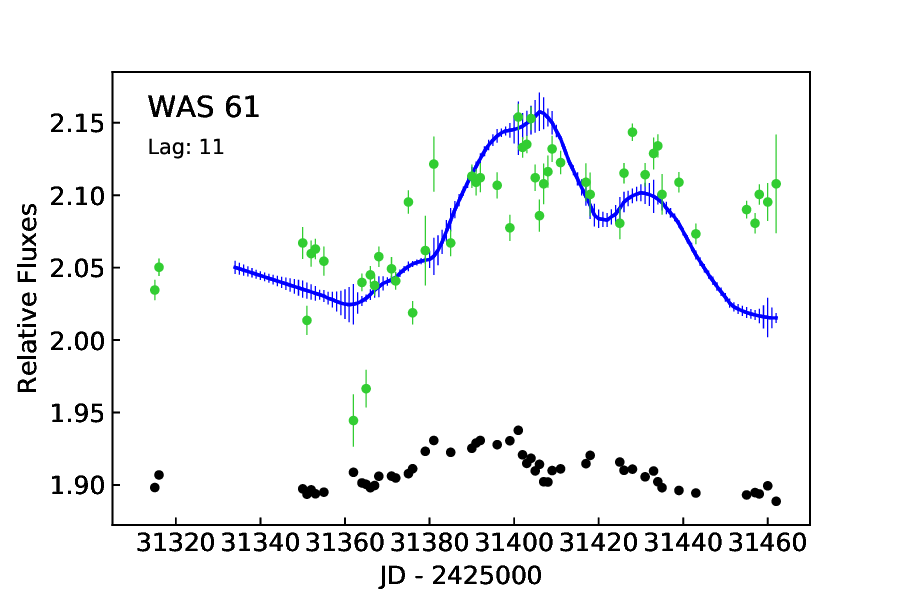}}\quad
    \subfloat{\includegraphics[width=.45\textwidth]{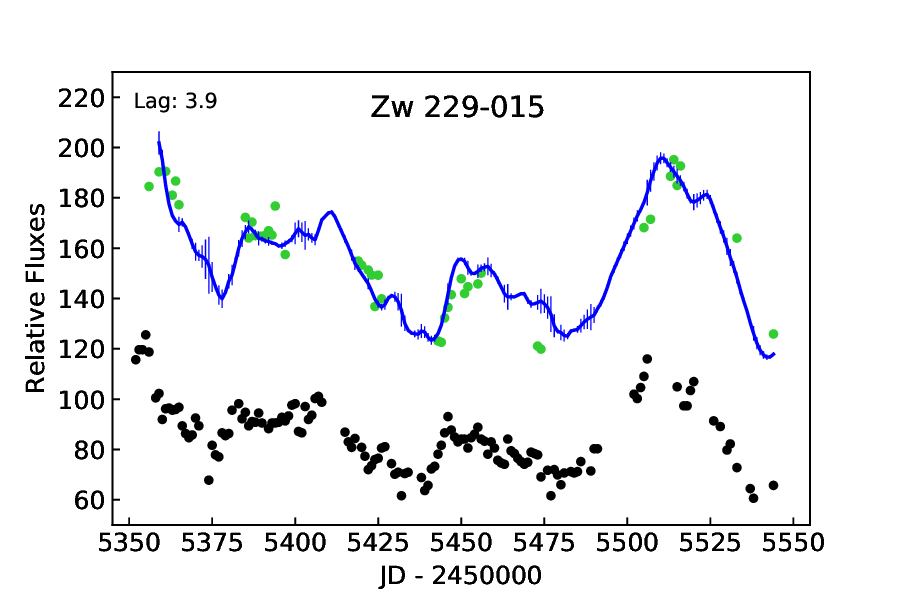}}\quad
    \subfloat{\includegraphics[width=.45\textwidth]{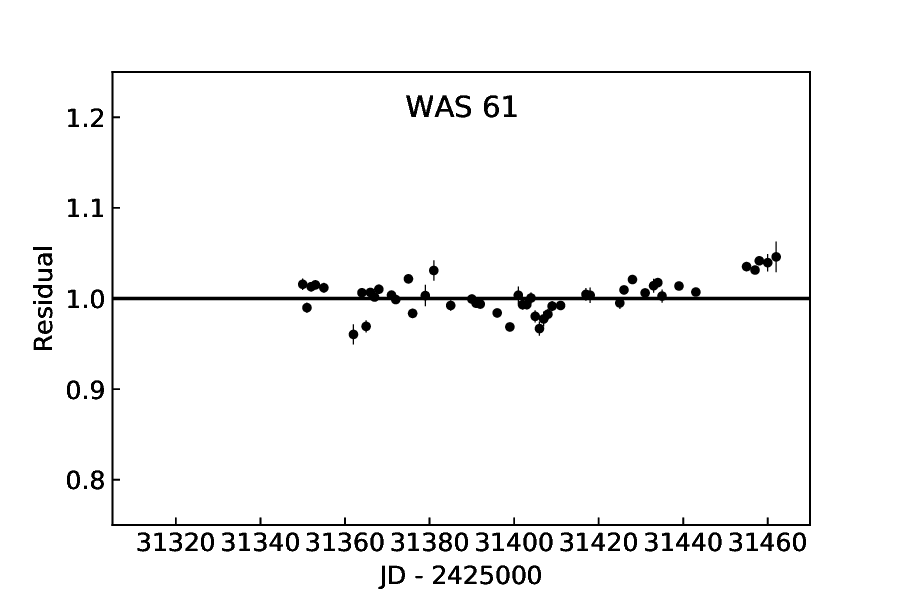}}\quad
    \subfloat{\includegraphics[width=.45\textwidth]{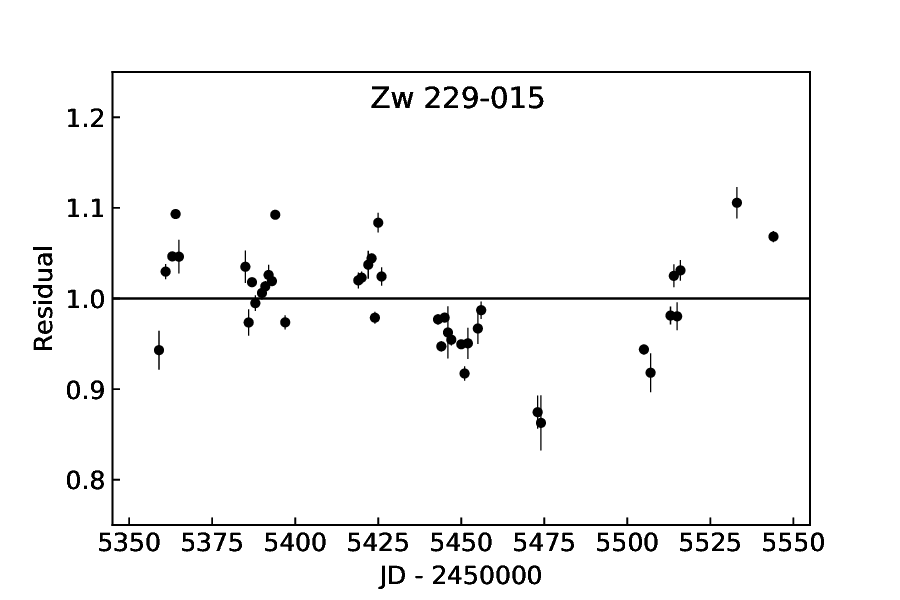}}\quad
   \caption{-- Concluded.}
\end{figure*}

%\appendix
%\section{~}
%\subsection{Notes on individual objects}

\end{document}